\documentclass{aa}  
\usepackage{graphicx}
\usepackage{xcolor}
\usepackage{scalefnt}

\usepackage{txfonts,textcomp}
\usepackage{hyperref}
\newbox\grsign \setbox\grsign=\hbox{$>$} \newdimen\grdimen \grdimen=\ht\grsign
\newbox\simlessbox \newbox\simgreatbox
\setbox\simgreatbox=\hbox{\raise.5ex\hbox{$>$}\llap
     {\lower.5ex\hbox{$\sim$}}}\ht1=\grdimen\dp1=0pt
\setbox\simlessbox=\hbox{\raise.5ex\hbox{$<$}\llap
     {\lower.5ex\hbox{$\sim$}}}\ht2=\grdimen\dp2=0pt
\def\simgreat{\mathrel{\copy\simgreatbox}}
\def\simless{\mathrel{\copy\simlessbox}}
\newbox\simppropto
\setbox\simppropto=\hbox{\raise.5ex\hbox{$\sim$}\llap
     {\lower.5ex\hbox{$\propto$}}}\ht2=\grdimen\dp2=0pt

\begin{document}

\title{\fontsize{14.3}{10}\selectfont{Abundances in 78 metal-rich bulge spheroid stars from APOGEE}}
\titlerunning{Abundances in 78 metal-rich bulge spheroid stars from APOGEE}
 \authorrunning{H. Ernandes et al.}
 
\author{
H. Ernandes \inst{1,2}
\and B. Barbuy \inst{3} 
\and C. Chiappini\inst{4}
\and S. Feltzing\inst{1}
\and A. P\'erez-Villegas\inst{5}
\and A.C.S. Fria\c ca\inst{3} 
 \and S. O. Souza \inst{6} 
\and R.P. Nunes\inst{3} 
\and A.B.A. Queiroz\inst{7,8}
\and J. G. Fern\'andez-Trincado \inst{9,10}  
\and A.L. Rocha de Abreu\inst{3} 
\and A. Plotnikova\inst{1} 
}

\institute{
Lund Observatory, Department of Geology, Lund University, Sölvegatan 12, Lund, Sweden 
\and
Nicolaus Copernicus Astronomical Center, Polish Academy of Sciences, ul. Bartycka 18, 00-716 Warsaw, Poland
\and 
Universidade de S\~ao Paulo,  IAG, Departamento de Astronomia, 05508-090 S\~ao Paulo, Brazil
\and 
Astrophysikalisches Institut Potsdam, An der Sternwarte 16, Potsdam, 14482, Germany
\and
Instituto de Astronom\'ia, Universidad Nacional Aut\'onoma de M\'exico, A. P. 106, C.P. 22800, Ensenada, B. C., M\'exico
\and
Max Planck Institute for Astronomy, K\"onigstuhl 17, D-69117 Heidelberg, Germany
\and
Instituto de Astrof\'{\i}sica de Canarias, C/ Via Lactea s/n, 38205 La Laguna, Tenerife, Spain
\and
Departamento de Astrof\'isica, Universidad de La Laguna, E-38205 La Laguna, Tenerife, Spain
\and
Universidad Cat\'olica del Norte, N\'ucleo UCN en Arqueolog\'ia Gal\'actica - Inst. de Astronom\'ia, Av. Angamos 0610, Antofagasta, Chile
\and
Universidad Cat\'olica del Norte, Departamento de Ingenier\'ia de Sistemas y Computaci\'on, Av. Angamos 0610, Antofagasta, Chile
%
}
            
   \date{Received ....; accepted .....}
 
  \abstract
   {The inner Galaxy is the most complex region of the Milky Way, comprising the early bulge, 
  inner thin and thick discs, and inner halo stars; moreover, the formation of the bar caused transfer of
gas and stars from the disc to the inner Galaxy.
Moreover, accretion of dwarf galaxies took place along the Galaxy's lifetime, merging with the original bulge. In this work, we sought
to constrain the metal-rich stars of the earliest spheroidal bulge.}
{With the aim of studying the oldest bulge stars, which show a
distribution in a spheroid, we applied a selection based on kinematical and dynamical criteria, in the metal-rich range
[Fe/H] $> -$0.8. This analysis complements our previous work on
a symmetric sample with [Fe/H] $< -$0.8.}
  {We derived the individual abundances through spectral synthesis for the elements C, N, O, Al, P, S, K, Mn, and Ce using the stellar physical parameters available for our sample from Data Release 17 of the Apache Point Observatory Galactic Evolution Experiment (APOGEE DR17) project in the {\it H} band. We also compared the present results, together with literature data, with chemical-evolution models. }
   {The abundances of the alpha elements Mg Si, and Ca, and iron-peak elements V, Cr, Co, and Ni from APOGEE DR17 follow the expected behaviour as compared with the chemical-evolution models. Mn shows the expected secondary behaviour. S and K show a large star-to-star spread, but remain broadly compatible with the model predictions. Phosphorus and cerium display a clear abundance excess around [Fe/H]$\sim -$0.7 that is more pronounced than in our metal-poor sample, suggesting a distinctive chemical signature for the earliest bulge population. Diagnostic diagrams involving [Mg/Mn] versus [Al/Fe] and [Ni/Fe] versus [(C+N)/O] indicate an in situ origin of the bulk of the sample. At super-solar metallicities, a subset of stars shows enhanced K and Mn (and possibly S) together with low [Ce/Fe] ratios, hinting at enrichment processes linked to the nuclear disc and bar. These stars may therefore trace a chemically distinct population shaped by the unique dynamical and star formation conditions of the innermost Galaxy.}
  {}
   \keywords{Galaxy: bulge  -- Stars: abundances }
   \maketitle

\section{Introduction}

The Galactic bulge plays a fundamental role in our understanding of the formation and early evolution of the Milky Way, as it is expected to host a large fraction of the Galaxy’s oldest stellar populations. Its central location, high stellar density, and wide metallicity distribution make the bulge a key laboratory for testing early chemical enrichment and assembly scenarios. Indeed, metallicities of the bulk of the bulge stellar populations span a broad range of approximately $-$1.5 $<$ [Fe/H] $<$ +0.5 \citep{mcwilliamrich94,barbuy18a,queiroz21}, reflecting a complex formation history involving rapid early star formation and subsequent dynamical evolution.

Strong age constraints further support the bulge as an ancient Galactic component. The bulk of bulge stars have an estimated age of about 11 Gyr \citep{clarkson08}, while the oldest component of, commonly associated with a spheroidal bulge, is inferred to be $\sim$13 Gyr. This is indicated by the ages of bulge globular clusters (GCs) with moderate metallicities \citep[see Table 3 in][]{souza24a}. These findings are consistent with early formation scenarios in which a significant fraction of the oldest stars in the Galaxy formed rapidly and are now concentrated in the central regions \citep[e.g.][]{gao10,tumlinson10}.

We note that at the metal-poor end of the metallicity distribution, the nature of stellar populations in the inner Galaxy becomes increasingly ambiguous. In particular, it is not always clear whether stars with [Fe/H] $< -$1.5 should be considered bona fide bulge members, part of a metal-poor inner halo, or remnants of accreted systems \citep[e.g.][]{howes16,geisler23,Ardern-Arentsen2024,nepal25}. Despite this ambiguity, metal-poor stars are of exceptional importance, as they are commonly interpreted as tracers of the earliest phases of star formation and chemical enrichment in the Milky Way. Disentangling their origin is therefore both challenging and crucial for reconstructing the bulge’s formation history.

Using a combination of photometric, kinematical, and dynamical diagnostics, \citet{queiroz21} identified distinct stellar populations in the inner Galaxy and proposed selection criteria aimed at isolating the spheroidal bulge component. Their analysis showed that these criteria are particularly effective at identifying moderately metal-poor bulge stars, which are often associated with the earliest stellar populations formed in the Milky Way. The identification of the spheroidal bulge is nevertheless difficult, especially given that the bulk of stars in \citet{queiroz21} were metal-rich. Building on the selection strategy introduced by \citet{queiroz21} and the calculation of stellar orbits, we carried out a series of studies focusing on the chemical properties of the most metal-poor spheroidal bulge candidates, with [Fe/H] $< -$0.8 \citep{razera22,barbuy23,barbuy24,barbuy25a}. The abundance patterns of these stars were interpreted in the context of rapid early chemical enrichment.

While metal-poor stars provide a direct window to the earliest bulge formation, extending the analysis to higher metallicities offers a complementary and necessary perspective. The metal-rich regime may retain chemical signatures of the early bulge while also reflecting later evolutionary processes, and thus it provides an open pathway to understanding the continuity and full metallicity distribution of the spheroidal bulge population. In the present work, we applied the same kinematical and dynamical selection criteria to explore this more metal-rich regime, targeting spheroidal bulge candidates with [Fe/H] $> -$0.8. This extension allowed us to investigate whether the chemical characteristics identified at lower metallicities persist into the metal-rich tail.

Recent studies have emphasised that particular care is required when interpreting metal-rich bulge samples. In particular, \citet{nepal25} identified a spheroidal bulge component with a metallicity distribution function peaking at [Fe/H] $\simeq -$0.7, but also cautioned that contamination from bar stars on hotter orbits cannot be entirely excluded at the metal-rich end. This highlights the importance of a detailed chemical analysis based on a broad set of elemental abundances, which can provide additional constraints on the nature and origin of these stars beyond kinematical selection alone.
Therefore, we selected stars from the reduced proper-motion (RPM) sample of \citet{queiroz20,queiroz21}, cross-matched with stars observed in the H band by the Apache Point Observatory Galactic Evolution Experiment (APOGEE; \citealt{majewski17}). We further required effective temperatures of T$_{\rm eff}$ $>$ 3900 K and additional quality control requirements on the spectra, in particular S/N$>$80. This selection resulted in a sample of 78 stars. For these stars, we analysed the spectra and derived chemical abundances, in particular revising the APOGEE abundances of C, N, O, Al, P, S, K, Mn, and Ce. The abundances of Mg, Si, Ca, V, Cr, Co, and Ni are adopted from the APOGEE Stellar Parameters and Chemical Abundances Pipeline (ASPCAP).

In Sect. \ref{sample} the sample is described. In  Sect. \ref{calculations} the analysis is presented. In Sect. \ref{results} the results are reported and chemical evolution models
are described. In Sect. \ref{discussion} the results are compared with chemical evolution models and discussed. Conclusions are drawn in Sect. \ref{conclusions}.

\section{Data and selection sample}\label{sample}

Our sample selection closely follows that described in \citet{razera22}, with the additional requirement that stars satisfy a metallicity cut of [Fe/H] $> -0.8$. The criteria adopted to define the sample are as follows:
\begin{enumerate}
    \item Azimuthal velocity, V$_{\phi} < 0$, selecting stars counter-rotating with respect to the bulge rotation.
    \item Non-bar-following orbits.
    \item Metallicity: [Fe/H] $> -0.8$.
    \item Apogalactic distance: Apo $< 4$~kpc.
    \item Orbital eccentricity: ecc $> 0.7$.
\end{enumerate}

The adopted cut in azimuthal velocity (V$_{\phi} < 0$) is an effective criterion 
to select stars counter-rotating with respect to the bulge. While stars with 
low orbital eccentricities (ecc $< 0.7$) may also belong to the spheroidal 
component, disc stars are not expected to preferentially populate the 
V$_{\phi} < 0$ regime. We therefore imposed an eccentricity cut of ecc $> 0.7$, 
acknowledging that this requirement introduces a selection bias. This choice is 
intended to minimise the inclusion of stars on resonant orbits with ambiguous dynamical origins.

The sample includes dynamical information from the studies by \citet{queiroz20,queiroz21}, which combined distances derived with the \texttt{StarHorse} code \citep{santiago16,queiroz18}, proper motions in equatorial coordinates (RA, Dec.) from \textit{Gaia} Early Data Release 3 \citep[EDR3;][]{gaia21}, and radial velocities and chemical abundances from the APOGEE survey. 
Applying the above selection criteria yields an initial sample of 349 stars. 
We further refined this sample by requiring effective temperatures of T$_{\rm eff} \geq 3900$~K and APOGEE spectra with signal-to-noise ratios (S/Ns) $> 80$, resulting in a sample of 75 stars. Three stars with S/N $< 80$ (a18, a20, and a72) were also included because they are of interest due to their [Fe/H] $\sim$ -0.5 permitting a better
metallicity coverage; so, the final sample consists of 78 stars. 

We note that four stars in the final sample (2M17384300-3904130 = a27, 
2M17363449-3926379 = a25, 2M18034066-3002198 = a54, and 2M18075097-3202528 = a68) reported values of \texttt{VSCATTER} in the range of 2-5~km~s$^{-1}$. Such values may indicate the presence of binarity and/or pulsation variability. Interestingly, star a27 is classified as P-rich, while a54 is classified as N-rich, suggesting that their chemical abundance patterns may be related to the observed \texttt{VSCATTER}.The kinematical and dynamical characteristics of the sample stars, including coordinates, proper motions, distance,
peri- and apo-Galactic distances, maximum height from the Galactic plane, and eccentricity are given in Table \ref{dynamics}. 

\begin{figure}
   \centering
    \includegraphics[width=8.5cm]{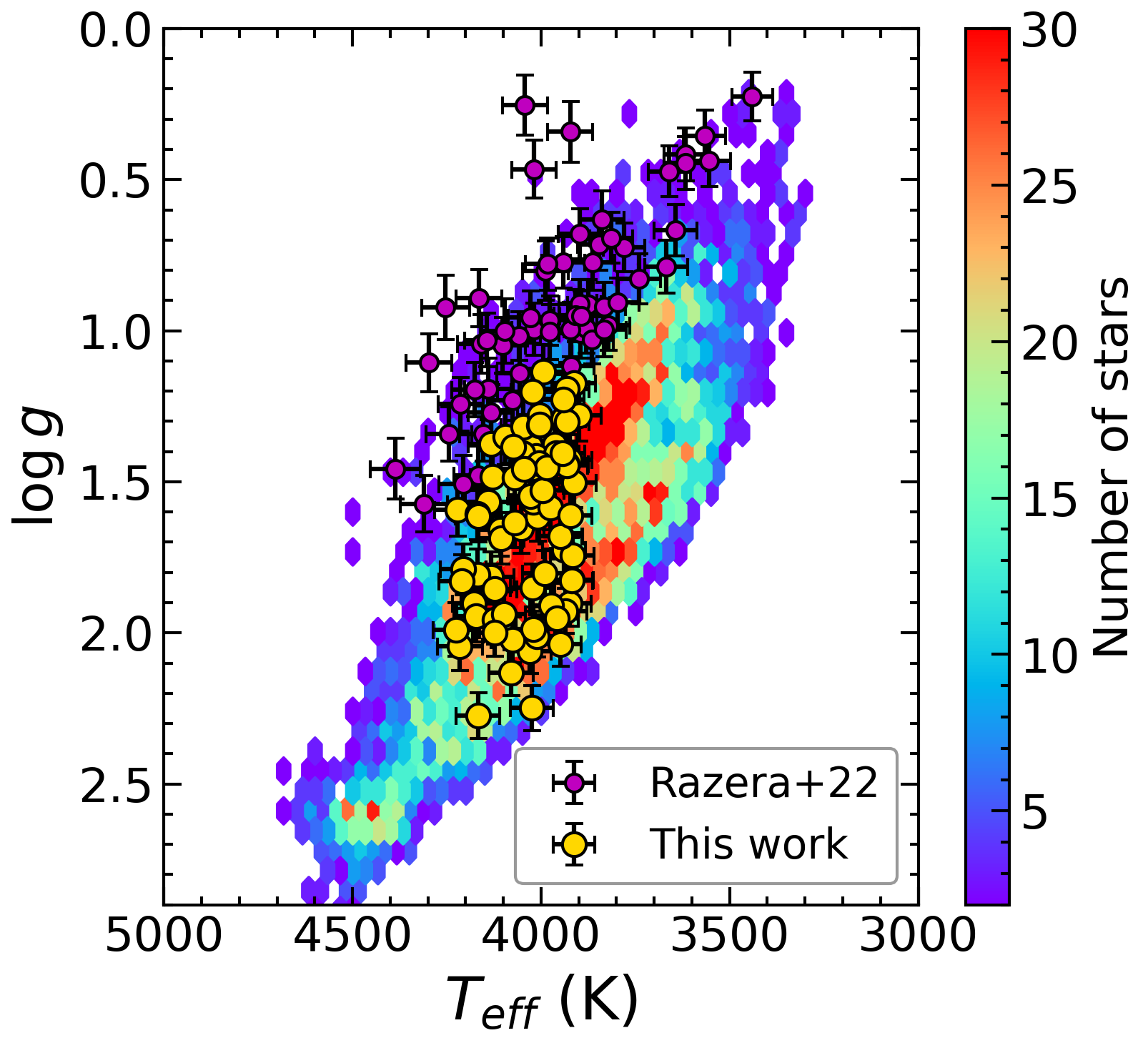}
    \caption{ Kiel diagram, based on {\tt TEFF\_SPEC} and {\tt LOGG\_SPEC} from APOGEE DR17 of the present sample, compared to \citet[purple]{razera22}.\ The RPM sample from \citet{queiroz21} is shown in the background as a hexagonally binned density distribution.}
    \label{cmd}
\end{figure}

In Fig. \ref{cmd} a Kiel diagram of the sample stars is plotted with the  effective temperature and gravity log~g from APOGEE-ASPCAP, 
and compared with the RPM sample of \citet{queiroz21}.
We note that metal-rich red giant branch (RGB) stars are cool, as can be seen through
the curved and cooler RGB of metal-rich GCs \citep[e.g.][]{ortolani25}.

The APOGEE survey is part of the Sloan Digital Sky Survey IV \citep[SDSS-IV/V;][]{blanton17}, offering high resolution ($R \sim$ 22,500)  and
high S/Ns in the \textit{H} band  (15140-16940 {\rm \AA}; \citealt{wilson19}) and including  about 7$\times$10$^{5}$ stars.
 APOGEE-1  and APOGEE-2 used the 2.5m Sloan Foundation Telescope at the Apache Point Observatory in New Mexico 
 \citep{gunn06}, and the 2.5m Ir\'en\'ee du Pont Telescope at the Las Campanas Observatory in Chile \citep{bowen73}, respectively. 
 \citet{santana21} and \citet{beaton21} reported the target selection
for the Southern and Northern Hemispheres, respectively. 
The detectors are the H2RG (2048 x 2048) Near-Infrared HgCdTe Detectors with 18$\mu$ pixels.

A great advantage of the APOGEE project is the 
stellar parameter derivation, together with chemical abundances, through a Nelder-Mead algorithm  \citep{nelder65} with a
simultaneous  fit of the stellar parameter's effective temperature (T$_{\rm eff}$), gravity (log~g), 
metallicity ([Fe/H]), and microturbulence velocity (v$_{\rm t}$), together with the abundances of carbon, nitrogen, and $\alpha$ elements. The 
ASPCAP pipeline \citep{garcia-perez16} is based on the FERRE code \citep{allende-prieto06} and the APOGEE line list \citet{smith21}. 
In the present work, we use APOGEE DR 17 \citep{abdurro22} uncalibrated (spectroscopic) stellar parameters.

\begin{table}
\centering
\caption[5]{Stellar parameters and S/Ns from DR 17.}
\resizebox{0.45\textwidth}{!}{
\begin{tabular}{lccccc|c}
\noalign{\smallskip}
\hline
\noalign{\smallskip}
ID & \hbox{APOGEE ID}  & T$_{\rm eff}$ (K) & log~g & [Fe/H] & v$_{t}$ (km/s) & S/N  \\  
\hline
\noalign{\smallskip}
%
a1& 2M16252511-4103018  &4215.4 &2.04&  -0.36   & 1.39 & 116 \\
a2& 2M17031887-3421087  &4027.6 &1.38&  -0.73   & 1.78 & 109 \\
a3& 2M17043110-3433265  &3913.5 &1.17&  -0.71   & 1.93 & 137 \\
a4& 2M17055616-3439207  &4132.1 &1.37&  -0.73   & 1.83 & 105 \\
a5& 2M17133444-3041065  &4024.1 &2.25&  0.40    & 1.20 & 92 \\
a6& 2M17140245-3106280  &4177.1 &1.90&  -0.28   & 1.59 & 121 \\
a7& 2M17153186-2426492  &3963.7 &1.38&  -0.63   & 1.57 & 97 \\
a8& 2M17163330-2808396  &4093.7 &1.35&  -0.73   & 2.08 & 84 \\
a9& 2M17163774-2824023  &4012.4 &1.42&  -0.71   & 1.64 & 131 \\
a10& 2M17173212-2407066 &4046.6 &1.32&  -0.77   & 1.69 & 118 \\
a11& 2M17192106-2638374 &4127.9 &1.48&  -0.56   & 1.58 & 112 \\
a12& 2M17192497-2650234 &3912.6 &1.50&  -0.38   & 1.56 & 110 \\
a13& 2M17193684-2748495 &4046.9 &1.44&  -0.66   & 1.66 & 143 \\
a14& 2M17251032-2636183 &4052.2 &1.65&  -0.26   & 1.23 & 91 \\
a15& 2M17263539-2035044 &4048.6 &1.41&  -0.72   & 1.79 & 238 \\
a16& 2M17280115-2829180 &3932.5 &1.19&  -0.79   & 2.11 & 106 \\
a17& 2M17285838-2956022 &3939.6 &1.73&  0.03    & 1.64 & 107 \\
a18& 2M17322148-2419153 &4108.5 &1.66&  -0.48   & 1.58 & 52 \\
a19& 2M17324991-3019162 &3959.8 &1.41&  -0.61   & 1.81 & 90 \\
a20& 2M17344080-3326314 &4006   &1.31&  -0.55   & 1.89 & 79 \\
a21& 2M17344126-2652250 &4171.6 &1.95&  -0.31   & 1.39 & 98 \\
a22& 2M17345655-2646140 &3899.8 &1.28&  -0.64   & 1.77 & 111 \\
a23& 2M17351303-1836250 &3941.9 &1.29&  -0.72   & 1.76 & 176 \\
a24& 2M17351307-2717212 &3923.9 &1.43&  -0.21   & 1.77 & 86 \\
a25& 2M17363449-3926379 &4131.6 &1.81&  -0.29   & 1.36 & 99 \\
a26& 2M17383707-2056139 &4023.1 &1.85&  0.01    & 1.29 & 86 \\
a27& 2M17384300-3904130 &4105.1 &1.69&  -0.50   & 1.50 & 99 \\
a28& 2M17394883-2208105 &4074.1 &1.38&  -0.81   & 1.83 & 130 \\
a29& 2M17410295-2228195 &4029.9 &2.06&  0.24    & 1.27 & 132 \\
a30& 2M17412628-2218383 &4167.7 &1.81&  -0.58   & 1.44 & 118 \\
a31& 2M17414603-3335116 &3923.1 &1.90&  0.26    & 1.31 & 104 \\
a32& 2M17422805-3410327 &3949.5 &2.04&  0.51    & 1.31 & 96 \\
a33& 2M17430247-3503236 &3931.4 &1.44&  -0.21   & 1.68 & 144 \\
a34& 2M17440199-2238149 &3917.1 &1.74&  0.22    & 1.36 & 85 \\
a35& 2M17463610-2336189 &4123.8 &1.96&  -0.25   & 1.46 & 111 \\
a36& 2M17474925-2211497 &3935.8 &1.93&  0.42    & 1.33 & 86 \\
a37& 2M17503867-3706497 &4023.1 &1.46&  -0.65   & 1.66 & 89 \\
a38& 2M17504075-3704209 &4009.8 &1.62&  -0.17   & 1.63 & 93 \\
a39& 2M17505373-1803210 &4140.7 &1.57&  -0.71   & 1.62 & 103 \\
a40& 2M17520538-1822315 &3946.9 &1.68&  0.02    & 1.35 & 137 \\
a41& 2M17525971-1912041 &3973.1 &1.91&  0.32    & 1.33 & 120 \\
a42& 2M17530359-2258461 &3918.3 &1.82&  0.41    & 1.32 & 83 \\
a43& 2M17531682-2104467 &4002.9 &1.28&  -0.74   & 2.15 & 96 \\
a44& 2M17540360-3638369 &4020.3 &1.57&  -0.27   & 1.69 & 87 \\
a45& 2M17591751-3116052 &4027.7 &1.55&  -0.52   & 1.70 & 81 \\
a46& 2M18000162-3047385 &4075   &2.02&  0.14    & 1.28 & 108 \\
a47& 2M18011040-2844047 &3931.1 &1.30&  -0.66   & 1.60 & 100 \\
a48& 2M18013822-2727488 &3942   &1.23&  -0.78   & 1.83 & 172 \\
a49& 2M18014518-2752453 &4004.3 &1.31&  -0.73   & 1.64 & 132 \\
a50& 2M18014962-3002151 &4167.4 &2.27&  0.22    & 1.33 & 104 \\
a51& 2M18022293-2850057 &4206.8 &1.79&  -0.36   & 1.51 & 90 \\
a52& 2M18031846-3000497 &3976.1 &1.58&  -0.07   & 1.46 & 146 \\
a53& 2M18032107-2959127 &3948.9 &1.68&  0.07    & 1.40 & 124 \\
a54& 2M18034066-3002198 &4011.8 &2.01&  0.32    & 1.33 & 159 \\
a55& 2M18034548-3122086 &3917.5 &1.83&  0.29    & 1.30 & 160 \\
a56& 2M18034880-3002202 &4098.6 &1.94&  0.15    & 1.48 & 135 \\
a57& 2M18041554-3001431 &4209.5 &1.83&  -0.61   & 1.40 & 85 \\
a58& 2M18041724-3132539 &4123.1 &1.86&  -0.50   & 1.34 & 103 \\
a59& 2M18044803-2752467 &4220.7 &1.59&  -0.53   & 1.86 & 89 \\
a60& 2M18050144-3005149 &4006.3 &1.44&  -0.45   & 1.78 & 172 \\
a61& 2M18050235-3002348 &3996   &1.53&  -0.13   & 1.61 & 146 \\
a62& 2M18051997-2923441 &4079.2 &2.13&  0.18    & 0.78 & 97 \\
a63& 2M18052433-3036305 &4020.6 &1.99&  0.28    & 1.37 & 111 \\
a64& 2M18061581-2748521 &4069.9 &1.64&  -0.57   & 1.59 & 145 \\
a65& 2M18073289-3156045 &4123.3 &2.00&  -0.17   & 1.29 & 125 \\
a66& 2M18074173-3153200 &4069.6 &1.48&  -0.70   & 1.80 & 108 \\
a67& 2M18074493-3145155 &3932.1 &1.45&  -0.21   & 1.59 & 154 \\
a68& 2M18075097-3202528 &3957.6 &1.95&  0.15    & 1.07 & 121 \\
a69& 2M18080123-3143583 &4224.7 &1.99&  -0.57   & 1.46 & 104 \\
a70& 2M18084466-2529453 &3984.7 &1.45&  -0.45   & 1.66 & 227 \\
a71& 2M18084995-3124422 &4168.7 &1.61&  -0.69   & 1.53 & 107 \\
a72& 2M18093136-2559533 &3994.5 &1.14&  -0.73   & 1.85 & 66 \\
a73& 2M18112625-2527164 &3942.2 &1.40&  -0.51   & 1.57 & 111 \\
a74& 2M18125891-2610195 &3991.2 &1.80&  0.17    & 1.33 & 104 \\
a75& 2M18163725-3253337 &4022.5 &1.20&  -0.79   & 1.86 & 399 \\
a76& 2M18191892-2203579 &3921.3 &1.61&  0.00    & 1.32 & 136 \\
a77& 2M18203837-2235146 &4167.4 &1.61&  -0.55   & 1.88 & 115 \\
a78& 2M18365761-2322171 &4045.5 &1.46&  -0.72   & 1.79 & 108 \\

\hline
\noalign{\smallskip}
\label{param}
\end{tabular}}
\end{table}

\section{Spectroscopic analysis}\label{calculations}

From our previous work \citep{razera22,barbuy23,barbuy24,barbuy25a}, we verified that the lines of the following elements
were well analysed with the ASPCAP software within the APOGEE collaboration: 
Mg, Si, Ca, V, Cr, Co, and Ni.
In this work, we derived abundances for the elements that are not fully treated with the APOGEE ASPCAP software; they are
C, N, O, P, S, K, V, Mn, Ce, and Yb. We also preferred to revise Al. On the other hand,
the Na and Ti lines were too blended to be reliable. We derived the  abundances in the \textit{H} band using the {\tt TURBOSPECTRUM} code from \citet{alvarez98} and \citet{plez12}
 to compute synthetic spectra, which were compared with the observed spectra line-by-line and star-by-star.

The model atmospheres are interpolated within the CN-mild MARCS grids from \citet{gustafsson08}. 
The solar abundances of the elements studied are from \citet{asplund21}.

We adopted the uncalibrated, or `spectroscopic', stellar parameter's 
effective temperature (T$_{\rm eff}$), gravity (log~g), metallicity ([Fe/H]), and microturbulence velocity (v$_{t}$) from the APOGEE DR17 {\tt TURBOSPECTRUM} results. 
These parameters are reported in Table \ref{param}, together with 
the S/Ns of the DR17 spectra.

Table \ref{linelist} reports the lines in the \textit{H} band that we used to measure the abundances
of the elements in the spectra of the sample stars. 
Oscillator strengths were adopted from the line list of the APOGEE collaboration
\citep{smith21}.
The  full atomic line list employed is that from the APOGEE collaboration,
together with the molecular lines described in \citet{smith21}. 

\begin{table}
\small
\scalefont{0.5}
\centering
\caption[4]{Line list and oscillator strengths. }
\resizebox{0.4\textwidth}{!}{
\begin{tabular}{l@{}ccrr@{}rcccccccc}
\hline
\hline
\noalign{\smallskip}
\hbox{Ion} & \hbox{$\lambda$} & \hbox{$\chi_{ex}$}  &\hbox{log~gf}  &\\
& \hbox{(\AA)} &\hbox{(eV)} & \hbox{APOGEE}  &  \\ 
\noalign{\smallskip}
\hline
\noalign{\smallskip}
\hbox{AlI} & 16718.957 & 4.085 & 0.220 &   \\
& 16750.539 & 4.088 &  0.408 &   \\
& 16763.359 & 4.087 &  -0.480 &   \\
\hline
\noalign{\smallskip}
\hbox{PI} 
& 15711.522 & 7.176 & $-$0.404 &\\
& 16482.932 & 7.213 & $-$0.273 &\\ 
\hline
\noalign{\smallskip}
\hbox{SI~} & 15469.816 & $-$0.199 & \\
& 15475.616 & 8.047 & $-$0.744 &\\ 
& 15478.482 & 8.047 & $-$0.040 &\\
\hline
\noalign{\smallskip}
\hbox{KI~} & 15163.067 & 2.670 & 0.630 &\\
& 15168.377 & 2.670 &  0.481 &\\
\hline
\noalign{\smallskip}
\hbox{MnI~}
& 15159.260 & 4.889 & 0.606 (hfs) & \\
& 15217.700 & 4.889 & 0.520 (hfs) & \\
& 15262.330 & 4.889 & 0.379 (hfs) &  \\

\hline
\noalign{\hrule\vskip 0.1cm} 
\hbox{CeII} 
& 15784.750 & 0.318 & -1.54 &   \\
& 15958.400 & 0.470 & -1.71 &  \\
& 16327.320 & 0.561 & -2.40 &    \\
& 16376.480 & 0.122 & -1.79 &    \\
& 16595.180 & 0.122 & -2.19 &   \\
& 16722.510 & 0.470 & -1.65 &   \\
\hline
\hbox{YbII} & 16498.430 & 3.017  & -0.640 &  \\
\hline
\noalign{\hrule\vskip 0.1cm} 
\hline                 
\label{linelist}
\end{tabular}}
\begin{minipage}{8cm}
\vspace{0.1cm}
\end{minipage}
\end{table} 

\section{Results}
\label{results}

\begin{figure}
    \centering
    \includegraphics[width=8.5cm]{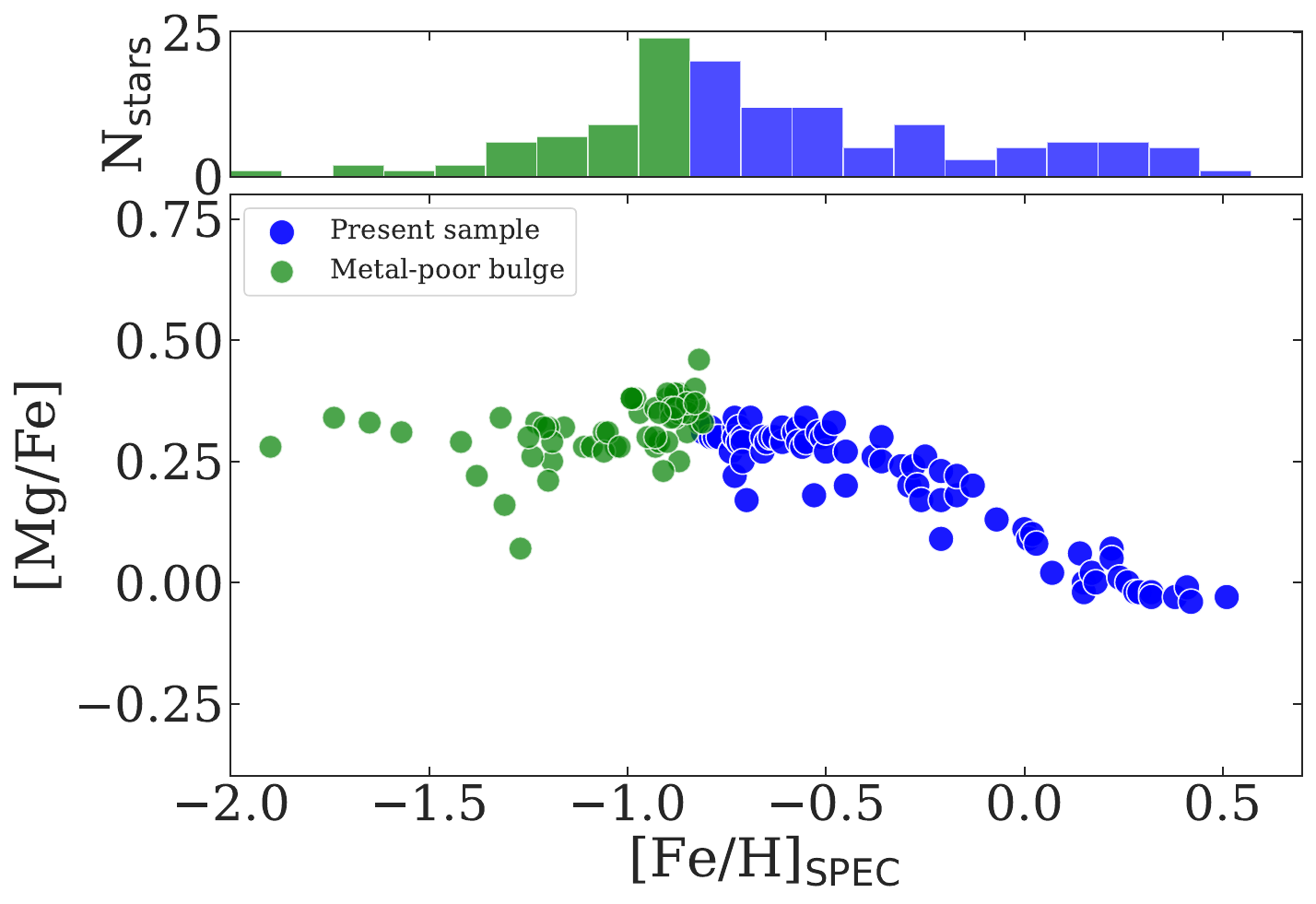}
    \caption{[Mg/Fe] versus [Fe/H] with the histogram of metallicities for the present sample and the \citet{razera22} metal-poor sample.}
    \label{mgfe}
\end{figure}

Figure \ref{mgfe} shows the ASPCAP [Mg/Fe] versus [Fe/H] for the sample stars. 
It is showing the suitability of the sample as an old bulge one and as having a well-covered metallicity range.

\subsection{Abundances}

In this subsection we describe the derivation of the abundances from the different spectral lines. These are listed below.

\begin{figure*}
    \centering
    \includegraphics[width=14.5cm]{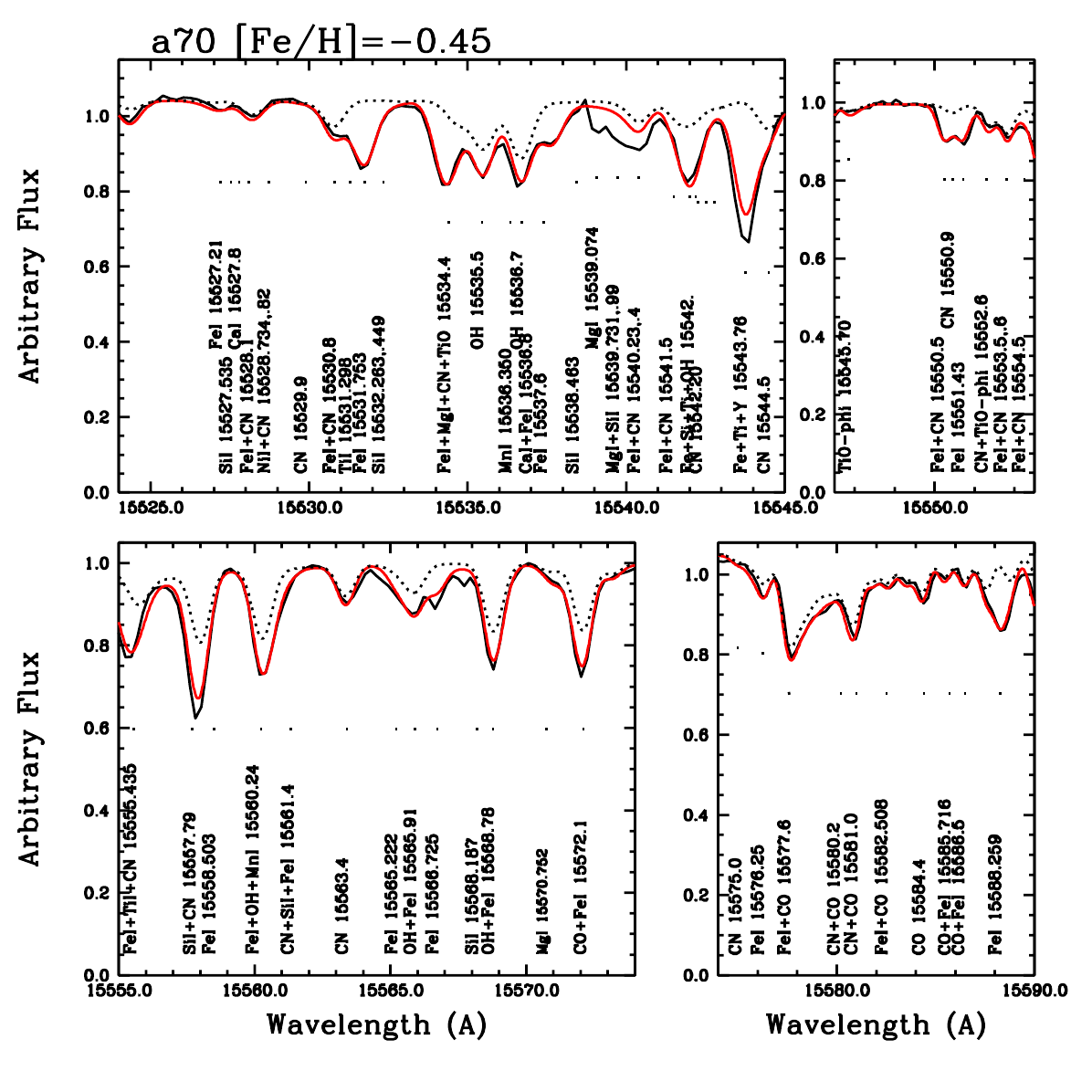}
    \caption{Fit to the 15525--15590 {\rm \AA} region in star a70. The main lines are indicated.}
    \label{mra70cno}
\end{figure*}

{\it C,N,O}: For the C, N, O abundance derivation, we used the region 15525–15590 {\rm \AA}, which contains lines of
CO, OH, and CN,  as described and illustrated in \citet{barbuy21b} and \citet{razera22}.
The fit is illustrated in Fig. \ref{mra70cno} for star a70, where the location of CO, OH, and CN lines are indicated.

\begin{figure}
    \centering
    \includegraphics[width=8.5cm]{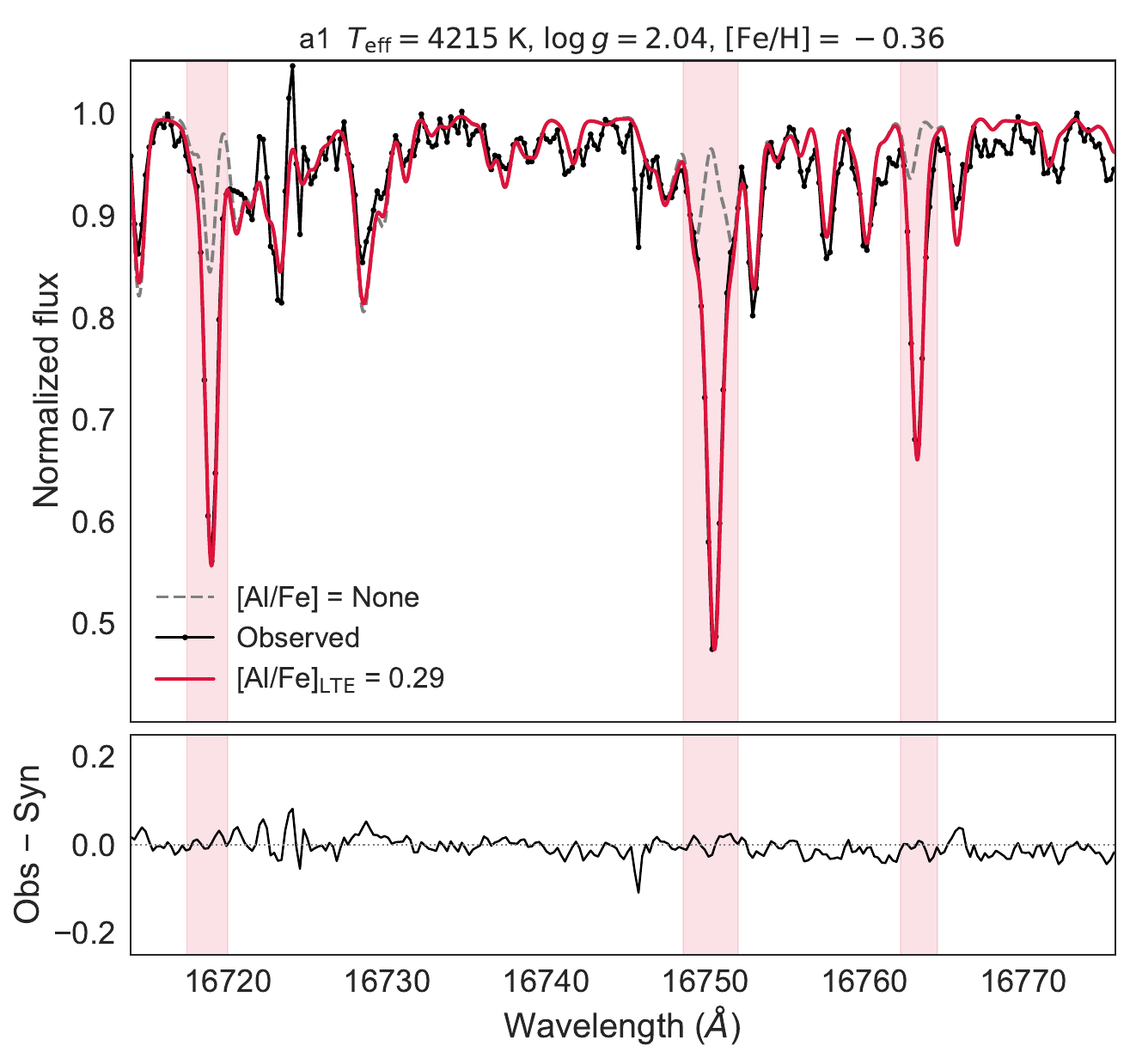}
    \caption{Fit to \ion{Al}{I} 16718.957, 16750.539, and 16763.359 {\rm \AA\ }lines. The synthetic spectra were computed with [Al/Fe] = 0.29 and are shown in red; we compare them with the observed spectrum, which is shown in black. }
    \label{alfit}
\end{figure}

{\it Aluminium}: Al has three strong and rather clean lines in the 
{\it H} band, located at 16718.957,  16750.539, and 16763.359 {\rm \AA}. 
The bluest line has a blend in its left wing, and the 16750.539 {\rm \AA} line has two blended lines in its wings that are far enough from the centre to make its abundance reliable. The 16763.359 {\rm \AA} line has a small blend in its left side, as seen in Fig. \ref{alfit}. 

\begin{figure}
    \centering
    \includegraphics[width=8.5cm]{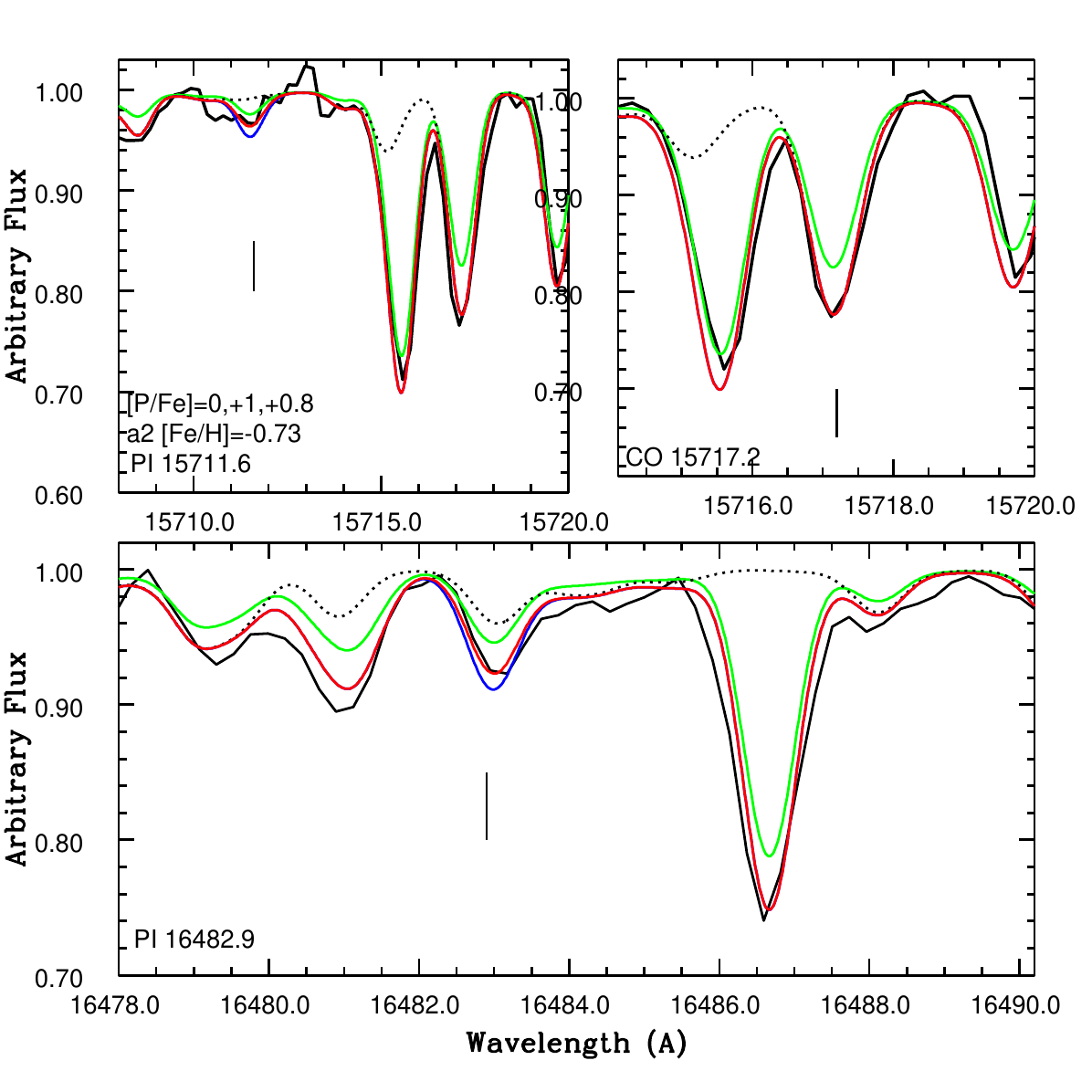}
    \caption{Fit to \ion{P}{I} 15711.522 {\rm \AA} and \ion{P}{I} 16482.932 {\rm \AA} lines and the
    nearby CO 15717.2 {\rm \AA} line for star a2: 2M17031887-3421087; this 
    confirms that the CO strength is reliable.
    Synthetic spectra were computed with [P/Fe] = 0.0 (green), +1.0 (blue), and the final value +0.8 (red),
    compared with the observed spectrum (black). 
    The dotted line corresponds to molecular lines only.}
    \label{mra2pco}
\end{figure}

{\it Phosphorus}: We carried out the fitting of the lines 
\ion{P}{I} 15711.522 {\rm \AA} and \ion{P}{I} 16482.932 {\rm \AA}
with more weight on the stronger latter line. Only for the more
metal-poor stars among the sample were we able to measure a P abundance
higher than a solar ratio.
 In a number of stars, the P measurement is not possible in the spectra where the 16482.932 {\rm \AA} line is too noisy, has strong artefacts, or
falls in a gap near the HgCdTe 2048x2048  detector edge. 
Figure \ref{mra2pco} illustrates the fit to both \ion{P}{I} lines and the nearby CO 15717.2 {\rm \AA} line,
which attests the reliability of the strength of a CO feature that coincides with the stronger and more important \ion{P}{I} line.

\begin{figure}
    \centering
    \includegraphics[width=8.5cm]{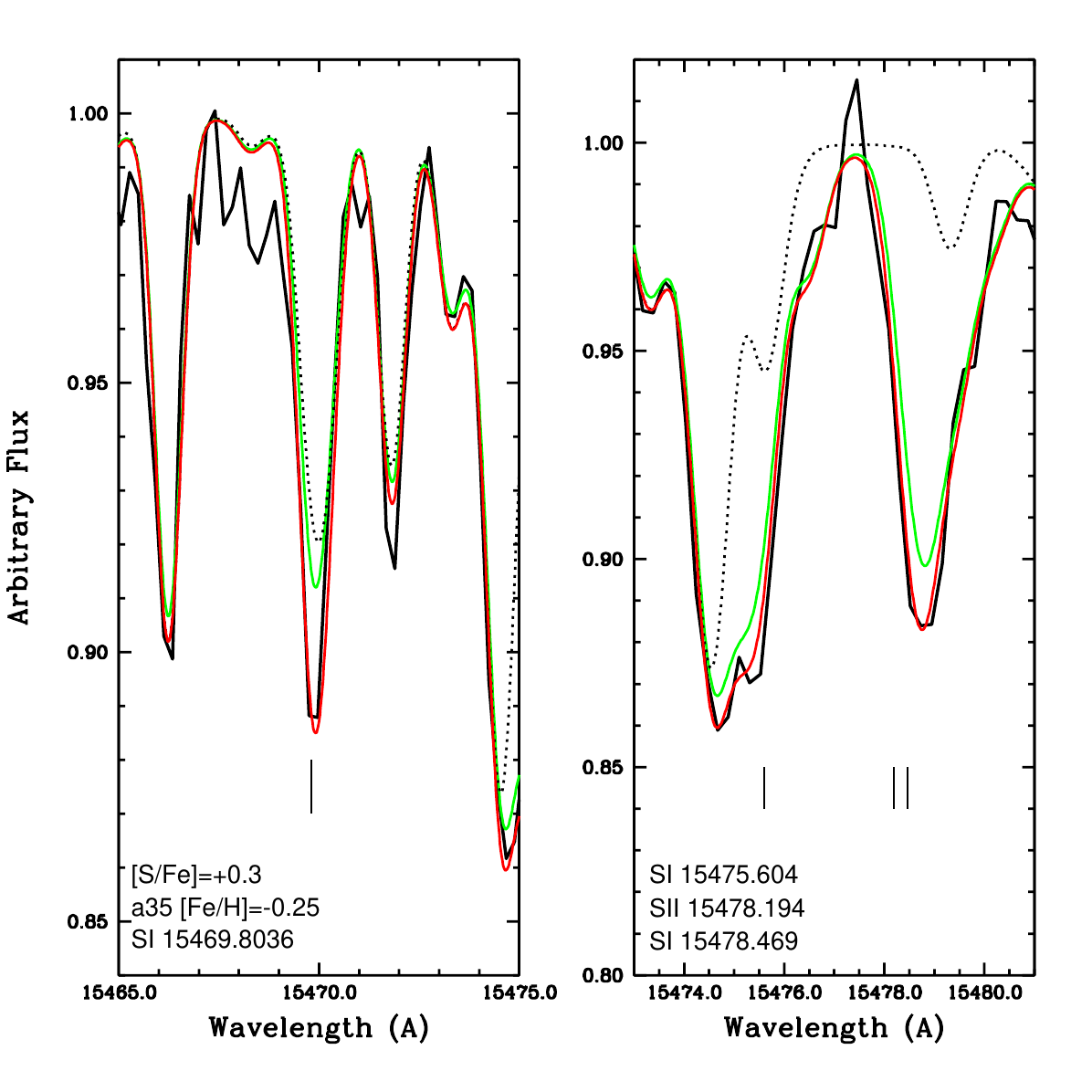}
    \caption{Fit to \ion{S}{I} 15469.8036 {\rm \AA} line (left panel) and 
    \ion{S}{I} 15475.604, 
    \ion{S}{II}  15478.194, and  \ion{S}{I} 15478.469 {\rm \AA}  lines
    (right panel) for star a35: 2M17463610-2336189.
    Synthetic spectra were computed with [S/Fe] = 0.0 (green lines),
    and the final S/Fe] = +0.3 abundance (red lines) is
    compared with the observed spectrum (black). 
    The dotted black line was computed with molecular lines only.}
    \label{a35snew}
\end{figure}

{\it Sulphur}: 
We fitted the three measurable lines:
\ion{S}{I} 15469.826, 15475.624, and 15478.406 {\rm \AA}.
The first two were blended with other atomic and molecular lines; therefore, we derived the S abundances from
the third line \citep[see also][]{hayes22,barbuy25a}.
For the metal-rich stars ([Fe/H]$>$0), the line is often dominated
by other blends and is not very sensitive to the S abundances.
Figure \ref{a35snew} shows the fit to the S lines, including
the contribution from molecular lines.

\begin{figure}
    \centering
    \includegraphics[width=8.5cm]{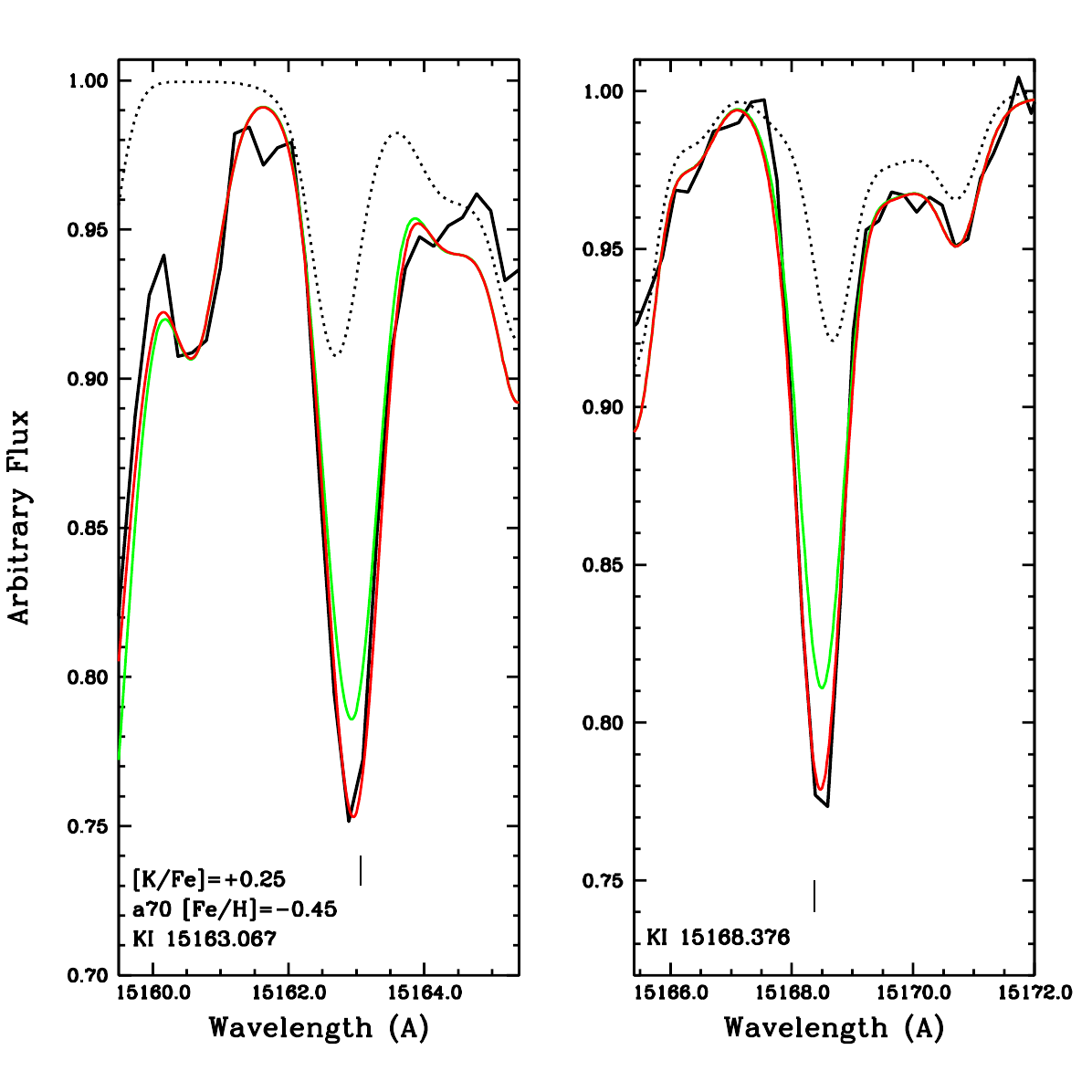}
    \caption{Fit to \ion{K}{I} 15163.067 and 15168.376 {\rm \AA} lines
    for star a70: 2M18084466-2529453.
    Synthetic spectra were computed with [K/Fe] = 0.0 (green lines); the
    final value of [K/Fe] = +0.25 (red lines) is 
    compared with the observed spectrum (black). }
    \label{a70knew}
\end{figure}

{\it Potassium:}
We simultaneously fitted the two available \ion{K}{I} 15163.067 and 15168.376 {\rm \AA\ lines}, which are plotted in two different
panels to better adapt their continua.
In most cases, the two lines are well fit with the same K abundance, and in cases where they differ, we adopted a mean value.
Figure \ref{a70knew} shows the fit to the \ion{K}{I} lines for star a70.

\begin{figure}
    \centering
    \includegraphics[width=8.5cm]{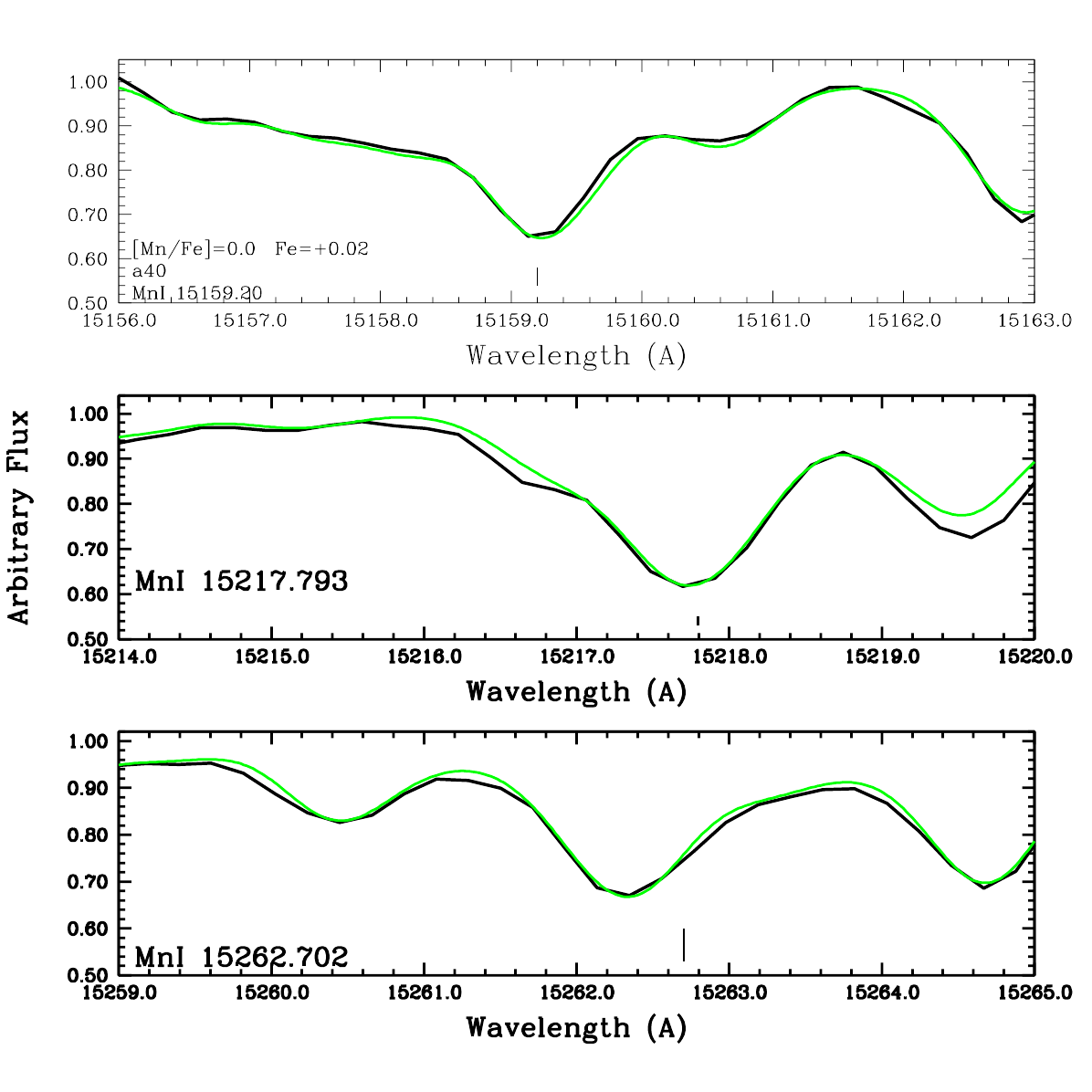}
    \caption{Fit to \ion{Mn}{I} 15159.26, 15217.70 and 15262.33 {\rm \AA} lines
    for star a40: 2M17520538-1822315.
    Synthetic spectra were computed with [P/Fe] = 0.0 (green lines), and are
    compared with the observed spectrum (black). }
    \label{a40mn}
\end{figure}

{\it Manganese:}
The three lines \ion{Mn}{I} 15159.200,  15217.793, and 15262.702 {\rm \AA}
were measured in all stars.
Figure \ref{a40mn} illustrates the fit to the three lines for star a40:
2M17520538-1822315.

\begin{figure}
    \centering
    \includegraphics[width=8.5cm]{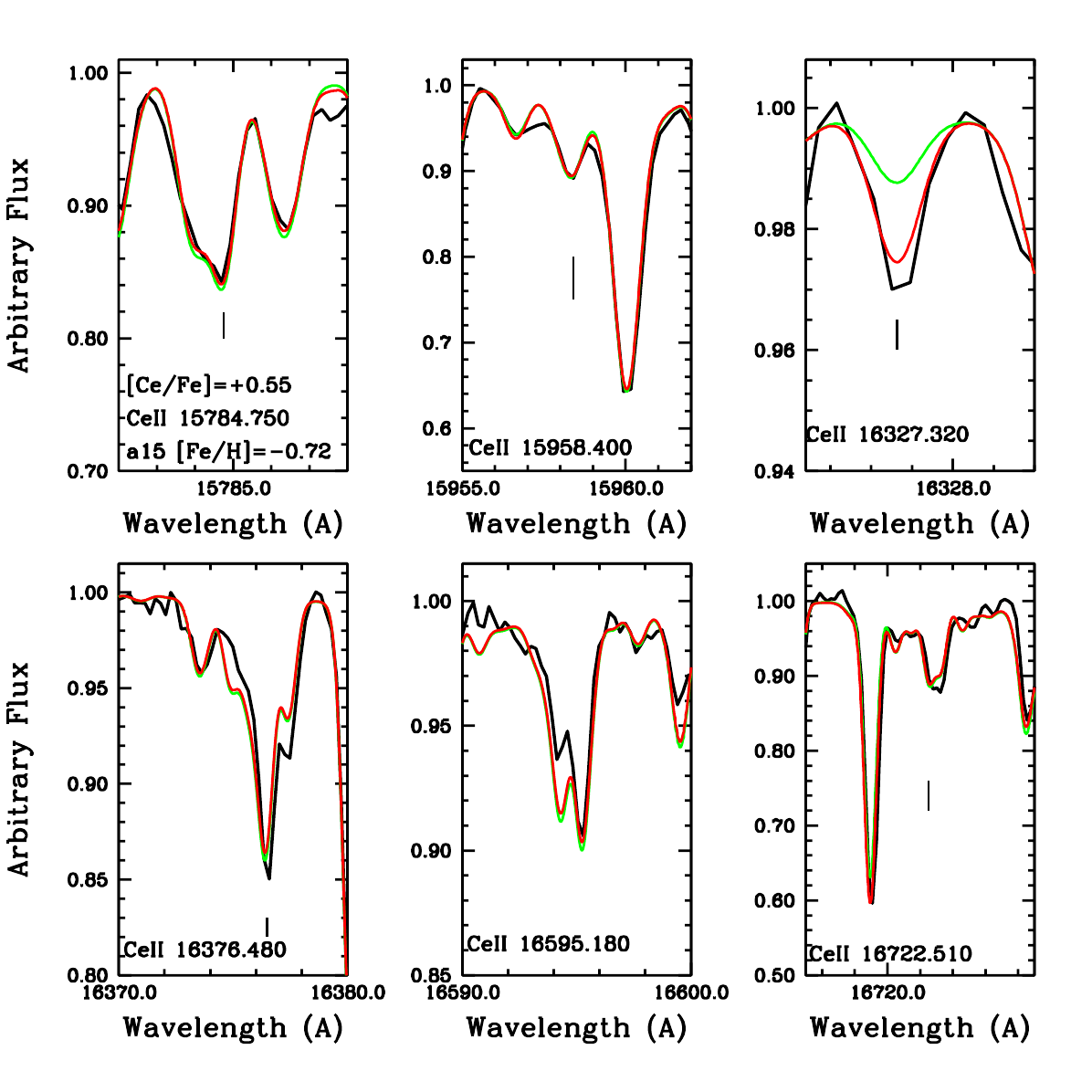}
    \caption{Fit to six Ce lines for star a15:  2M17263539-2035044.
    Synthetic spectra were computed with [Ce/Fe] = 0.0 (green), and the final value +0.8 (red) is    compared with the observed spectrum (black). }
    \label{a15ce}
\end{figure}

{\it Cerium:}
We simultaneously fitted the six Ce II lines indicated in Table \ref{linelist}.
Enhanced Ce was measured in particular for the moderately metal-poor stars with -0.8 $<$ [Fe/H] $<$ -0.7,
and this is clear from all six lines. For the metal-rich stars, we relied on
the \ion{Ce}{II} 16722.51 {\rm \AA}, which is clean, although it is next to
another line; the left side of this broad line is a good measure of the
Ce abundance.
Figure \ref{a15ce} shows the fits for star a15:  2M17263539-2035044, 
which has  [Fe/H]=-0.72 and enhanced Ce. 

{\it Ytterbium}:
The unique Yb line at \ion{Yb}{I} 16498.420 {\rm \AA} can only be measured reliably in the 
warmer and higher S/N spectra. As shown by \citet{montelius22}, the line itself is blended with CO,
but even more difficult is the continuum placement due to the many molecular bands near the Yb line.
Due to the low S/N of the spectra in the Yb line region, we do not
report their derived abundances.

\subsection{Uncertainties}\label{uncertainties}

Typical systematic uncertainties due to stellar parameters were evaluated by varying the parameters by fixed amounts of $\Delta$T$_{\rm eff}$ = 100\,K, $\Delta$log \textit{g} $=$ 0.2\,dex, 
$\Delta v_{\rm t}$ = 0.2 km $s^{-1}$, shown in Table \ref{errors}. This was applied to the star a2. Note that the uncertainties in C,N,O are intertwined, as well as the P abundance,
which is given by assuming the C,N,O derived for the modified parameters.
From Table \ref{errors} it is clear that stellar parameters are crucial for having
reliable abundances. Fortunately, it appears that the ASPCAP-derived
parameters are suitable due to the simultaneous fitting of the parameters
plus C,N,O abundances, as discussed in \citet{dasilva24}; however, for 
stars with over-solar metallicities the errors are larger.

\begin{table}
  \caption{Abundance uncertainties for star a2 due to changes in stellar parameters of $\Delta$T$_{\rm eff}$ = 100 K, $\Delta$log \textit{g} $=$ 0.2\,dex, and $\Delta$v$_{\rm t}$ = 0.2 km s$^{-1}$. 
} 
\label{errors}
\begin{flushleft}
\small
\tabcolsep 0.5cm
\begin{tabular}{lc@{}c@{}c@{}c@{}c@{}}
\noalign{\smallskip}
\hline
\noalign{\smallskip}
\hline
\noalign{\smallskip}
\hbox{Element} & \hbox{$\Delta$T} & \hbox{$\Delta$log $g$} & 
\phantom{-}\hbox{$\Delta$v$_{t}$} & \phantom{-}\hbox{($\sum$x$^{2}$)$^{1/2}$} \\
\hbox{} & \hbox{100\,K} & \phantom{-}\hbox{0.2\,dex} & \phantom{-}\hbox{0.2 kms$^{-1}$} & & \\
\noalign{\smallskip}
\hline
\noalign{\smallskip}
\noalign{\hrule\vskip 0.1cm}
\multicolumn{5}{c}{a2 = - T$_{\rm eff}$=4027.6\,K, log~g=1.38} \\
\noalign{\smallskip}
\hline
\noalign{\smallskip}
\hline
\noalign{\smallskip}
C &  0.07 & 0.00 & 0.00 &  0.07   \\
N &  0.01 & 0.00 & 0.00 &  0.01   \\
O &  0.13 & 0.05 & 0.01 &  0.14   \\
Al & 0.05 & 0.12 & 0.05 &  0.14   \\
P & 0.15  & 0.11 & 0.00 &  0.19  \\
S & 0.08 &  0.10 & 0.00 &  0.13  \\
K & 0.00 &  0.00 & 0.00 &  0.00 \\
Mn & 0.10 & 0.15 & 0.10 &  0.21   \\
Ce & 0.00 & 0.10 & 0.00 &  0.10   \\
\noalign{\smallskip} 
\hline 
\end{tabular}
\tablefoot{The corresponding total error is given in the last column.}
\end{flushleft}
 \end{table}

\begin{figure}
    \centering
    \includegraphics[width=8.5cm]{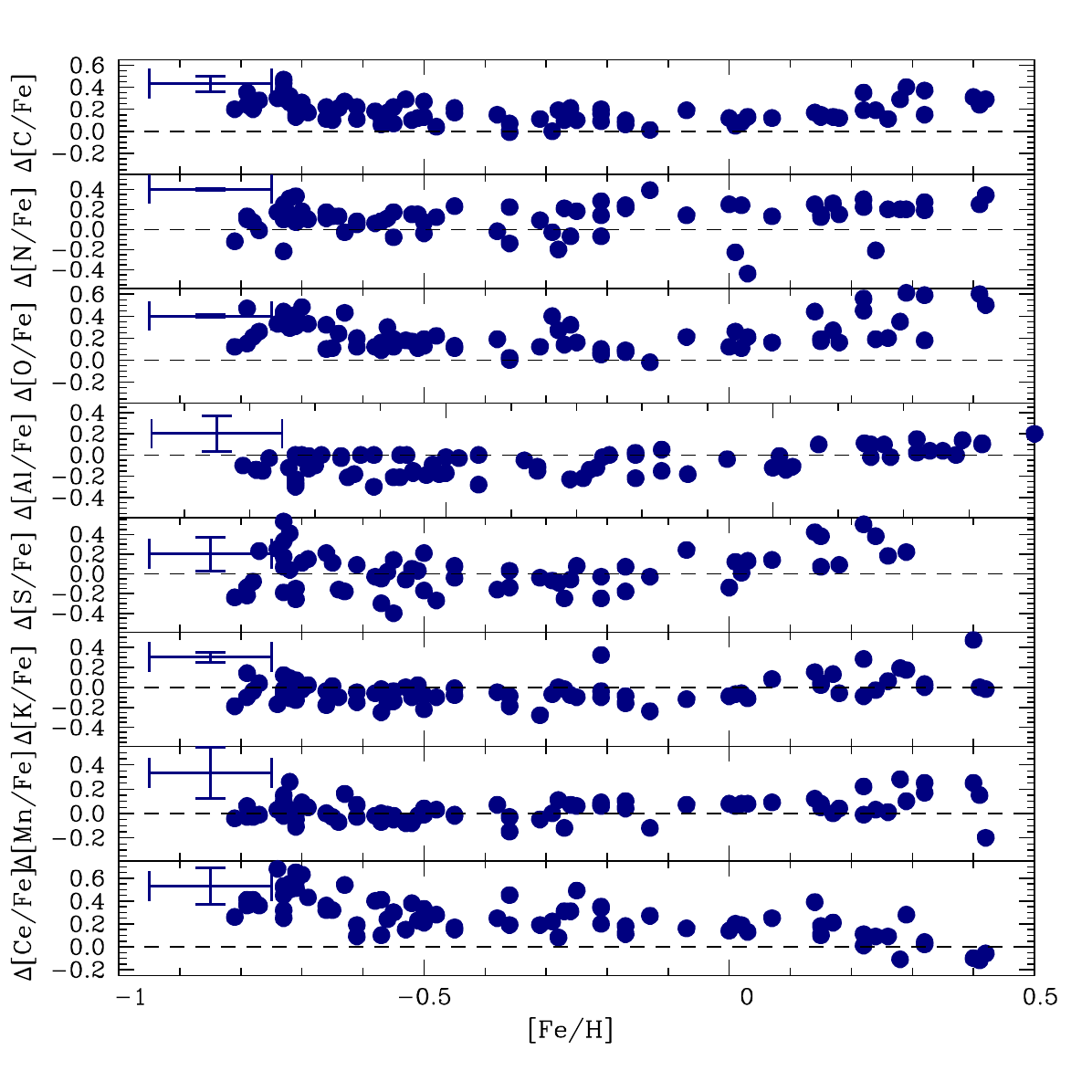}
    \caption{[C, N, O, Al, S, K, MN, Ce/Fe] versus [Fe/H], plotting the differences between the present results and the APOGEE DR17 values. Error bars correspond to the errors for the cool star reported in Table \ref{errors}.}
    \label{plotdiff}
\end{figure}

Figure \ref{plotdiff} shows the difference between
our derived abundances of C, N, O, Al, S, K, Mn, and Ce minus the ones from APOGEE-ASPCAP DR17.  The mean difference between the present results and ASPCAP DR17 are given in Eq. (1):
\begin{flushleft}
\begin{equation}\label{eq1}
\begin{array}{l}
{\rm [C/Fe]}_{\rm present}-{\rm [C/Fe]}_{\rm ASPCAP}=+0.20\\
{\rm [N/Fe]}_{\rm present}-{\rm [N/Fe]}_{\rm ASPCAP}=-0.01\\
{\rm [O/Fe]}_{\rm present}-{\rm [O/Fe]}_{\rm ASPCAP}=+0.26\\
{\rm [Al/Fe]}_{\rm present}-{\rm [Al/Fe]}_{\rm ASPCAP}=-0.13\\
{\rm [S/Fe]}_{\rm present}-{\rm [S/Fe]}_{\rm ASPCAP}=-0.03\\
{\rm [K/Fe]}_{\rm present}-{\rm [K/Fe]}_{\rm ASPCAP}=+0.03\\
{\rm [Mn/Fe]}_{\rm present}-{\rm [Mn/Fe]}_{\rm ASPCAP}=+0.04\\
{\rm [Ce/Fe]}_{\rm present}-{\rm [Ce/Fe]}_{\rm ASPCAP}=+0.25
\end{array}
.\end{equation}
\end{flushleft}

\subsection{Non-LTE effects}\label{NLTE}

For Al, we computed abundances in local thermodynamic equilibrium (LTE) and non-LTE.
For that analysis, we computed the synthetic spectra using {\tt TURBOSPECTRUM} \citep{TurboNLTE}, which allowed us to switch from LTE and non-LTE computations using the same code. The non-LTE aluminium calculations were based on the grids presented by \citet{Ezzeddine18}, which provide a prescription for inelastic hydrogen-atom collisions on their 1D non-LTE calculations. The same procedure was applied in \citet{Ernandes25}.
It is found that the non-LTE calculation, using {\tt TURBOSPECTRUM}
version 2020 and non-LTE predictions tends to give too-low values of Al.
Our Al LTE abundances have a mean value of -0.13, which is lower than the ASPCAP ones (see Eq. 1).
By applying the non-LTE corrections, as available with {\tt TURBOSPECTRUM}-2020, Al becomes much lower; it is
probably maximised non-LTE corrections. 
As discussed in \citet{Ernandes25} and \citet{feltzing23} based on calculations by \citet{nordlanderlind17}, for a giant star
a mean correction of $-$0.2 dex is expected.
There are no calculations of non-LTE deviations for the lines of 
\ion{P}{I}, and \ion{Ce}{I} analysed in this work; therefore, these
uncertainties cannot be evaluated. 

For sulphur, \citet{korotin25} presented grids of non-LTE corrections
for the lines studied here. The corrections range from $-$0.06 dex
for T$_{\rm eff}$ $<$ 4000 K up to about $-$0.25 dex for T$_{\rm eff}$ $\sim$ 4200 K
and [S/Fe]=+0.4 to +0.8. The sample stars have rather cool temperatures in
the range of 4225 $>$ T$_{\rm eff}$ $>$ 3900 K, where the corrections are
not higher than $-$0.25 dex. Therefore, the non-LTE effects partly explain the
S-excesses found for our sample.
For \ion{K}{I,} the differences between synspec calculations from APOGEE DR17 
that take into account non-LTE, and LTE calculations give
differences below 0.1 dex.

For manganese, corrections were provided for the lines at 15217.793
and 15262.702 {\rm \AA} by \citet{bergemann08Mn}, but with
corrections of 0.1 and 0.5 dex in the mean, respectively.
Given this disparity and no available corrections for the third line,
we chose not to apply non-LTE corrections to Mn, likewise in \citet{barbuy24}.
Finally, abundances of Mg, Si, Ca, Ti, V, Cr, Co, and Ni adopted from ASPCAP, no non-LTE corrections were applied.

\subsection{Chemical-evolution models}

The chemical-evolution model first applied to elliptical galaxies \citep{friaca98}
is a multi-zone chemical evolution coupled with a hydrodynamical code,
allowing the inflow and outflow of gas.
The Galactic bulge is assumed to be a classical spheroid with a baryonic mass of 2$\times$10$^{9}$ M$_{\odot}$
and a dark-halo mass of 1.3$\times$10$^{10}$ M$_{\odot}$.
Cosmological parameters from the \citet{planck20} were adopted:
$\Omega_{m}$ = 0.31, $\Omega_{\Lambda}$ = 0.69, 
a Hubble constant of H$_{0}$= 68 km s$^{-1}$Mpc$^{-1}$,
and an age of the Universe of 13.801$\pm$0.024 Gyr.

For the nucleosynthesis yields, we adopted (i)
the metallicity-dependent yields from core-collapse supernovae and type II supernovae (SNe II) from \citet{woosley95} for massive stars, 
and for low metallicities (Z $<$ 0.01 Z$_{\odot}$ , or [Fe/H] $< -$2.5), the yields
are from high-explosion-energy hypernovae from \citet{nomoto13};
(ii) type Ia supernovae (SNe Ia) yields from \citet{iwamoto99} –their models W7 
(progenitor star of initial metallicity
Z = Z$_{\odot}$) and W70 (zero initial metallicity); and (iii) yields from \citet{vandenhoek97}  with variable $\eta$ (the asymptotic giant branch case) for intermediate-mass
stars (0.8–8 M$_{\odot}$) with initial Z = 0.001, 0.004, 0.008, 0.02, and
0.4.
For the metal-poor stars, neutrino-interaction processes
were also considered, adopting yields from \citet{yoshida08}. For more details, see \citet[][]{barbuy25a}.

Corrections to yields from WW95 follow recommendations from
\citet{timmes95}; these are a factor-of-two multiplication for Co and P, and a factor
of up to four for K measurements \citep[see more details in][]{barbuy24,barbuy25a}.
Models were computed for radii of r $< 0.5$, $0.5 <$ r $< 1$, $1 <$ r $< 2$,
and $2 <$ r $< 3$ kpc from the Galactic centre and for specific star-formation rate
values of $\nu = 1$ and $3$ Gyr$^{-1}$. The final abundances are reported in Table A.1.

\section{Discussion}
\label{discussion}

\begin{figure}
    \centering
\includegraphics[width=9cm]{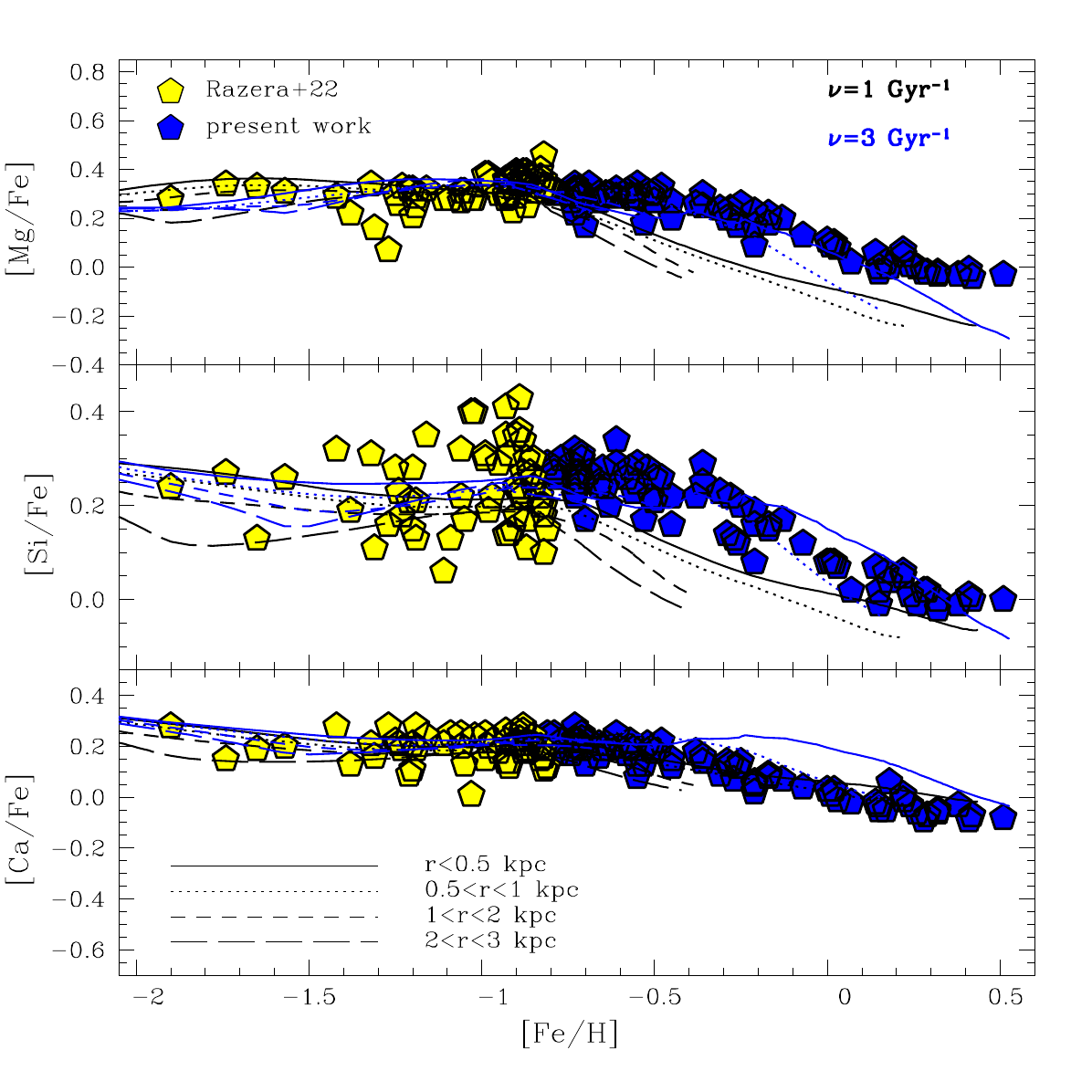}
    \caption{[Mg,Si,Ca/Fe] versus [Fe/H] for bulge spheroid stars.
 Filled blue pentagons show the ASPCAP results for the present sample; filled yellow pentagons the ASPCAP results for the sample described in \citet{razera22}. 
 For the chemical-evolution models, different model lines correspond to the outputs of models computed for radii of r $<$ 0.5, 0.5 $<$ r $<$ 1, 1 $<$ r $<$ 2,  and 2 $<$ r $<$ 3 kpc from the Galactic centre. Black lines correspond to
    specific star formation of $\nu$ = 1 Gyr$^{-1}$, and blue lines refer to $\nu$ = 3 Gyr$^{-1}$.    }
\label{mgsica}
\end{figure}

Figure \ref{mgsica} shows the fit to the alpha elements Mg, Si, and Ca, which were derived in the ASPCAP procedure both for the present sample and the metal-poor spheroid sample described in \citet{razera22}. 
For these elements, the lines are clean and the ASPCAP results 
are fully reliable, although for Si there is a non-negligible
spread on the metal-poor side. Our chemical evolution model was applied to these alpha elements in \cite{razera22},
and they were adopted here. The behaviour of these alpha elements is as expected from
an old stellar population enriched by SNe II,
where the lowering values of the [alpha/Fe] elements is due
to the enrichment in Fe by SNe Ia over time; these have a time lag relative to the earliest chemical enrichment by massive stars. The chemical-evolution models form the bulk of the bulge in about 1-0.3 Gyr, showing the abundance ratios of alpha elements to be compatible with a fast chemical enrichment. Note that the more metal-rich stars (from [Fe/H] $\simeq -$0.7 onwards) appear to require models with higher star-formation efficiency to reproduce the Mg/Fe and Si/Fe ratios, whereas for Ca the same models over-predict the Ca/Fe ratios for metallicities larger than [Fe/H] $\simeq -$0.4.

\begin{figure}
    \centering
    \includegraphics[width=1.\linewidth]{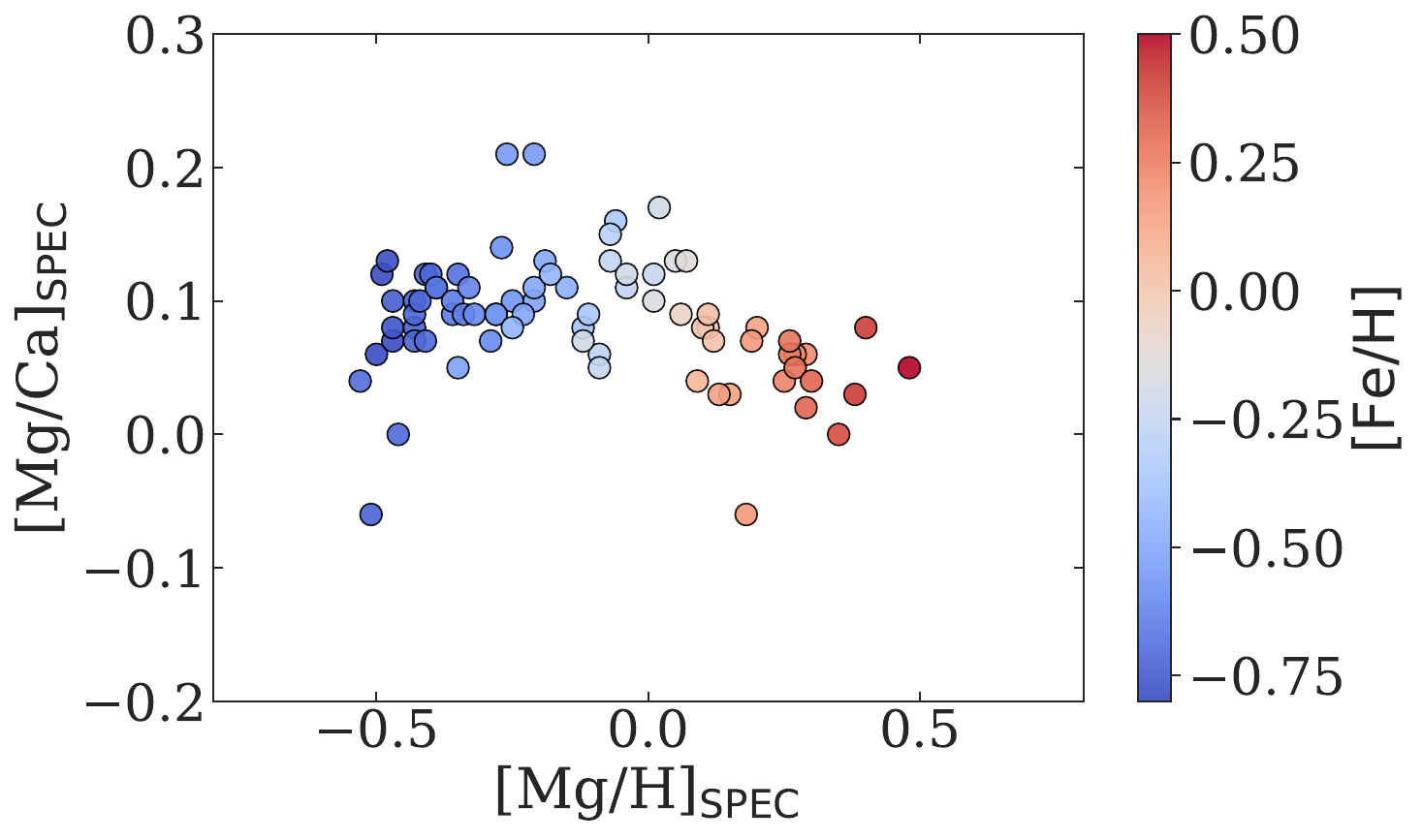}
    \caption{[Mg/Ca] versus [Mg/H] indicating the nucleosynthesis proxies of hydrostatic phases (Mg) compared with a product of
    explosive phases (Ca). These are colour-coded by their [Fe/H] abundances using the non-calibrated APOGEE DR17 abundances. }
    \label{fig:hexplot}
\end{figure}

Further evidence of alpha elements is shown in Fig. \ref{fig:hexplot}, which illustrates the magnesium excess relative to calcium in the metal-poor stars. Mg abundance is higher as a consequence of  it being formed in the hydrostatic phases of massive stars, while calcium is produced in the explosive phases of the SNe II and SNe Ia \citep[e.g.][]{mcwilliam16}.

\begin{figure}
\includegraphics[angle=0,width=9cm]{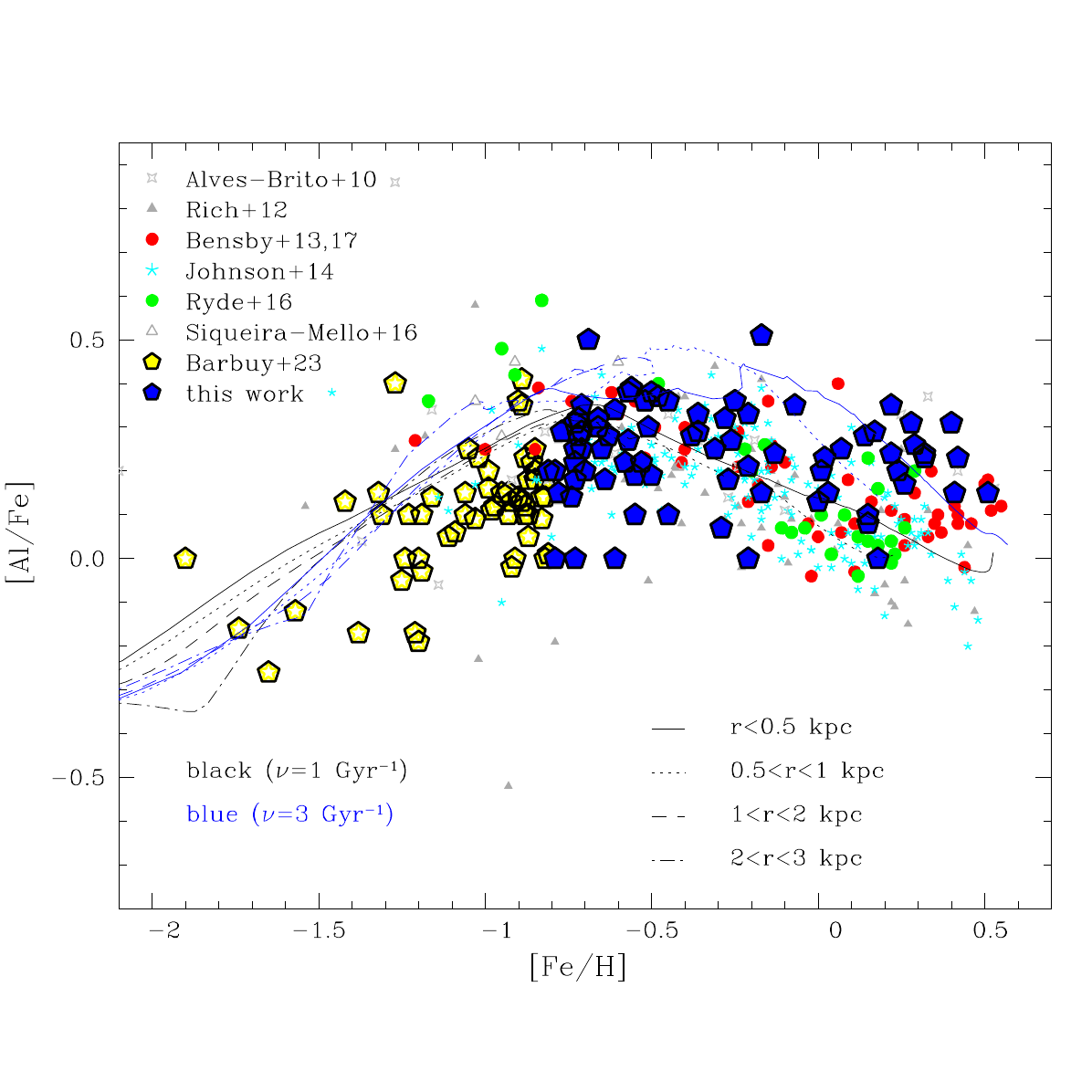}
\caption{[Al/Fe] versus [Fe/H]  for the present results compared with literature bulge samples and chemical evolution models.
Blue pentagons refer to this work; four-pointed grey stars to \citet{alves-brito10};
filled red circles to \citet{bensby17};
filled grey triangles \citet{}and 
cyan stars to \citet{johnson14}; filled green circles to \citet{ryde16}; open grey triangles to \citet{siqueira-mello16};
and yellow pentagons to \citet{barbuy23}.
Chemodynamical-evolution models with star formation rate of $\nu$ = 1 and 3 Gyr$^{-1}$  or formation timescale 
of 1 and 0.3 Gyr are over-plotted.}
\label{plotal78}
\end{figure}

In Fig. \ref{plotal78}, we plot the present results for [Al/Fe],
compared with literature values, and those for chemical-evolution models.
The LTE Al abundances fit the chemical-evolution models,
whereas the non-LTE results are too low. 

The literature data include results from 
\citet{alves-brito10}, \citet{bensby17}, 
\citet{johnson14},  \citet{ryde16}, \citet{siqueira-mello16}, and
\citet{barbuy23}.  The different datasets trace a consistent picture and are compatible with model predictions; Al/Fe increases with metallicity, reaches a peak, and then declines as the bulk of SNe Ia contributions to iron set in. The models also predict a substantial scatter at higher metallicities, highlighting the importance of this regime between [Fe/H] $-$0.8 to solar and beyond when it comes to constraining chemodynamical models of the bulge. The rather high Al abundance at
the metal-rich end may be due to metallicity-dependent yields from
SNe II \citep{curtis19} and SNe Ia \citep{bravo19a,keegans23}.

\begin{figure}
    \centering
    \includegraphics[width=1.0\linewidth]{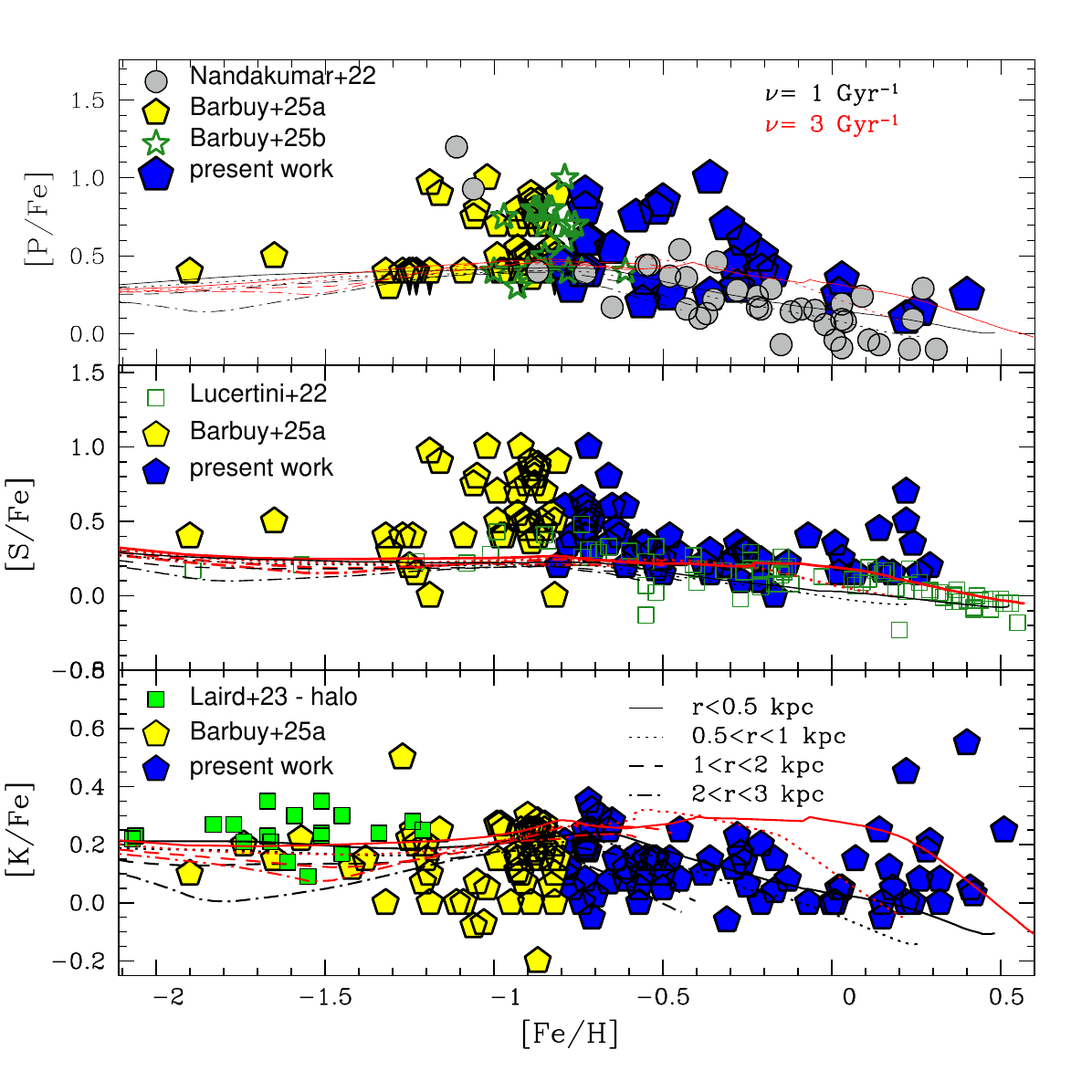}
    \caption{[P/Fe] versus [Fe/H] (upper panel),  [S/Fe] versus [Fe/H] (middle panel), and [K/Fe] versus [Fe/H] (lower panel) for the present data compared with literature and chemical evolution models.
 Blue pentagons show the present results, and yellow pentagons show those of \citet{barbuy25a}.
For P, dark green open stars are for \citet{barbuy25b}, filled light grey circles for \citet{nandakumar22}.
 For S, open forest green squares are for \citet{lucertini22}.
 For K, filled green circles show the \citet{reinhard24} results.
                Different model lines correspond to the outputs of models computed for radii r $<$ 0.5, 0.5 $<$ r $<$ 1, 1 $<$ r $<$ 2, and 2 $<$ r $<$ 3 kpc from the Galactic centre. Black lines correspond to
                a specific star formation of $\nu$ = 1 Gyr$^{-1}$, while red lines show $\nu$ = 3 Gyr$^{-1}$.     
    }
    \label{psk78}
\end{figure}

The most interesting result of the present work is the excess abundances of P around the  $-$0.9 $\simless$ [Fe/H] $\simless -$0.6 metallicity range. This effect was already pointed out by \citet{masseron20} and \citet{brauner23,brauner24}, and it was also detected
by \citet{barbuy25a,barbuy25b}. In Fig. \ref{psk78} we show the new data compared with values from previous works on the bulge;
these include, in particular, our results for bulge stars with [Fe/H] $< -$0.8, which are complementary
to the present data. 
Also included are stars from moderately metal-rich
GCs by \citet{barbuy25b}. We added the data of 
\citet{nandakumar22} for disc stars
because their sample also shows a P excess at the
same metallicities of our sample.
It confirms that the higher P excesses are found at metallicities of $-$1.1 $<$ [Fe/H] $< -$0.5,
as already pointed out by \citet{brauner23}. Here, we found P excess down to [Fe/H] $\sim -$1.2.
We suggest that this is a characteristic of the oldest bulge stars or the inner disc.
In Fig. \ref{psk78}, we over-plot the chemical-evolution models for specific star-formation rates
of $\nu$ = 1 and 3 Gyr$^{-1}$ for the phosphorus data. The models were
adopted from \citet{barbuy25a}. It is clear that models assuming standard nucleosynthesis for phosphorus are unable to reproduce the observed P excess, suggesting that this discrepancy may be related to the nature of the stars contributing to the very early phases of bulge enrichment (see discussions in \citealt{chiappini11} and \citealt{barbuy18a}).
The evidence showing that it occurs in the limited-metallicity range cited above for our selected sample with characteristics of bulge spheroid stars coincides with the recent identification of the old spheroidal bulge from a large sample of 
APOGEE and \textit{Gaia} RVS (Radial Velocity Spectrometer) by \citet{nepal25}; this sample shows a peak metallicity at [Fe/H] $\sim -$0.7.
The P excess in part of the stars could be seen as an effect
of second-generation stars stripped from GCs or, otherwise, due to an early enrichment by massive stars \citet[see discussion]{barbuy25b}.

Figure \ref{psk78} (middle panel) shows the present sulfur abundances compared with literature data by
\citet{lucertini22} and \citet{barbuy25a}.
We see an S excess particularly around a metallicity of [Fe/H]$\sim -$0.7, and the same is found from the ASPCAP results. The non-LTE effects, as
computed by \citet{korotin25}, partly explain this excess. 

Figure \ref{psk78} (lower panel) shows the K abundances for our sample compared with literature data, including those of  \citet{barbuy25a}
and \citet{reinhard24} for halo stars, which we included to show the
 level of K abundances on the metal-poor side.
There is a considerable spread in the K abundances, but it is compatible with the models. 
The K excess of the two very metal-rich and K-rich stars is confirmed, and the spectra cannot be
reproduced with the lower ASPCAP K abundances.
The yields of $^{39}$K
are over-estimated by the classic models of SNe II by a (high) factor of ten; following \citet{timmes95},
we adopted a factor of four for Z$\leq$Z$_{\odot}$. Ingredients that can increase the yields would be the rotation of massive stars
\citep{limongi18} or shell mergers in the late evolution of massive stars \citep{rauscher02,ritter18}.
In fact, the abundances can inform us of the presence of these processes.

\begin{figure}
    \centering
    \includegraphics[width=9cm]{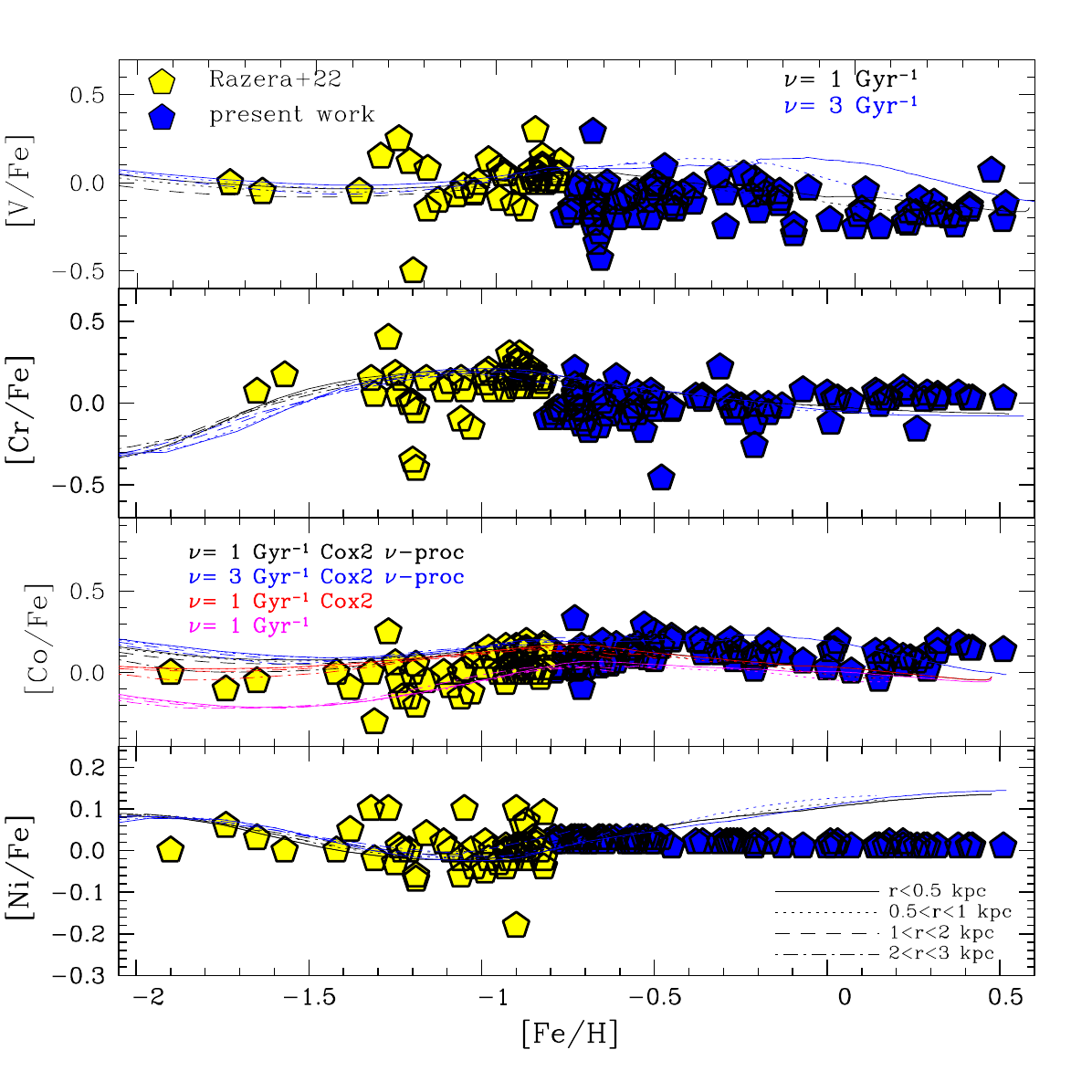}
    \caption{[V, Cr, Co, Ni/Fe] versus [Fe/H] for bulge spheroid stars.
 Symbols: Blue pentagons show ASPCAP results for the present sample, and yellow pentagons show ASPCAP results for the sample described in \citet{razera22}. Chemical-evolution models
 Different model lines correspond to the outputs of models computed for radii of r $<$ 0.5, 0.5 $<$ r $<$ 1, 1 $<$ r $<$ 2,  and 2 $<$ r $<$ 3 kpc from the Galactic centre. Black lines correspond to a specific star formation of $\nu$ = 1 Gyr$^{-1}$, and blue lines show $\nu$ = 3 Gyr$^{-1}$;  the red lines are models for
 $\nu$ = 1 Gyr$^{-1}$ and original Co yields from WW95 
 multiplied by 2.
    }
    \label{plotironpeak}
\end{figure}

The iron-peak elements V, Cr, Co, and Ni are plotted in
Fig. \ref{plotironpeak}. The abundances for the present sample
are the results from ASPCAP, which are combined with results from \citet{razera22} for the
metal-poor spheroid sample.
Our chemical-evolution models were presented in \citet{ernandes20} and \citet{barbuy24}, and the same models are over-plotted on the data.
The models fit the data very well, except for the metal-rich end
of Ni, where the models tend to give higher Ni than observed for [Fe/H]$>$ $-$0.5. This is due to yields
from SNe Ia from \citet{iwamoto99}, which over-estimated Ni. As commented in \citet{barbuy24}, 
the W7 model over-produces $^{58}$Ni, the main Ni isotope. 
We also needed to take into account the impact of the uncertainties of the weak interaction rates on the yields predicted by Chandrasekhar-mass and sub-Chandrasekhar models of SNe Ia \citep{bravo19a,bravo19b}. As an example, for sub-Chandrasekhar SNe Ia, the yields of $^{58}$Ni could increase by a factor of two.

\begin{figure}
    \centering
    \includegraphics[width=9.0cm]{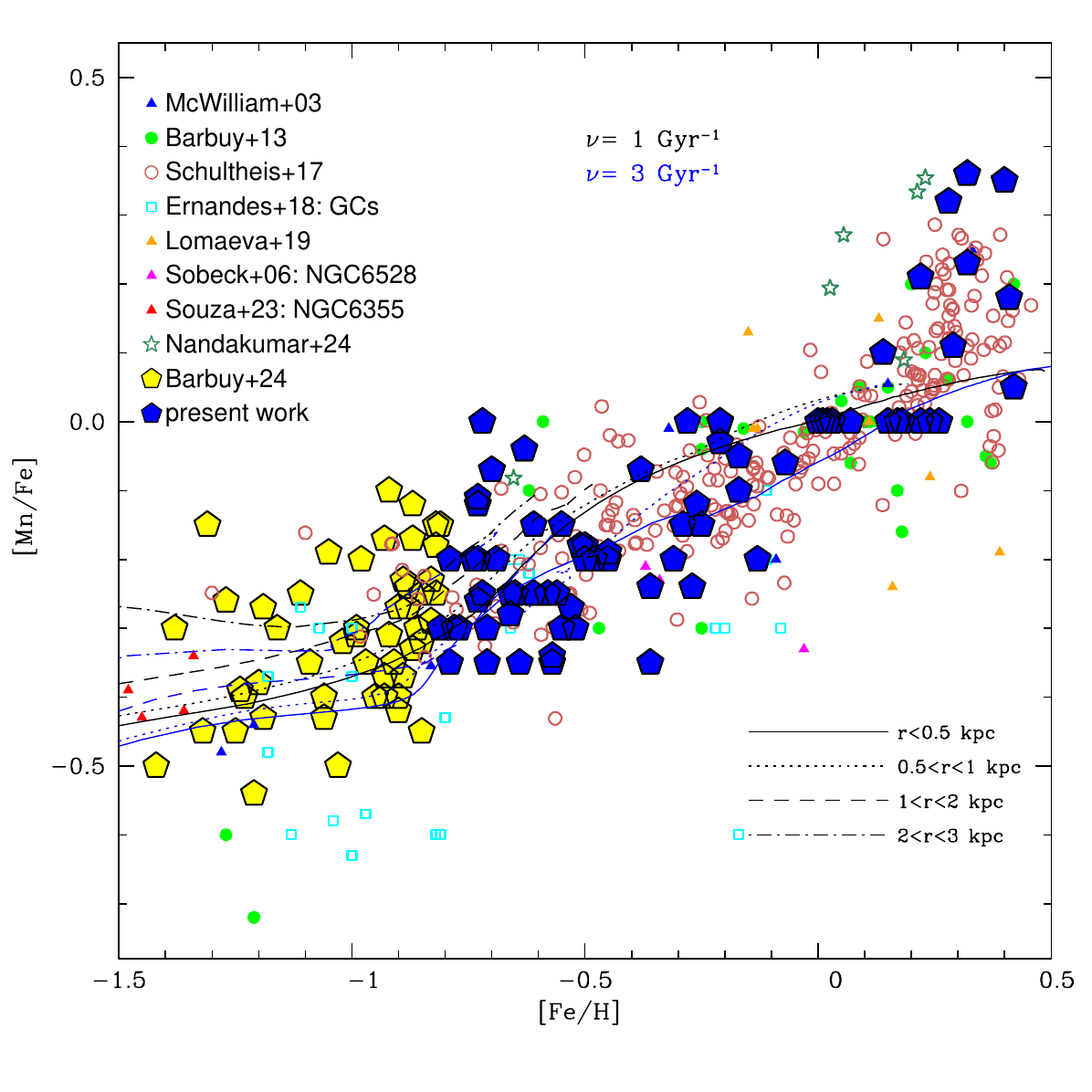}
    \caption{Lower panel: [Mn/Fe] versus [Fe/H]. Chemical-evolution models with star formation rates of $\nu$ = 1  and 3 Gyr $^{-1}$ (black and blue lines, respectively) are over-plotted on the results of the present study (blue pentagons) and literature data.
    Bulge GCs are from \citet[blue squares]{ernandes18}, 
    bulge field stars are from \citet[sea green open stars]{nandakumar24}, \citet[filled orange triangles]{lomaeva19},  \citet[open red circles]{schultheis17}, \citet[filled green circles]{barbuy13}, \citet[filled blue triangles]{mcwilliam03}, \citet[yellow pentagons]{barbuy24}, and GCs NGC~6528 by \citet[filled magenta triangles]{sobeck06}, NGC~6355 by \citet[filled red triangles]{souza23}.  
    Different model lines correspond to the outputs of models computed for radii r $<$ 0.5, 0.5 $<$ r $<$ 1, 1 $<$ r $<$ 2, and 2 $<$ r $<$ 3 kpc from the Galactic centre.} 
    \label{mn78star}
\end{figure}

For the iron-peak element Mn, in Fig. \ref{mn78star}  the abundances are shown after being re-derived in this work for the sample stars; we added the \citet{barbuy24} results for the metal-poor sample and compared them with literature results for bulge stars, 
 including the  GCs from \citealt{ernandes18}, the metal-rich cluster NGC~6528 from \citet{sobeck06}, and the  relatively 
 metal-poor cluster NGC~6355 from \citet{souza23}. We also considered bulge field stars 
\citep{nandakumar24,lomaeva19,schultheis17,barbuy13}; \citet{mcwilliam03} data are from the inner galaxy Sagittarius.
We note that although the values of \citet{schultheis17} are based on the early results from the APOGEE Data Release 13 (DR13), their dataset is still compatible with our new analysis.
The models are the same as in \citet{barbuy24}, and they clearly show the secondary behaviour of Mn. Overall, the models reproduce most of the data well. Interestingly, the most metal-rich stars exhibit an excess in [Mn/Fe] compared to the model predictions, which may indicate that these stars belong to a distinct stellar population and have experienced a different chemical evolution history. In addition, as discussed previously, a similar behaviour was also shown for K, where some of the most metal-rich stars clearly display enhanced values. Some of the over-abundances found here for the most metal-rich stars remind us of those found recently by \citet{Ryde2025} for nuclear stellar disc (NSD) giants and \citet{nandakumar2025} for stars in an NSD cluster. An interesting feature is the presence of Mn-rich, metal-rich stars. The increase of [Mn/Fe] with [Fe/H] was already shown by \citet{mcwilliam16}. A possible explanation is metallicity-dependent yields, similar to those proposed above for Al. This hypothesis was tested, for instance, in \citet{cescutti08}, which computed chemical-evolution models that explicitly include metallicity-dependent SN Ia yields for Mn.

\begin{figure}
    \centering
    \includegraphics[width=1.0\linewidth]{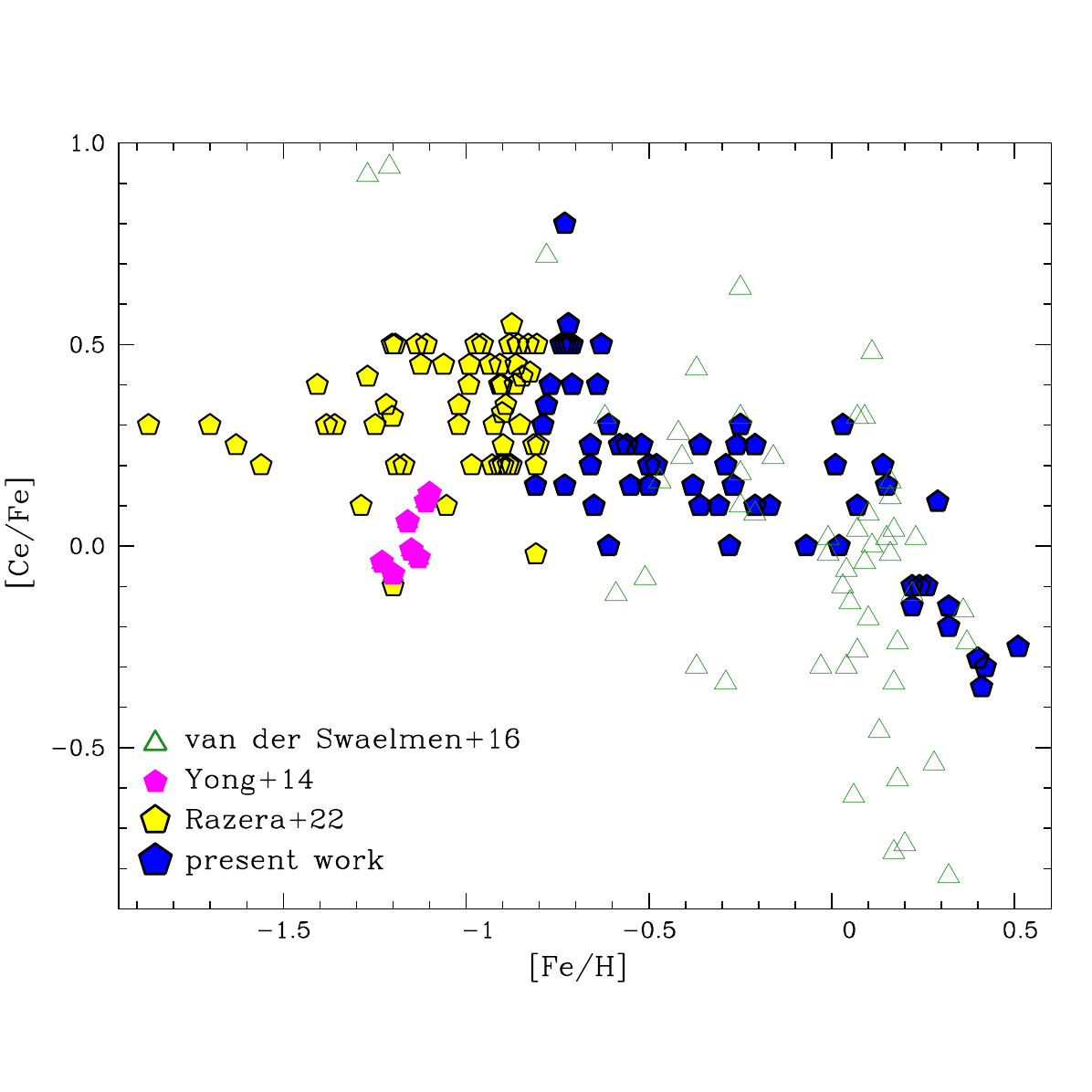}
    \caption{[Ce/Fe] versus [Fe/H] for the present data (open blue stars) compared with literature data: \citet[open red stars]{barbuy25a}, \citet[filled dark green triangles]{vanderswaelmen16}, and \citet[filled magenta pentagons]{yong14}. }
    \label{plotcex}
\end{figure}

Another result of interest is the behaviour of Ce abundances, shown in Fig. \ref{plotcex}, with enhancements at the same metallicity range as the P enhancements.
The correlation with P and the heavy elements of the
ﬁfirst peak (Sr, Y, and Zr) and second peak (Ba, La, Ce, and Nd), which are predominantly s elements was suggested by \citet{masseron20b}.
The present results, together with those from \citet{razera22},
might suggest such a correlation, but with a large spread. For the most-metal rich stars, we find [Ce/Fe] to be below solar; this is again similar to what was found in the NSD by \cite{Ryde2025} and \citet{nandakumar2025}.
In other words, the higher values of [Ce/Fe] at [Fe/H] $\sim$ -0.7 appear to
be an effect that we also noted with phosphorus, which is typical of the oldest bulge stellar population; this should
be further investigated in future studies.
The marked decline at the higher metallicities is another striking feature, and it could be explained
by the fast star formation rate in the bulge, leading to solar-metallicity stars not having had time
to be enriched by asymptotic giant branch stars.

\begin{figure}
    \centering
    \includegraphics[width=1.0\linewidth]{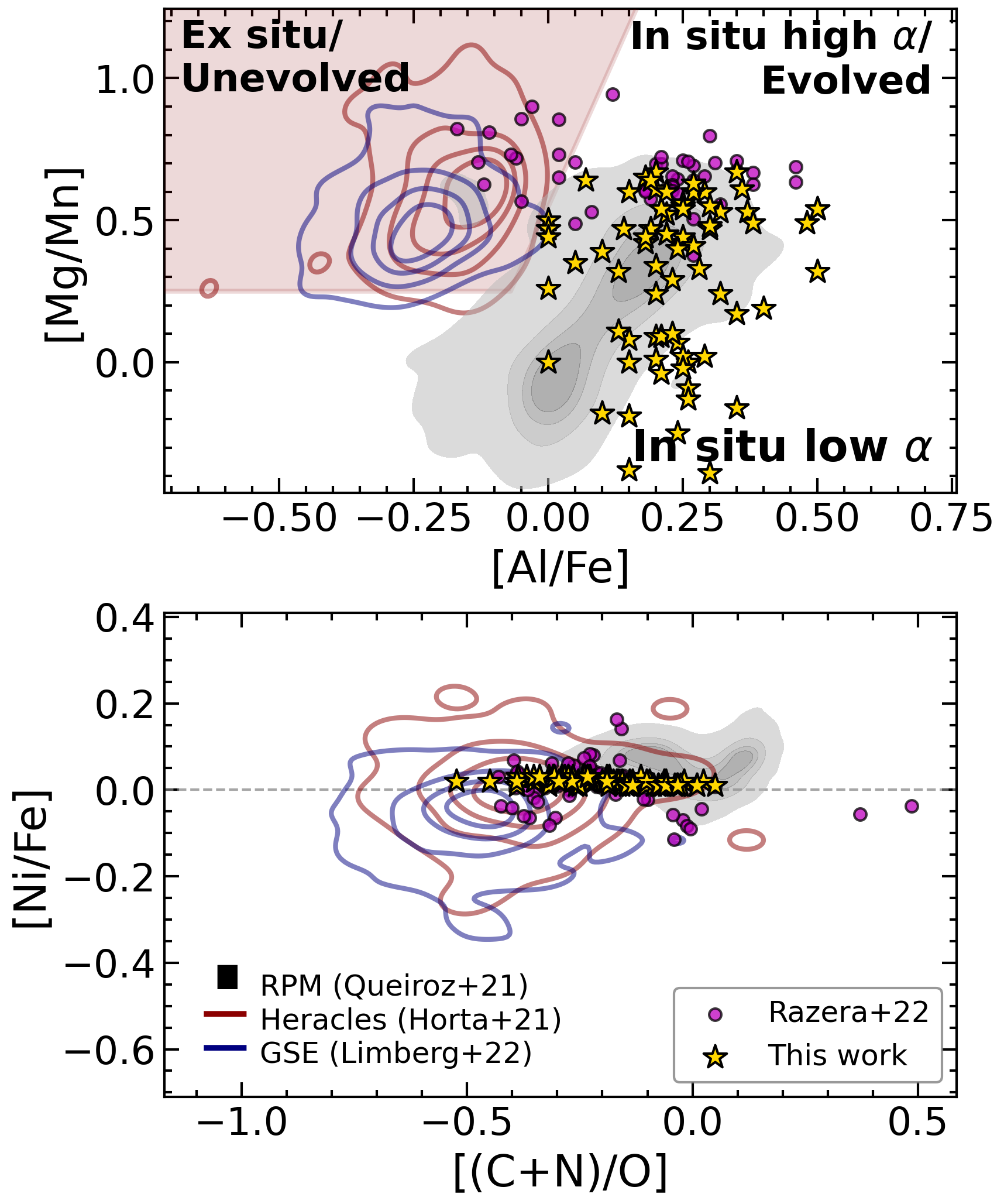}
    \caption{Plotting distribution of our sample stars in the [Mg/Mn] versus [Al/Fe] parameter space,
    and [Ni/Fe] versus [C+N/O], compared with the loci of the RPM sample and the Heracles and Gaia-Sausage-Enceladus  structures.
    The grey background points show the RPM sample from \citet{queiroz21}.
    }
    \label{fig:mgmnal}
\end{figure}

An important verification is given in 
Fig. \ref{fig:mgmnal} displaying the [Mg/Mn] versus [Al/Fe], which indicates the location of the in situ and ex situ stars as well as high-alpha and low-alpha ones. In the present sample, the more metal-rich stars show a low alpha, as expected, even if they were selected to be part of the spheroidal bulge \citep[see also discussion in][]{nepal25}. All our stars have [Al/Fe] $\simgreat$ 0.0, which is typical of in situ stars. The sample was compared with the loci of the RPM sample by \citet{queiroz21}, the Gaia--Enceladus--Sausage \citep{limberg22}, and Heracles \citep{horta21}. All sample stars are in the in situ origin region. As for those showing low-alpha stars, this is expected since they are metal-rich ([Fe/H]$\geq$0.0), as can be seen in Fig. \ref{mgsica}. Some of the most metal-rich ones also show large enhancement of Mn (see Fig. \ref{mn78star}), further lowering the [Mg/Mn] ratio. The Ni abundances are all [Ni/Fe]$\geq$0.0, therefore once more indicating the in situ origin of the sample stars. 

\section{Conclusions}\label{conclusions}

We studied bulge sample stars selected from kinematical and
dynamical criteria and with a metallicity of [Fe/H] $> -$0.8, thus
completing the work by \citet{razera22} and \citet{barbuy23,barbuy24,barbuy25a}
on a symmetric sample with [Fe/H] $< -$0.8.
APOGEE abundances of the alpha elements Mg, Si, and Ca and the iron-peak
elements V, Cr, Co, and Ni from the APOGEE-ASPCAP DR17 were adopted,
and these elements follow the chemical-evolution models very well.

We re-derived C, N, O, even if the ASPCAP results were reliable, 
due to the need to have these molecular lines fitted adequately in
order to rely on the derivation of the P abundances, for which the
unique line is blended with CO.
Al abundances were used as indicators of an in situ or ex situ origin of stars.
It is important to stress that the indicator is Al in LTE, given that the
relative abundances are the reference, as explained in \citet{Ernandes25}.
In the present work, essentially all the sample stars show [Al/Fe]$\geq$0.0.
S, K, Mn, and Ce were also re-derived, given our previous experience
showing that these lines are susceptible to blends or the delicate 
fitting of profiles.

A first important result is represented by the excesses of P and, possibly, of Ce,
with a peak at metallicity [Fe/H] $\sim -$0.7 to $-$0.8. 
This is particularly interesting given that the selection of 
bulge spheroidal stars by \citet{nepal25} shows a metallicity peak
exactly at [Fe/H]$=-$0.7. Therefore, the behaviour of P and Ce
might be a particular characteristic of the earliest stellar populations
in the Galaxy.
S and K show a large spread, but they are compatible with models.
For S in particular, non-LTE corrections is rather important.

Finally, the [Mg/Mn] versus [Al/Fe] plot suggested by \citet{Hawkins15}
to be a discriminator between in situ and ex situ stars
indicates that our sample appears to have characteristics of an in situ
origin.
The complementary plot [Ni/Fe] versus [(C+N)/O]  \citep{montalban21,ortigoza-urdaneta23}
also indicates the in situ origin of the sample stars.

Another interesting result, which needs to be confirmed with more data, is related to the most metal-rich stars in our sample. Some of these stars show clear enhancements of K and Mn (and maybe S) and low Ce/Fe ratios. Some of the abundance behaviour observed in the super-solar metallicity regime may be connected to enrichment processes specific to the nuclear disc and the Galactic bar (see discussions in \citealt{nepal25} and \citealt{Ryde2025}). In particular, the enhanced abundance ratios seen in the most metal-rich stars could reflect a distinct chemical evolution pathway influenced by the high star-formation efficiency, gas inflows, and bar formation in the innermost regions. If so, these stars may not simply represent the metal-rich extension of inner-disc or bulge populations, but instead trace stellar populations shaped by the unique dynamical and chemical environment of the nuclear disc and bar. Further observational and modelling efforts will be required to disentangle these effects and to assess the relative contributions of different enrichment channels at the highest metallicities. 

\begin{acknowledgements}
H.E. and S.F. were supported by a project grant from the Knut and
Alice Wallenberg Foundation (KAW 2020.0061 Galactic Time
Machine, PI Feltzing).
B.B., A.C.S.F., and C.C. acknowledge grants from FAPESP, Conselho Nacional de Desenvolvimento Cient\'ifico e Tecnol\'ogico (CNPq) and Coordena\c{c}\~ao de Aperfei\c{c}oamento de Pessoal de N\'ivel Superior (CAPES) - Financial code 001. H.E. acknowledges a post-doctoral fellowship at Lund Observatory.
S.O.S. acknowledges the support from Dr. Nadine Neumayer's Lise Meitner grant from the Max Planck Society.
A.P.-V., B.B., and S.O.S. acknowledge the DGAPA-PAPIIT grants IA103224 and IN112526
R.P.N. and A.L.R.A. acknowledge the FAPESP Inicia\c c\~ao
Cient\'{\i}fica fellowships no. 2025/05122-1 and 2025/00600-5. 
BB, HE, PS and SOS are part of the Brazilian Participation Group (BPG) in the Sloan Digital Sky Survey (SDSS), from the
Laborat\'orio Interinstitucional de e-Astronomia – LIneA, Brazil.
J.G.F-T gratefully acknowledges the grants support provided by ANID Fondecyt Postdoc No. 3230001 (Sponsoring researcher), the Joint Committee ESO-Government of Chile under the agreement 2023 ORP 062/2023, and the support of the Doctoral Program in Artificial Intelligence, DISC-UCN.
Apogee project: Funding for the Sloan Digital Sky Survey IV has been provided by the Alfred P. Sloan Foundation, the U.S. Department of Energy Office of Science, and the Participating Institutions. SDSS acknowledges support and resources from the Center for High-Performance Computing at the University of Utah. The SDSS website is www.sdss.org. 
SDSS is managed by the Astrophysical Research Consortium for the Participating Institutions of the SDSS Collaboration including the Brazilian Participation Group, the Carnegie Institution for Science, Carnegie Mellon University, Center for Astrophysics | Harvard \& Smithsonian (CfA), the Chilean Participation Group, the French Participation Group, Instituto de Astrof\'isica de Canarias, The Johns Hopkins University, Kavli Institute for the Physics and Mathematics of the Universe (IPMU) / University of Tokyo, the Korean Participation Group, Lawrence Berkeley National Laboratory, Leibniz Institut f\"ur Astrophysik Potsdam (AIP), Max-Planck-Institut f\"ur Astronomie (MPIA Heidelberg), Max-Planck-Institut f\"ur Astrophysik (MPA Garching), Max-Planck-Institut f\"ur Extraterrestrische Physik (MPE), National Astronomical Observatories of China, New Mexico State University, New York University, University of Notre Dame, Observat\'orio Nacional / MCTI, The Ohio State University, Pennsylvania State University, Shanghai Astronomical Observatory, United Kingdom Participation Group, Universidad Nacional Aut\'onoma de M\'exico, University of Arizona, University of Colorado Boulder, University of Oxford, University of Portsmouth, University of Utah, University of Virginia, University of Washington, University of Wisconsin, Vanderbilt University, and Yale University.
\end{acknowledgements}


\bibliographystyle{aa} 
\bibliography{bibliogpsk}

@ARTICLE{abdurro22,
       author = {{Abdurro'uf} and {Accetta}, Katherine and {Aerts}, Conny and {Silva Aguirre}, V{\'\i}ctor and {Ahumada}, Romina and {Ajgaonkar}, Nikhil and {Filiz Ak}, N. and {Alam}, Shadab and {Allende Prieto}, Carlos and {Almeida}, Andr{\'e}s and {Anders}, Friedrich and {Anderson}, Scott F. and {Andrews}, Brett H. and {Anguiano}, Borja and {Aquino-Ort{\'\i}z}, Erik and {Arag{\'o}n-Salamanca}, Alfonso and {Argudo-Fern{\'a}ndez}, Maria and {Ata}, Metin and {Aubert}, Marie and {Avila-Reese}, Vladimir and et al.},
        title = "{The Seventeenth Data Release of the Sloan Digital Sky Surveys: Complete Release of MaNGA, MaStar, and APOGEE-2 Data}",
      journal = {\apjs},
         year = 2022,
        month = apr,
       volume = {259},
       number = {2},
          eid = {35},
        pages = {35},
          doi = {10.3847/1538-4365/ac4414},
archivePrefix = {arXiv},
       eprint = {2112.02026},
 primaryClass = {astro-ph.GA},
       adsurl = {https://ui.adsabs.harvard.edu/abs/2022ApJS..259...35A}
}

@ARTICLE{allende-prieto06,
       author = {{Allende Prieto}, Carlos and {Beers}, Timothy C. and {Wilhelm}, Ronald and {Newberg}, Heidi Jo and {Rockosi}, Constance M. and {Yanny}, Brian and {Lee}, Young Sun},
        title = "{A Spectroscopic Study of the Ancient Milky Way: F- and G-Type Stars in the Third Data Release of the Sloan Digital Sky Survey}",
      journal = {\apj},
         year = 2006,
        month = jan,
       volume = {636},
       number = {2},
        pages = {804-820},
          doi = {10.1086/498131},
archivePrefix = {arXiv},
       eprint = {astro-ph/0509812},
 primaryClass = {astro-ph},
       adsurl = {https://ui.adsabs.harvard.edu/abs/2006ApJ...636..804A}
}

@ARTICLE{alvarez98,
       author = {{Alvarez}, R. and {Plez}, B.},
        title = "{Near-infrared narrow-band photometry of M-giant and Mira stars: models meet observations}",
      journal = {\aap},
         year = 1998,
        month = feb,
       volume = {330},
        pages = {1109-1119},
          doi = {10.48550/arXiv.astro-ph/9710157},
archivePrefix = {arXiv},
       eprint = {astro-ph/9710157},
 primaryClass = {astro-ph},
       adsurl = {https://ui.adsabs.harvard.edu/abs/1998A&A...330.1109A}
}

@ARTICLE{alves-brito10,
       author = {{Alves-Brito}, A. and {Mel{\'e}ndez}, J. and {Asplund}, M. and {Ram{\'\i}rez}, I. and {Yong}, D.},
        title = "{Chemical similarities between Galactic bulge and local thick disk red giants: O, Na, Mg, Al, Si, Ca, and Ti}",
      journal = {\aap},
         year = 2010,
        month = apr,
       volume = {513},
          eid = {A35},
        pages = {A35},
          doi = {10.1051/0004-6361/200913444},
archivePrefix = {arXiv},
       eprint = {1001.2521},
 primaryClass = {astro-ph.SR},
       adsurl = {https://ui.adsabs.harvard.edu/abs/2010A&A...513A..35A}
}

@ARTICLE{asplund21,
       author = {{Asplund}, M. and {Amarsi}, A.~M. and {Grevesse}, N.},
        title = "{The chemical make-up of the Sun: A 2020 vision}",
      journal = {\aap},
         year = 2021,
        month = sep,
       volume = {653},
          eid = {A141},
        pages = {A141},
          doi = {10.1051/0004-6361/202140445},
archivePrefix = {arXiv},
       eprint = {2105.01661},
 primaryClass = {astro-ph.SR},
       adsurl = {https://ui.adsabs.harvard.edu/abs/2021A&A...653A.141A}
}

@ARTICLE{barbuy13,
       author = {{Barbuy}, B. and {Hill}, V. and {Zoccali}, M. and {Minniti}, D. and {Renzini}, A. and {Ortolani}, S. and {G{\'o}mez}, A. and {Trevisan}, M. and {Dutra}, N.},
        title = "{Manganese abundances in Galactic bulge red giants}",
      journal = {\aap},
         year = 2013,
        month = nov,
       volume = {559},
          eid = {A5},
        pages = {A5},
          doi = {10.1051/0004-6361/201322380},
archivePrefix = {arXiv},
       eprint = {1310.4108},
 primaryClass = {astro-ph.GA},
       adsurl = {https://ui.adsabs.harvard.edu/abs/2013A&A...559A...5B}
}

@ARTICLE{barbuy18a,
       author = {{Barbuy}, Beatriz and {Chiappini}, Cristina and {Gerhard}, Ortwin},
        title = "{Chemodynamical History of the Galactic Bulge}",
      journal = {\araa},
         year = 2018,
        month = sep,
       volume = {56},
        pages = {223-276},
          doi = {10.1146/annurev-astro-081817-051826},
archivePrefix = {arXiv},
       eprint = {1805.01142},
 primaryClass = {astro-ph.GA},
       adsurl = {https://ui.adsabs.harvard.edu/abs/2018ARA&A..56..223B}
}

@ARTICLE{barbuy21b,
       author = {{Barbuy}, B. and {Ernandes}, H. and {Souza}, S.~O. and {Razera}, R. and {Moura}, T. and {Mel{\'e}ndez}, J. and {P{\'e}rez-Villegas}, A. and {Zoccali}, M. and {Minniti}, D. and {Dias}, B. and {Ortolani}, S. and {Bica}, E.},
        title = "{Gemini/Phoenix H-band analysis of the globular cluster AL 3}",
      journal = {\aap},
         year = 2021,
        month = apr,
       volume = {648},
          eid = {A16},
        pages = {A16},
          doi = {10.1051/0004-6361/202039761},
archivePrefix = {arXiv},
       eprint = {2102.12674},
 primaryClass = {astro-ph.SR},
       adsurl = {https://ui.adsabs.harvard.edu/abs/2021A&A...648A..16B}
}

@ARTICLE{barbuy23,
       author = {{Barbuy}, B. and {Fria{\c{c}}a}, A.~C.~S. and {Ernandes}, H. and {Moura}, T. and {Masseron}, T. and {Cunha}, K. and {Smith}, V.~V. and {Souto}, D. and {P{\'e}rez-Villegas}, A. and {Souza}, S.~O. and {Chiappini}, C. and {Queiroz}, A.~B.~A. and {Fern{\'a}ndez-Trincado}, J.~G. and {da Silva}, P. and {Santiago}, B.~X. and {Anders}, F. and {Schiavon}, R.~P. and {Valentini}, M. and {Minniti}, D. and {Geisler}, D. and {Placco}, V.~M. and {Zoccali}, M. and {Schultheis}, M. and {Nitschelm}, C. and {Beers}, T.~C. and {Razera}, R.},
        title = "{Light elements Na and Al in 58 bulge spheroid stars from APOGEE}",
      journal = {\mnras},
         year = 2023,
        month = dec,
       volume = {526},
       number = {2},
        pages = {2365-2376},
          doi = {10.1093/mnras/stad2888},
       adsurl = {https://ui.adsabs.harvard.edu/abs/2023MNRAS.526.2365B}
}

@ARTICLE{barbuy24,
       author = {{Barbuy}, B. and {Fria{\c{c}}a}, A.~C.~S. and {Ernandes}, H. and {da Silva}, P. and {Souza}, S.~O. and {Fern{\'a}ndez-Trincado}, J.~G. and {Cunha}, K. and {Smith}, V.~V. and {Masseron}, T. and {P{\'e}rez-Villegas}, A. and {Chiappini}, C. and {Queiroz}, A.~B.~A. and {Santiago}, B.~X. and {Beers}, T.~C. and {Anders}, F. and {Schiavon}, R.~P. and {Valentini}, M. and {Minniti}, D. and {Geisler}, D. and {Souto}, D. and {Placco}, V.~M. and {Zoccali}, M. and {Feltzing}, S. and {Schultheis}, M. and {Nitschelm}, C.},
        title = "{Abundances of iron-peak elements in 58 bulge spheroid stars from APOGEE}",
      journal = {\aap},
         year = 2024,
        month = nov,
       volume = {691},
          eid = {A296},
        pages = {A296},
          doi = {10.1051/0004-6361/202452235},
archivePrefix = {arXiv},
       eprint = {2410.13751},
 primaryClass = {astro-ph.SR},
       adsurl = {https://ui.adsabs.harvard.edu/abs/2024A&A...691A.296B}
}

@ARTICLE{barbuy25a,
       author = {{Barbuy}, B. and {Ernandes}, H. and {Fria{\c{c}}a}, A.~C.~S. and {Camargo}, M.~S. and {da Silva}, P. and {Souza}, S.~O. and {Masseron}, T. and {Brauner}, M. and {Garc{\'\i}a-Hern{\'a}ndez}, D.~A. and {Fern{\'a}ndez-Trincado}, J.~G. and {Cunha}, K. and {Smith}, V.~V. and {P{\'e}rez-Villegas}, A. and {Chiappini}, C. and {Queiroz}, A.~B.~A. and {Santiago}, B.~X. and {Beers}, T.~C. and {Anders}, F. and {Schiavon}, R.~P. and {Valentini}, M. and {Minniti}, D. and {Geisler}, D. and {Souto}, D. and {Placco}, V.~M. and {Zoccali}, M. and {Feltzing}, S. and {Schultheis}, M. and {Nitschelm}, C.},
        title = "{Abundances of P, S, and K in 58 bulge spheroid stars from APOGEE}",
      journal = {\aap},
         year = 2025,
        month = aug,
       volume = {700},
          eid = {A184},
        pages = {A184},
          doi = {10.1051/0004-6361/202555713},
archivePrefix = {arXiv},
       eprint = {2507.11667},
 primaryClass = {astro-ph.SR},
       adsurl = {https://ui.adsabs.harvard.edu/abs/2025A&A...700A.184B}
}

@ARTICLE{barbuy25b,
       author = {{Barbuy}, Beatriz and {Fern{\'a}ndez-Trincado}, Jos{\'e} G. and {Camargo}, Morgan S. and {Geisler}, Doug and {Brauner}, Maren and {Villanova}, Sandro and {Minniti}, Dante and {Garc{\'\i}a-Hern{\'a}ndez}, Domingo Anibal and {Souza}, Stefano O. and {Ernandes}, Heitor and {Fria{\c{c}}a}, Am{\^a}ncio and {Pignatari}, Marco},
        title = "{Investigating Phosphorus Abundances in a Sample of APOGEE-2 Bulge Globular Clusters}",
      journal = {\aj},
         year = 2025,
        month = oct,
       volume = {170},
       number = {4},
          eid = {245},
        pages = {245},
          doi = {10.3847/1538-3881/ae046b},
archivePrefix = {arXiv},
       eprint = {2509.11207},
 primaryClass = {astro-ph.SR},
       adsurl = {https://ui.adsabs.harvard.edu/abs/2025AJ....170..245B}
}

@ARTICLE{beaton21,
       author = {{Beaton}, Rachael L. and {Oelkers}, Ryan J. and {Hayes}, Christian R. and {Covey}, Kevin R. and {Chojnowski}, S.~D. and {De Lee}, Nathan and {Sobeck}, Jennifer S. and {Majewski}, Steven R. and {Cohen}, Roger E. and {Fern{\'a}ndez-Trincado}, Jos{\'e} and {Longa-Pe{\~n}a}, Pen{\'e}lope and {O'Connell}, Julia E. and {Santana}, Felipe A. and {Stringfellow}, Guy S. and {Zasowski}, Gail and {Aerts}, Conny and et al.},
        title = "{Final Targeting Strategy for the Sloan Digital Sky Survey IV Apache Point Observatory Galactic Evolution Experiment 2 North Survey}",
      journal = {\aj},
         year = 2021,
        month = dec,
       volume = {162},
       number = {6},
          eid = {302},
        pages = {302},
          doi = {10.3847/1538-3881/ac260c},
archivePrefix = {arXiv},
       eprint = {2108.11907},
 primaryClass = {astro-ph.GA},
       adsurl = {https://ui.adsabs.harvard.edu/abs/2021AJ....162..302B}
}

@ARTICLE{bensby17,
       author = {{Bensby}, T. and {Feltzing}, S. and {Gould}, A. and {Yee}, J.~C. and {Johnson}, J.~A. and {Asplund}, M. and {Mel{\'e}ndez}, J. and {Lucatello}, S. and {Howes}, L.~M. and {McWilliam}, A. and {Udalski}, A. and {Szyma{\'n}ski}, M.~K. and {Soszy{\'n}ski}, I. and {Poleski}, R. and {Wyrzykowski}, {\L}. and {Ulaczyk}, K. and {Koz{\l}owski}, S. and {Pietrukowicz}, P. and {Skowron}, J. and {Mr{\'o}z}, P. and {Pawlak}, M. and {Abe}, F. and {Asakura}, Y. and {Bhattacharya}, A. and {Bond}, I.~A. and {Bennett}, D.~P. and {Hirao}, Y. and {Nagakane}, M. and {Koshimoto}, N. and {Sumi}, T. and {Suzuki}, D. and {Tristram}, P.~J.},
        title = "{Chemical evolution of the Galactic bulge as traced by microlensed dwarf and subgiant stars. VI. Age and abundance structure of the stellar populations in the central sub-kpc of the Milky Way}",
      journal = {\aap},
         year = 2017,
        month = sep,
       volume = {605},
          eid = {A89},
        pages = {A89},
          doi = {10.1051/0004-6361/201730560},
archivePrefix = {arXiv},
       eprint = {1702.02971},
 primaryClass = {astro-ph.GA},
       adsurl = {https://ui.adsabs.harvard.edu/abs/2017A&A...605A..89B}
}

@ARTICLE{bergemann08Mn,
       author = {{Bergemann}, M. and {Gehren}, T.},
        title = "{NLTE abundances of Mn in a sample of metal-poor stars}",
      journal = {\aap},
         year = 2008,
        month = dec,
       volume = {492},
       number = {3},
        pages = {823-831},
          doi = {10.1051/0004-6361:200810098},
archivePrefix = {arXiv},
       eprint = {0811.0681},
 primaryClass = {astro-ph},
       adsurl = {https://ui.adsabs.harvard.edu/abs/2008A&A...492..823B}
}

@ARTICLE{blanton17,
       author = {{Blanton}, Michael R. and {Bershady}, Matthew A. and {Abolfathi}, Bela and {Albareti}, Franco D. and {Allende Prieto}, Carlos and {Almeida}, Andres and {Alonso-Garc{\'\i}a}, Javier and {Anders}, Friedrich and {Anderson}, Scott F. and {Andrews}, Brett and {Aquino-Ort{\'\i}z}, Erik and et al.},
        title = "{Sloan Digital Sky Survey IV: Mapping the Milky Way, Nearby Galaxies, and the Distant Universe}",
      journal = {\aj},
         year = 2017,
        month = jul,
       volume = {154},
       number = {1},
          eid = {28},
        pages = {28},
          doi = {10.3847/1538-3881/aa7567},
archivePrefix = {arXiv},
       eprint = {1703.00052},
 primaryClass = {astro-ph.GA},
       adsurl = {https://ui.adsabs.harvard.edu/abs/2017AJ....154...28B}
}

@ARTICLE{bowen73,
       author = {{Bowen}, I.~S. and {Vaughan}, A.~H., Jr.},
        title = "{The optical design of the 40-in. telescope and of the Ir{\'e}n{\'e}e DuPont telescope at Las Campanas Observatory, Chile.}",
      journal = {\ao},
         year = 1973,
        month = jan,
       volume = {12},
        pages = {1430-1434},
          doi = {10.1364/AO.12.001430},
       adsurl = {https://ui.adsabs.harvard.edu/abs/1973ApOpt..12.1430B}
}

@ARTICLE{brauner23,
       author = {{Brauner}, Maren and {Masseron}, Thomas and {Garc{\'\i}a-Hern{\'a}ndez}, Domingo A. and {Pignatari}, Marco and {Womack}, Kate A. and {Lugaro}, Maria and {Hayes}, Christian R.},
        title = "{Unveiling the chemical fingerprint of phosphorus-rich stars. I. In the infrared region of APOGEE-2}",
      journal = {\aap},
         year = 2023,
        month = may,
       volume = {673},
          eid = {A123},
        pages = {A123},
          doi = {10.1051/0004-6361/202346048},
archivePrefix = {arXiv},
       eprint = {2303.12590},
 primaryClass = {astro-ph.SR},
       adsurl = {https://ui.adsabs.harvard.edu/abs/2023A&A...673A.123B}
}

@ARTICLE{brauner24,
       author = {{Brauner}, Maren and {Pignatari}, Marco and {Masseron}, Thomas and {Garc{\'\i}a-Hern{\'a}ndez}, D.~A. and {Lugaro}, Maria},
        title = "{Unveiling the chemical fingerprint of phosphorus-rich stars: II. Heavy-element abundances from UVES/VLT spectra}",
      journal = {\aap},
         year = 2024,
        month = oct,
       volume = {690},
          eid = {A262},
        pages = {A262},
          doi = {10.1051/0004-6361/202451327},
archivePrefix = {arXiv},
       eprint = {2408.12938},
 primaryClass = {astro-ph.SR},
       adsurl = {https://ui.adsabs.harvard.edu/abs/2024A&A...690A.262B}
}

@ARTICLE{bravo19a,
       author = {{Bravo}, E.},
        title = "{Sensitivity of Type Ia supernovae to electron capture rates}",
      journal = {\aap},
         year = 2019,
        month = apr,
       volume = {624},
          eid = {A139},
        pages = {A139},
          doi = {10.1051/0004-6361/201935095},
archivePrefix = {arXiv},
       eprint = {1903.08344},
 primaryClass = {astro-ph.SR},
       adsurl = {https://ui.adsabs.harvard.edu/abs/2019A&A...624A.139B}
}

@ARTICLE{bravo19b,
       author = {{Bravo}, Eduardo and {Badenes}, Carles and {Mart{\'\i}nez-Rodr{\'\i}guez}, H{\'e}ctor},
        title = "{SNR-calibrated Type Ia supernova models}",
      journal = {\mnras},
         year = 2019,
        month = feb,
       volume = {482},
       number = {4},
        pages = {4346-4363},
          doi = {10.1093/mnras/sty2951},
archivePrefix = {arXiv},
       eprint = {1810.13123},
 primaryClass = {astro-ph.SR},
       adsurl = {https://ui.adsabs.harvard.edu/abs/2019MNRAS.482.4346B}
}

@ARTICLE{cescutti08,
       author = {{Cescutti}, G. and {Matteucci}, F. and {Lanfranchi}, G.~A. and {McWilliam}, A.},
        title = "{The chemical evolution of manganese in different stellar systems}",
      journal = {\aap},
         year = 2008,
        month = nov,
       volume = {491},
       number = {2},
        pages = {401-405},
          doi = {10.1051/0004-6361:200810537},
archivePrefix = {arXiv},
       eprint = {0807.1463},
 primaryClass = {astro-ph},
       adsurl = {https://ui.adsabs.harvard.edu/abs/2008A&A...491..401C}
}

@ARTICLE{chiappini11,
       author = {{Chiappini}, Cristina and {Frischknecht}, Urs and {Meynet}, Georges and {Hirschi}, Raphael and {Barbuy}, Beatriz and {Pignatari}, Marco and {Decressin}, Thibaut and {Maeder}, Andr{\'e}},
        title = "{Imprints of fast-rotating massive stars in the Galactic Bulge}",
      journal = {\nat},
         year = 2011,
        month = apr,
       volume = {472},
       number = {7344},
        pages = {454-457},
          doi = {10.1038/nature10000},
       adsurl = {https://ui.adsabs.harvard.edu/abs/2011Natur.472..454C}
}

@ARTICLE{clarkson08,
       author = {{Clarkson}, Will and {Sahu}, Kailash and {Anderson}, Jay and {Smith}, T. Ed and {Brown}, Thomas M. and {Rich}, R. Michael and {Casertano}, Stefano and {Bond}, Howard E. and {Livio}, Mario and {Minniti}, Dante and {Panagia}, Nino and {Renzini}, Alvio and {Valenti}, Jeff and {Zoccali}, Manuela},
        title = "{Stellar Proper Motions in the Galactic Bulge from Deep Hubble Space Telescope ACS WFC Photometry}",
      journal = {\apj},
         year = 2008,
        month = sep,
       volume = {684},
       number = {2},
        pages = {1110-1142},
          doi = {10.1086/590378},
archivePrefix = {arXiv},
       eprint = {0809.1682},
 primaryClass = {astro-ph},
       adsurl = {https://ui.adsabs.harvard.edu/abs/2008ApJ...684.1110C}
}

@ARTICLE{curtis19,
       author = {{Curtis}, Sanjana and {Ebinger}, Kevin and {Fr{\"o}hlich}, Carla and {Hempel}, Matthias and {Perego}, Albino and {Liebend{\"o}rfer}, Matthias and {Thielemann}, Friedrich-Karl},
        title = "{PUSHing Core-collapse Supernovae to Explosions in Spherical Symmetry. III. Nucleosynthesis Yields}",
      journal = {\apj},
         year = 2019,
        month = jan,
       volume = {870},
       number = {1},
          eid = {2},
        pages = {2},
          doi = {10.3847/1538-4357/aae7d2},
archivePrefix = {arXiv},
       eprint = {1805.00498},
 primaryClass = {astro-ph.SR},
       adsurl = {https://ui.adsabs.harvard.edu/abs/2019ApJ...870....2C}
}

@ARTICLE{dasilva24,
       author = {{da Silva}, Patr{\'\i}cia and {Barbuy}, B. and {Ernandes}, H. and {Souza}, S.~O. and {Fern{\'a}ndez-Trincado}, J.~G. and {Gonz{\'a}lez-D{\'\i}az}, D.},
        title = "{Abundances in eight bulge stars from the optical and near-infrared}",
      journal = {\aap},
         year = 2024,
        month = jul,
       volume = {687},
          eid = {A66},
        pages = {A66},
          doi = {10.1051/0004-6361/202449342},
archivePrefix = {arXiv},
       eprint = {2405.06153},
 primaryClass = {astro-ph.SR},
       adsurl = {https://ui.adsabs.harvard.edu/abs/2024A&A...687A..66D}
}

@ARTICLE{ernandes18,
       author = {{Ernandes}, H. and {Barbuy}, B. and {Alves-Brito}, A. and {Fria{\c{c}}a}, A. and {Siqueira-Mello}, C. and {Allen}, D.~M.},
        title = "{Iron-peak elements Sc, V, Mn, Cu, and Zn in Galactic bulge globular clusters}",
      journal = {\aap},
         year = 2018,
        month = aug,
       volume = {616},
          eid = {A18},
        pages = {A18},
          doi = {10.1051/0004-6361/201731708},
archivePrefix = {arXiv},
       eprint = {1801.06157},
 primaryClass = {astro-ph.SR},
       adsurl = {https://ui.adsabs.harvard.edu/abs/2018A&A...616A..18E}
}

@ARTICLE{ernandes20,
       author = {{Ernandes}, H. and {Barbuy}, B. and {Fria{\c{c}}a}, A.~C.~S. and {Hill}, V. and {Zoccali}, M. and {Minniti}, D. and {Renzini}, A. and {Ortolani}, S.},
        title = "{Cobalt and copper abundances in 56 Galactic bulge red giants}",
      journal = {\aap},
         year = 2020,
        month = aug,
       volume = {640},
          eid = {A89},
        pages = {A89},
          doi = {10.1051/0004-6361/202037869},
archivePrefix = {arXiv},
       eprint = {2007.00397},
 primaryClass = {astro-ph.SR},
       adsurl = {https://ui.adsabs.harvard.edu/abs/2020A&A...640A..89E}
}

@ARTICLE{feltzing23,
       author = {{Feltzing}, Sofia and {Feuillet}, Diane},
        title = "{The Metal-weak Milky Way Stellar Disk Hidden in the Gaia-Sausage-Enceladus Debris: The APOGEE DR17 View}",
      journal = {\apj},
         year = 2023,
        month = aug,
       volume = {953},
       number = {2},
          eid = {143},
        pages = {143},
          doi = {10.3847/1538-4357/ace185},
archivePrefix = {arXiv},
       eprint = {2303.00016},
 primaryClass = {astro-ph.GA},
       adsurl = {https://ui.adsabs.harvard.edu/abs/2023ApJ...953..143F}
}

@ARTICLE{friaca98,
       author = {{Friaca}, Amancio C.~S. and {Terlevich}, Roberto J.},
        title = "{Formation and evolution of elliptical galaxies and QSO activity}",
      journal = {\mnras},
         year = 1998,
        month = aug,
       volume = {298},
       number = {2},
        pages = {399-415},
          doi = {10.1046/j.1365-8711.1998.01626.x},
archivePrefix = {arXiv},
       eprint = {astro-ph/9803283},
 primaryClass = {astro-ph},
       adsurl = {https://ui.adsabs.harvard.edu/abs/1998MNRAS.298..399F}
}

@ARTICLE{gaia21,
       author = {{Gaia Collaboration} and {Brown}, A.~G.~A. and {Vallenari}, A. and {Prusti}, T. and {de Bruijne}, J.~H.~J. and {Babusiaux}, C. and {Biermann}, M. and {Creevey}, O.~L. and {Evans}, D.~W. and {Eyer}, L. and {Hutton}, A. and {Jansen}, F. and {Jordi}, C. and {Klioner}, S.~A. and {Lammers}, U. and {Lindegren}, L. and {Luri}, X. and {Mignard}, F. and {Panem}, C. and {Pourbaix}, D. and {Randich}, S. and {Sartoretti}, P. and {Soubiran}, C. and {Walton}, N.~A. and {Arenou}, F. and {Bailer-Jones}, C.~A.~L. and et al.},
        title = "{Gaia Early Data Release 3. Summary of the contents and survey properties}",
      journal = {\aap},
         year = 2021,
        month = may,
       volume = {649},
          eid = {A1},
        pages = {A1},
          doi = {10.1051/0004-6361/202039657},
archivePrefix = {arXiv},
       eprint = {2012.01533},
 primaryClass = {astro-ph.GA},
       adsurl = {https://ui.adsabs.harvard.edu/abs/2021A&A...649A...1G}
}

@ARTICLE{gao10,
       author = {{Gao}, L. and {Theuns}, Tom and {Frenk}, C.~S. and {Jenkins}, A. and {Helly}, J.~C. and {Navarro}, J. and {Springel}, V. and {White}, S.~D.~M.},
        title = "{The earliest stars and their relics in the Milky Way}",
      journal = {\mnras},
         year = 2010,
        month = apr,
       volume = {403},
       number = {3},
        pages = {1283-1295},
          doi = {10.1111/j.1365-2966.2009.16225.x},
archivePrefix = {arXiv},
       eprint = {0909.1593},
 primaryClass = {astro-ph.CO},
       adsurl = {https://ui.adsabs.harvard.edu/abs/2010MNRAS.403.1283G}
}

@ARTICLE{geisler23,
       author = {{Geisler}, D. and {Parisi}, M.~C. and {Dias}, B. and {Villanova}, S. and {Mauro}, F. and {Saviane}, I. and {Cohen}, R.~E. and {Moni Bidin}, C. and {Minniti}, D.},
        title = "{Ca triplet metallicities and velocities for 12 globular clusters toward the galactic bulge}",
      journal = {\aap},
         year = 2023,
        month = jan,
       volume = {669},
          eid = {A115},
        pages = {A115},
          doi = {10.1051/0004-6361/202244959},
archivePrefix = {arXiv},
       eprint = {2210.02193},
 primaryClass = {astro-ph.GA},
       adsurl = {https://ui.adsabs.harvard.edu/abs/2023A&A...669A.115G}
}

@ARTICLE{gunn06,
       author = {{Gunn}, James E. and {Siegmund}, Walter A. and {Mannery}, Edward J. and {Owen}, Russell E. and {Hull}, Charles L. and {Leger}, R. French and {Carey}, Larry N. and {Knapp}, Gillian R. and {York}, Donald G. and {Boroski}, William N. and {Kent}, Stephen M. and {Lupton}, Robert H. and {Rockosi}, Constance M. and et al.},
        title = "{The 2.5 m Telescope of the Sloan Digital Sky Survey}",
      journal = {\aj},
         year = 2006,
        month = apr,
       volume = {131},
       number = {4},
        pages = {2332-2359},
          doi = {10.1086/500975},
archivePrefix = {arXiv},
       eprint = {astro-ph/0602326},
 primaryClass = {astro-ph},
       adsurl = {https://ui.adsabs.harvard.edu/abs/2006AJ....131.2332G}
}

@ARTICLE{garcia-perez16,
       author = {{Garc{\'\i}a P{\'e}rez}, Ana E. and {Allende Prieto}, Carlos and {Holtzman}, Jon A. and {Shetrone}, Matthew and {M{\'e}sz{\'a}ros}, Szabolcs and {Bizyaev}, Dmitry and {Carrera}, Ricardo and {Cunha}, Katia and {Garc{\'\i}a-Hern{\'a}ndez}, D.~A. and {Johnson}, Jennifer A. and {Majewski}, Steven R. and {Nidever}, David L. and {Schiavon}, Ricardo P. and {Shane}, Neville and {Smith}, Verne V. and {Sobeck}, Jennifer and {Troup}, Nicholas and {Zamora}, Olga and {Weinberg}, David H. and {Bovy}, Jo and {Eisenstein}, Daniel J. and {Feuillet}, Diane and {Frinchaboy}, Peter M. and {Hayden}, Michael R. and {Hearty}, Fred R. and {Nguyen}, Duy C. and {O'Connell}, Robert W. and {Pinsonneault}, Marc H. and {Wilson}, John C. and {Zasowski}, Gail},
        title = "{ASPCAP: The APOGEE Stellar Parameter and Chemical Abundances Pipeline}",
      journal = {\aj},
         year = 2016,
        month = jun,
       volume = {151},
       number = {6},
          eid = {144},
        pages = {144},
          doi = {10.3847/0004-6256/151/6/144},
archivePrefix = {arXiv},
       eprint = {1510.07635},
 primaryClass = {astro-ph.SR},
       adsurl = {https://ui.adsabs.harvard.edu/abs/2016AJ....151..144G}
}

@ARTICLE{gustafsson08,
       author = {{Gustafsson}, B. and {Edvardsson}, B. and {Eriksson}, K. and {J{\o}rgensen}, U.~G. and {Nordlund}, {\r{A}}. and {Plez}, B.},
        title = "{A grid of MARCS model atmospheres for late-type stars. I. Methods and general properties}",
      journal = {\aap},
         year = 2008,
        month = aug,
       volume = {486},
       number = {3},
        pages = {951-970},
          doi = {10.1051/0004-6361:200809724},
archivePrefix = {arXiv},
       eprint = {0805.0554},
 primaryClass = {astro-ph},
       adsurl = {https://ui.adsabs.harvard.edu/abs/2008A&A...486..951G}
}

@ARTICLE{hayes22,
       author = {{Hayes}, Christian R. and {Masseron}, Thomas and {Sobeck}, Jennifer and {Garc{\'\i}a-Hern{\'a}ndez}, D.~A. and {Allende Prieto}, Carlos and {Beaton}, Rachael L. and {Cunha}, Katia and {Hasselquist}, Sten and {Holtzman}, Jon A. and {J{\"o}nsson}, Henrik and {Majewski}, Steven R. and {Shetrone}, Matthew and {Smith}, Verne V. and {Almeida}, Andr{\'e}s},
        title = "{BACCHUS Analysis of Weak Lines in APOGEE Spectra (BAWLAS)}",
      journal = {\apjs},
         year = 2022,
        month = sep,
       volume = {262},
       number = {1},
          eid = {34},
        pages = {34},
          doi = {10.3847/1538-4365/ac839f},
archivePrefix = {arXiv},
       eprint = {2208.00071},
 primaryClass = {astro-ph.GA},
       adsurl = {https://ui.adsabs.harvard.edu/abs/2022ApJS..262...34H}
}

@ARTICLE{horta21,
       author = {{Horta}, Danny and {Schiavon}, Ricardo P. and {Mackereth}, J. Ted and {Pfeffer}, Joel and {Mason}, Andrew C. and {Kisku}, Shobhit and {Fragkoudi}, Francesca and {Allende Prieto}, Carlos and {Cunha}, Katia and {Hasselquist}, Sten and {Holtzman}, Jon and {Majewski}, Steven R. and {Nataf}, David and {O'Connell}, Robert W. and {Schultheis}, Mathias and {Smith}, Verne V.},
        title = "{Evidence from APOGEE for the presence of a major building block of the halo buried in the inner Galaxy}",
      journal = {\mnras},
         year = 2021,
        month = jan,
       volume = {500},
       number = {1},
        pages = {1385-1403},
          doi = {10.1093/mnras/staa2987},
archivePrefix = {arXiv},
       eprint = {2007.10374},
 primaryClass = {astro-ph.GA},
       adsurl = {https://ui.adsabs.harvard.edu/abs/2021MNRAS.500.1385H}
}

@ARTICLE{howes16,
       author = {{Howes}, Louise M. and {Asplund}, Martin and {Keller}, Stefan C. and {Casey}, Andrew R. and {Yong}, David and {Lind}, Karin and {Frebel}, Anna and {Hays}, Austin and {Alves-Brito}, Alan and {Bessell}, Michael S. and {Casagrande}, Luca and {Marino}, Anna F. and {Nataf}, David M. and {Owen}, Christopher I. and {Da Costa}, Gary S. and {Schmidt}, Brian P. and {Tisserand}, Patrick},
        title = "{The EMBLA survey - metal-poor stars in the Galactic bulge}",
      journal = {\mnras},
         year = 2016,
        month = jul,
       volume = {460},
       number = {1},
        pages = {884-901},
          doi = {10.1093/mnras/stw1004},
archivePrefix = {arXiv},
       eprint = {1604.07834},
 primaryClass = {astro-ph.GA},
       adsurl = {https://ui.adsabs.harvard.edu/abs/2016MNRAS.460..884H}
}

@ARTICLE{iwamoto99,
       author = {{Iwamoto}, Koichi and {Brachwitz}, Franziska and {Nomoto}, Ken'ICHI and {Kishimoto}, Nobuhiro and {Umeda}, Hideyuki and {Hix}, W. Raphael and {Thielemann}, Friedrich-Karl},
        title = "{Nucleosynthesis in Chandrasekhar Mass Models for Type IA Supernovae and Constraints on Progenitor Systems and Burning-Front Propagation}",
      journal = {\apjs},
         year = 1999,
        month = dec,
       volume = {125},
       number = {2},
        pages = {439-462},
          doi = {10.1086/313278},
archivePrefix = {arXiv},
       eprint = {astro-ph/0002337},
 primaryClass = {astro-ph},
       adsurl = {https://ui.adsabs.harvard.edu/abs/1999ApJS..125..439I}
}

@ARTICLE{johnson14,
       author = {{Johnson}, Christian I. and {Rich}, R. Michael and {Kobayashi}, Chiaki and {Kunder}, Andrea and {Koch}, Andreas},
        title = "{Light, Alpha, and Fe-peak Element Abundances in the Galactic Bulge}",
      journal = {\aj},
         year = 2014,
        month = oct,
       volume = {148},
       number = {4},
          eid = {67},
        pages = {67},
          doi = {10.1088/0004-6256/148/4/67},
archivePrefix = {arXiv},
       eprint = {1407.2282},
 primaryClass = {astro-ph.SR},
       adsurl = {https://ui.adsabs.harvard.edu/abs/2014AJ....148...67J}
}

@ARTICLE{keegans23,
       author = {{Keegans}, James D. and {Pignatari}, Marco and {Stancliffe}, Richard J. and {Travaglio}, Claudia and {Jones}, Samuel and {Gibson}, Brad K. and {Townsley}, Dean M. and {Miles}, Broxton J. and {Shen}, Ken J. and {Few}, Gareth},
        title = "{Type Ia Supernova Nucleosynthesis: Metallicity-dependent Yields}",
      journal = {\apjs},
         year = 2023,
        month = sep,
       volume = {268},
       number = {1},
          eid = {8},
        pages = {8},
          doi = {10.3847/1538-4365/ace102},
archivePrefix = {arXiv},
       eprint = {2306.12885},
 primaryClass = {astro-ph.HE},
       adsurl = {https://ui.adsabs.harvard.edu/abs/2023ApJS..268....8K}
}

@ARTICLE{korotin25,
       author = {{Korotin}, S.~A. and {Kiselev}, K.~O.},
        title = "{Influence of Departures from LTE on Determinations of the Sulfur Abundances in A{\textendash}K Type Stars}",
      journal = {Astronomy Reports},
         year = 2024,
        month = dec,
       volume = {68},
       number = {12},
        pages = {1159-1175},
          doi = {10.1134/S1063772924700987},
archivePrefix = {arXiv},
       eprint = {2501.00774},
 primaryClass = {astro-ph.SR},
       adsurl = {https://ui.adsabs.harvard.edu/abs/2024ARep...68.1159K}
}

@ARTICLE{limberg22,
       author = {{Limberg}, Guilherme and {Souza}, Stefano O. and {P{\'e}rez-Villegas}, Angeles and {Rossi}, Silvia and {Perottoni}, H{\'e}lio D. and {Santucci}, Rafael M.},
        title = "{Reconstructing the Disrupted Dwarf Galaxy Gaia-Sausage/Enceladus Using Its Stars and Globular Clusters}",
      journal = {\apj},
         year = 2022,
        month = aug,
       volume = {935},
       number = {2},
          eid = {109},
        pages = {109},
          doi = {10.3847/1538-4357/ac8159},
archivePrefix = {arXiv},
       eprint = {2206.10505},
 primaryClass = {astro-ph.GA},
       adsurl = {https://ui.adsabs.harvard.edu/abs/2022ApJ...935..109L}
}

@ARTICLE{limongi18,
       author = {{Limongi}, Marco and {Chieffi}, Alessandro},
        title = "{Presupernova Evolution and Explosive Nucleosynthesis of Rotating Massive Stars in the Metallicity Range -3 {\ensuremath{\leq}} [Fe/H] {\ensuremath{\leq}} 0}",
      journal = {\apjs},
         year = 2018,
        month = jul,
       volume = {237},
       number = {1},
          eid = {13},
        pages = {13},
          doi = {10.3847/1538-4365/aacb24},
archivePrefix = {arXiv},
       eprint = {1805.09640},
 primaryClass = {astro-ph.SR},
       adsurl = {https://ui.adsabs.harvard.edu/abs/2018ApJS..237...13L}
}

@ARTICLE{lomaeva19,
       author = {{Lomaeva}, M. and {J{\"o}nsson}, H. and {Ryde}, N. and {Schultheis}, M. and {Thorsbro}, B.},
        title = "{Abundances of disk and bulge giants from high-resolution optical spectra. III. Sc, V, Cr, Mn, Co, Ni}",
      journal = {\aap},
         year = 2019,
        month = may,
       volume = {625},
          eid = {A141},
        pages = {A141},
          doi = {10.1051/0004-6361/201834247},
archivePrefix = {arXiv},
       eprint = {1903.01476},
 primaryClass = {astro-ph.GA},
       adsurl = {https://ui.adsabs.harvard.edu/abs/2019A&A...625A.141L}
}

@ARTICLE{lucertini22,
       author = {{Lucertini}, F. and {Monaco}, L. and {Caffau}, E. and {Bonifacio}, P. and {Mucciarelli}, A.},
        title = "{Sulfur abundances in the Galactic bulge and disk}",
      journal = {\aap},
         year = 2022,
        month = jan,
       volume = {657},
          eid = {A29},
        pages = {A29},
          doi = {10.1051/0004-6361/202140947},
archivePrefix = {arXiv},
       eprint = {2109.06216},
 primaryClass = {astro-ph.GA},
       adsurl = {https://ui.adsabs.harvard.edu/abs/2022A&A...657A..29L}
}

@ARTICLE{majewski17,
       author = {{Majewski}, Steven R. and {Schiavon}, Ricardo P. and {Frinchaboy}, Peter M. and {Allende Prieto}, Carlos and {Barkhouser}, Robert and {Bizyaev}, Dmitry and {Blank}, Basil and {Brunner}, Sophia and {Burton}, Adam and {Carrera}, Ricardo and {Chojnowski}, S. Drew and {Cunha}, K{\'a}tia and {Epstein}, Courtney and {Fitzgerald}, Greg and {Garc{\'\i}a P{\'e}rez}, Ana E. and {Hearty}, Fred R. and et al.},
        title = "{The Apache Point Observatory Galactic Evolution Experiment (APOGEE)}",
      journal = {\aj},
         year = 2017,
        month = sep,
       volume = {154},
       number = {3},
          eid = {94},
        pages = {94},
          doi = {10.3847/1538-3881/aa784d},
archivePrefix = {arXiv},
       eprint = {1509.05420},
 primaryClass = {astro-ph.IM},
       adsurl = {https://ui.adsabs.harvard.edu/abs/2017AJ....154...94M}
}

@ARTICLE{masseron20,
       author = {{Masseron}, Thomas and {Garc{\'\i}a-Hern{\'a}ndez}, D.~A. and {Santove{\~n}a}, Ra{\'u}l and {Manchado}, Arturo and {Zamora}, Olga and {Manteiga}, Minia and {Dafonte}, Carlos},
        title = "{Phosphorus-rich stars with unusual abundances are challenging theoretical predictions}",
      journal = {Nature Communications},
         year = 2020,
        month = aug,
       volume = {11},
          eid = {3759},
        pages = {3759},
          doi = {10.1038/s41467-020-17649-9},
archivePrefix = {arXiv},
       eprint = {2008.01633},
 primaryClass = {astro-ph.SR},
       adsurl = {https://ui.adsabs.harvard.edu/abs/2020NatCo..11.3759M}
}

@ARTICLE{masseron20b,
       author = {{Masseron}, T. and {Garc{\'\i}a-Hern{\'a}ndez}, D.~A. and {Zamora}, O. and {Manchado}, A.},
        title = "{Heavy-element Abundances in P-rich Stars: A New Site for the s-process?}",
      journal = {\apjl},
         year = 2020,
        month = nov,
       volume = {904},
       number = {1},
          eid = {L1},
        pages = {L1},
          doi = {10.3847/2041-8213/abc6ac},
archivePrefix = {arXiv},
       eprint = {2011.00460},
 primaryClass = {astro-ph.SR},
       adsurl = {https://ui.adsabs.harvard.edu/abs/2020ApJ...904L...1M}
}

@ARTICLE{mcwilliamrich94,
       author = {{McWilliam}, Andrew and {Rich}, R. Michael},
        title = "{The First Detailed Abundance Analysis of Galactic Bulge K Giants in Baade's Window}",
      journal = {\apjs},
         year = 1994,
        month = apr,
       volume = {91},
        pages = {749},
          doi = {10.1086/191954},
       adsurl = {https://ui.adsabs.harvard.edu/abs/1994ApJS...91..749M}
}

@ARTICLE{mcwilliam03,
       author = {{McWilliam}, Andrew and {Rich}, R. Michael and {Smecker-Hane}, Tammy A.},
        title = "{Constraints on the Origin of Manganese from the Composition of the Sagittarius Dwarf Spheroidal Galaxy and the Galactic Bulge}",
      journal = {\apjl},
         year = 2003,
        month = jul,
       volume = {592},
       number = {1},
        pages = {L21-L24},
          doi = {10.1086/377441},
       adsurl = {https://ui.adsabs.harvard.edu/abs/2003ApJ...592L..21M}
}

@ARTICLE{mcwilliam16,
       author = {{McWilliam}, Andrew},
        title = "{The Chemical Composition of the Galactic Bulge and Implications for its Evolution}",
      journal = {\pasa},
         year = 2016,
        month = aug,
       volume = {33},
          eid = {e040},
        pages = {e040},
          doi = {10.1017/pasa.2016.32},
archivePrefix = {arXiv},
       eprint = {1607.05299},
 primaryClass = {astro-ph.GA},
       adsurl = {https://ui.adsabs.harvard.edu/abs/2016PASA...33...40M}
}

@ARTICLE{montalban21,
       author = {{Montalb{\'a}n}, Josefina and {Mackereth}, J. Ted and {Miglio}, Andrea and {Vincenzo}, Fiorenzo and {Chiappini}, Cristina and {Buldgen}, Gael and {Mosser}, Beno{\^\i}t and {Noels}, Arlette and {Scuflaire}, Richard and {Vrard}, Mathieu and {Willett}, Emma and {Davies}, Guy R. and {Hall}, Oliver J. and {Nielsen}, Martin Bo and {Khan}, Saniya and {Rendle}, Ben M. and {van Rossem}, Walter E. and {Ferguson}, Jason W. and {Chaplin}, William J.},
        title = "{Chronologically dating the early assembly of the Milky Way}",
      journal = {Nature Astronomy},
         year = 2021,
        month = jan,
       volume = {5},
        pages = {640-647},
          doi = {10.1038/s41550-021-01347-7},
archivePrefix = {arXiv},
       eprint = {2006.01783},
 primaryClass = {astro-ph.GA},
       adsurl = {https://ui.adsabs.harvard.edu/abs/2021NatAs...5..640M}
}

@ARTICLE{montelius22,
       author = {{Montelius}, M. and {Forsberg}, R. and {Ryde}, N. and {J{\"o}nsson}, H. and {Af{\c{s}}ar}, M. and {Johansen}, A. and {Kaplan}, K.~F. and {Kim}, H. and {Mace}, G. and {Sneden}, C. and {Thorsbro}, B.},
        title = "{Chemical evolution of ytterbium in the Galactic disk}",
      journal = {\aap},
         year = 2022,
        month = sep,
       volume = {665},
          eid = {A135},
        pages = {A135},
          doi = {10.1051/0004-6361/202243140},
archivePrefix = {arXiv},
       eprint = {2202.00691},
 primaryClass = {astro-ph.GA},
       adsurl = {https://ui.adsabs.harvard.edu/abs/2022A&A...665A.135M}
}

@ARTICLE{nandakumar22,
       author = {{Nandakumar}, G. and {Ryde}, N. and {Montelius}, M. and {Thorsbro}, B. and {J{\"o}nsson}, H. and {Mace}, G.},
        title = "{The Galactic chemical evolution of phosphorus observed with IGRINS}",
      journal = {\aap},
         year = 2022,
        month = dec,
       volume = {668},
          eid = {A88},
        pages = {A88},
          doi = {10.1051/0004-6361/202244724},
archivePrefix = {arXiv},
       eprint = {2210.04940},
 primaryClass = {astro-ph.SR},
       adsurl = {https://ui.adsabs.harvard.edu/abs/2022A&A...668A..88N}
}

@ARTICLE{nandakumar24,
 author = {{Nandakumar}, Govind and {Ryde}, Nils and {Mace}, Gregory and {Kaplan}, Kyle F. and {Nieuwmunster}, Niels and {Jaffe}, Daniel and {Rich}, R. Michael and {Schultheis}, Mathias and {Agertz}, Oscar and {Andersson}, Eric and {Sneden}, Christopher and {Strickland}, Emily and {Thorsbro}, Brian},
 title = "{Composition of Giants 1{\textdegree} North of the Galactic Center: Detailed Abundance Trends for 21 Elements Observed with IGRINS}",
 journal = {\apj},
 year = 2024,
 month = mar,
 volume = {964},
 number = {1},
 eid = {96},
 pages = {96},
 doi = {10.3847/1538-4357/ad22dc},
archivePrefix = {arXiv},
 eprint = {2401.13991},
 primaryClass = {astro-ph.GA},
 adsurl = {https://ui.adsabs.harvard.edu/abs/2024ApJ...964...96N}
}

@ARTICLE{nelder65,
    author = {{Nelder}, J. A. and {Mead}, H.},
    title = "{A simplex method for function minimization}",
    journal = {{The Computer Journal}},
    year= 1965,
    month =jan,
    volume={7},
    number = {4},
    pages = {308–313},
    doi = {10.1093/comjnl/7.4.308},
}

@ARTICLE{nepal25,
       author = {{Nepal}, Samir and {Chiappini}, Cristina and {P{\'e}rez-Villegas}, Angeles and {Queiroz}, Anna B. and {Souza}, Stefano and {Steinmetz}, Matthias and {Anders}, Friedrich and {Khalatyan}, Arman and {Barbuy}, Beatriz and {Guiglion}, Guillaume},
        title = "{The Spheroidal Bulge of the Milky Way: Chemodynamically Distinct from the Inner-Thick Disc and Bar}",
      journal = {arXiv e-prints},
         year = 2025,
        month = jul,
          eid = {arXiv:2507.06863},
        pages = {arXiv:2507.06863},
          doi = {10.48550/arXiv.2507.06863},
archivePrefix = {arXiv},
       eprint = {2507.06863},
 primaryClass = {astro-ph.GA},
       adsurl = {https://ui.adsabs.harvard.edu/abs/2025arXiv250706863N}
}

@ARTICLE{nomoto13,
       author = {{Nomoto}, Ken'ichi and {Kobayashi}, Chiaki and {Tominaga}, Nozomu},
        title = "{Nucleosynthesis in Stars and the Chemical Enrichment of Galaxies}",
      journal = {\araa},
         year = 2013,
        month = aug,
       volume = {51},
       number = {1},
        pages = {457-509},
          doi = {10.1146/annurev-astro-082812-140956},
       adsurl = {https://ui.adsabs.harvard.edu/abs/2013ARA&A..51..457N}
}

@ARTICLE{nordlanderlind17,
       author = {{Nordlander}, T. and {Lind}, K.},
        title = "{Non-LTE aluminium abundances in late-type stars}",
      journal = {\aap},
         year = 2017,
        month = nov,
       volume = {607},
          eid = {A75},
        pages = {A75},
          doi = {10.1051/0004-6361/201730427},
archivePrefix = {arXiv},
       eprint = {1708.01949},
 primaryClass = {astro-ph.SR},
       adsurl = {https://ui.adsabs.harvard.edu/abs/2017A&A...607A..75N}
}

@ARTICLE{ortigoza-urdaneta23,
       author = {{Ortigoza-Urdaneta}, Mario and {Vieira}, Katherine and {Fern{\'a}ndez-Trincado}, Jos{\'e} G. and {Queiroz}, Anna B.~A. and {Barbuy}, Beatriz and {Beers}, Timothy C. and {Chiappini}, Cristina and {Anders}, Friedrich and {Minniti}, Dante and {Tang}, Baitian},
        title = "{Galactic ArchaeoLogIcaL ExcavatiOns (GALILEO). II. t-SNE portrait of local fossil relics and structures}",
      journal = {A\&A},
         year = 2023,
        month = aug,
       volume = {676},
          eid = {A140},
        pages = {A140},
          doi = {10.1051/0004-6361/202346325},
archivePrefix = {arXiv},
       eprint = {2306.08677},
 primaryClass = {astro-ph.GA},
       adsurl = {https://ui.adsabs.harvard.edu/abs/2023A&A...676A.140O}
}

@ARTICLE{ortolani25,
       author = {{Ortolani}, S. and {Souza}, S.~O. and {Nardiello}, D. and {Barbuy}, B. and {Bica}, E.},
        title = "{Revisited parameters for the twin bulge globular clusters NGC 6528 and NGC 6553}",
      journal = {\aap},
         year = 2025,
        month = jun,
       volume = {698},
          eid = {A181},
        pages = {A181},
          doi = {10.1051/0004-6361/202554560},
archivePrefix = {arXiv},
       eprint = {2505.01861},
 primaryClass = {astro-ph.GA},
       adsurl = {https://ui.adsabs.harvard.edu/abs/2025A&A...698A.181O}
}

@ARTICLE{planck20,
       author = {{Planck Collaboration} and {Aghanim}, N. and {Akrami}, Y. and {Ashdown}, M. and {Aumont}, J. and {Baccigalupi}, C. and {Ballardini}, M. and {Banday}, A.~J. and {Barreiro}, R.~B. and {Bartolo}, N. and {Basak}, S. and {Battye}, R. and {Benabed}, K. and {Bernard}, J. -P. and {Bersanelli}, M. and {Bielewicz}, P. and {Bock}, J.~J. and et al.},
        title = "{Planck 2018 results. VI. Cosmological parameters}",
      journal = {\aap},
         year = 2020,
        month = sep,
       volume = {641},
          eid = {A6},
        pages = {A6},
          doi = {10.1051/0004-6361/201833910},
archivePrefix = {arXiv},
       eprint = {1807.06209},
 primaryClass = {astro-ph.CO},
       adsurl = {https://ui.adsabs.harvard.edu/abs/2020A&A...641A...6P}
}

@misc{plez12,
       author = {{Plez}, B.},
        title = "{Turbospectrum: Code for spectral synthesis}",
 howpublished = {Astrophysics Source Code Library, record ascl:1205.004},
         year = 2012,
        month = may,
          eid = {ascl:1205.004},
       adsurl = {https://ui.adsabs.harvard.edu/abs/2012ascl.soft05004P}
}

@ARTICLE{queiroz18,
       author = {{Queiroz}, A.~B.~A. and {Anders}, F. and {Santiago}, B.~X. and {Chiappini}, C. and {Steinmetz}, M. and {Dal Ponte}, M. and {Stassun}, K.~G. and {da Costa}, L.~N. and {Maia}, M.~A.~G. and {Crestani}, J. and {Beers}, T.~C. and {Fern{\'a}ndez-Trincado}, J.~G. and {Garc{\'\i}a-Hern{\'a}ndez}, D.~A. and {Roman-Lopes}, A. and {Zamora}, O.},
        title = "{StarHorse: a Bayesian tool for determining stellar masses, ages, distances, and extinctions for field stars}",
      journal = {\mnras},
         year = 2018,
        month = may,
       volume = {476},
       number = {2},
        pages = {2556-2583},
          doi = {10.1093/mnras/sty330},
archivePrefix = {arXiv},
       eprint = {1710.09970},
 primaryClass = {astro-ph.IM},
       adsurl = {https://ui.adsabs.harvard.edu/abs/2018MNRAS.476.2556Q}
}

@ARTICLE{queiroz20,
       author = {{Queiroz}, A.~B.~A. and {Anders}, F. and {Chiappini}, C. and {Khalatyan}, A. and {Santiago}, B.~X. and {Steinmetz}, M. and {Valentini}, M. and {Miglio}, A. and {Bossini}, D. and {Barbuy}, B. and {Minchev}, I. and {Minniti}, D. and {Garc{\'\i}a Hern{\'a}ndez}, D.~A. and {Schultheis}, M. and et al.},
        title = "{From the bulge to the outer disc: StarHorse stellar parameters, distances, and extinctions for stars in APOGEE DR16 and other spectroscopic surveys}",
      journal = {\aap},
         year = 2020,
        month = jun,
       volume = {638},
          eid = {A76},
        pages = {A76},
          doi = {10.1051/0004-6361/201937364},
archivePrefix = {arXiv},
       eprint = {1912.09778},
 primaryClass = {astro-ph.GA},
       adsurl = {https://ui.adsabs.harvard.edu/abs/2020A&A...638A..76Q}
}

@ARTICLE{queiroz21,
       author = {{Queiroz}, A.~B.~A. and {Chiappini}, C. and {Perez-Villegas}, A. and {Khalatyan}, A. and {Anders}, F. and {Barbuy}, B. and {Santiago}, B.~X. and {Steinmetz}, M. and {Cunha}, K. and {Schultheis}, M. and {Majewski}, S.~R. and et al.},
        title = "{The Milky Way bar and bulge revealed by APOGEE and Gaia EDR3}",
      journal = {\aap},
         year = 2021,
        month = dec,
       volume = {656},
          eid = {A156},
        pages = {A156},
          doi = {10.1051/0004-6361/202039030},
archivePrefix = {arXiv},
       eprint = {2007.12915},
 primaryClass = {astro-ph.GA},
       adsurl = {https://ui.adsabs.harvard.edu/abs/2021A&A...656A.156Q}
}

@ARTICLE{rauscher02,
       author = {{Rauscher}, T. and {Heger}, A. and {Hoffman}, R.~D. and {Woosley}, S.~E.},
        title = "{Nucleosynthesis in Massive Stars with Improved Nuclear and Stellar Physics}",
      journal = {\apj},
         year = 2002,
        month = sep,
       volume = {576},
       number = {1},
        pages = {323-348},
          doi = {10.1086/341728},
archivePrefix = {arXiv},
       eprint = {astro-ph/0112478},
 primaryClass = {astro-ph},
       adsurl = {https://ui.adsabs.harvard.edu/abs/2002ApJ...576..323R}
}

@ARTICLE{razera22,
       author = {{Razera}, R. and {Barbuy}, B. and {Moura}, T.~C. and {Ernandes}, H. and {P{\'e}rez-Villegas}, A. and {Souza}, S.~O. and {Chiappini}, C. and {Queiroz}, A.~B.~A. and {Anders}, F. and {Fern{\'a}ndez-Trincado}, J.~G. and {Fria{\c{c}}a}, A.~C.~S. and {Cunha}, K. and {Smith}, V.~V. and {Santiago}, B.~X. and {Schiavon}, R.~P. and {Valentini}, M. and {Minniti}, D. and {Schultheis}, M. and {Geisler}, D. and {Sobeck}, J. and {Placco}, V.~M. and {Zoccali}, M.},
        title = "{Abundance analysis of APOGEE spectra for 58 metal-poor stars from the bulge spheroid}",
      journal = {\mnras},
         year = 2022,
        month = dec,
       volume = {517},
       number = {3},
        pages = {4590-4606},
          doi = {10.1093/mnras/stac2136},
archivePrefix = {arXiv},
       eprint = {2208.06634},
 primaryClass = {astro-ph.SR},
       adsurl = {https://ui.adsabs.harvard.edu/abs/2022MNRAS.517.4590R}
}

@ARTICLE{reinhard24,
       author = {{Reinhard}, Michael V. and {Laird}, John B.},
        title = "{Alpha Element Populations Among Local Halo Stars}",
      journal = {\aj},
         year = 2024,
        month = jan,
       volume = {167},
       number = {1},
          eid = {6},
        pages = {6},
          doi = {10.3847/1538-3881/ad0a96},
       adsurl = {https://ui.adsabs.harvard.edu/abs/2024AJ....167....6R}
}

@ARTICLE{ritter18,
       author = {{Ritter}, C. and {Andrassy}, R. and {C{\^o}t{\'e}}, B. and {Herwig}, F. and {Woodward}, P.~R. and {Pignatari}, M. and {Jones}, S.},
        title = "{Convective-reactive nucleosynthesis of K, Sc, Cl and p-process isotopes in O-C shell mergers}",
      journal = {\mnras},
         year = 2018,
        month = feb,
       volume = {474},
       number = {1},
        pages = {L1-L6},
          doi = {10.1093/mnrasl/slx126},
archivePrefix = {arXiv},
       eprint = {1704.05985},
 primaryClass = {astro-ph.SR},
       adsurl = {https://ui.adsabs.harvard.edu/abs/2018MNRAS.474L...1R}
}

@ARTICLE{ryde16,
       author = {{Ryde}, N. and {Schultheis}, M. and {Grieco}, V. and {Matteucci}, F. and {Rich}, R.~M. and {Uttenthaler}, S.},
        title = "{Chemical Evolution of the Inner 2 Degrees of the Milky Way Bulge: [{\ensuremath{\alpha}}/Fe] Trends and Metallicity Gradients}",
      journal = {\aj},
         year = 2016,
        month = jan,
       volume = {151},
       number = {1},
          eid = {1},
        pages = {1},
          doi = {10.3847/0004-6256/151/1/1},
archivePrefix = {arXiv},
       eprint = {1510.02622},
 primaryClass = {astro-ph.GA},
       adsurl = {https://ui.adsabs.harvard.edu/abs/2016AJ....151....1R}
}

@ARTICLE{santana21,
       author = {{Santana}, Felipe A. and {Beaton}, Rachael L. and {Covey}, Kevin R. and {O'Connell}, Julia E. and {Longa-Pe{\~n}a}, Pen{\'e}lope and {Cohen}, Roger and {Fern{\'a}ndez-Trincado}, Jos{\'e} G. and {Hayes}, Christian R. and {Zasowski}, Gail and {Sobeck}, Jennifer S. and {Majewski}, Steven R. and et al.},
        title = "{Final Targeting Strategy for the SDSS-IV APOGEE-2S Survey}",
      journal = {\aj},
         year = 2021,
        month = dec,
       volume = {162},
       number = {6},
          eid = {303},
        pages = {303},
          doi = {10.3847/1538-3881/ac2cbc},
archivePrefix = {arXiv},
       eprint = {2108.11908},
 primaryClass = {astro-ph.GA},
       adsurl = {https://ui.adsabs.harvard.edu/abs/2021AJ....162..303S}
}

@ARTICLE{santiago16,
       author = {{Santiago}, Bas{\'\i}lio X. and {Brauer}, Doroth{\'e}e E. and {Anders}, Friedrich and {Chiappini}, Cristina and {Queiroz}, Anna B. and {Girardi}, L{\'e}o and {Rocha-Pinto}, Helio J. and {Balbinot}, Eduardo and {da Costa}, Luiz N. and {Maia}, Marcio A.~G. and {Schultheis}, Mathias and {Steinmetz}, Matthias and {Miglio}, Andrea and {Montalb{\'a}n}, Josefina and {Schneider}, Donald P. and {Beers}, Timothy C. and {Frinchaboy}, Peter M. and {Lee}, Young Sun and {Zasowski}, Gail},
        title = "{Spectro-photometric distances to stars: A general purpose Bayesian approach}",
      journal = {\aap},
         year = 2016,
        month = jan,
       volume = {585},
          eid = {A42},
        pages = {A42},
          doi = {10.1051/0004-6361/201323177},
archivePrefix = {arXiv},
       eprint = {1501.05500},
 primaryClass = {astro-ph.IM},
       adsurl = {https://ui.adsabs.harvard.edu/abs/2016A&A...585A..42S}
}

@ARTICLE{schultheis17,
       author = {{Schultheis}, M. and {Rojas-Arriagada}, A. and {Garc{\'\i}a P{\'e}rez}, A.~E. and {J{\"o}nsson}, H. and {Hayden}, M. and {Nandakumar}, G. and {Cunha}, K. and {Allende Prieto}, C. and {Holtzman}, J.~A. and {Beers}, T.~C. and {Bizyaev}, D. and {Brinkmann}, J. and {Carrera}, R. and {Cohen}, R.~E. and {Geisler}, D. and {Hearty}, F.~R. and {Fernandez-Tricado}, J.~G. and {Maraston}, C. and {Minnitti}, D. and {Nitschelm}, C. and {Roman-Lopes}, A. and {Schneider}, D.~P. and {Tang}, B. and {Villanova}, S. and {Zasowski}, G. and {Majewski}, S.~R.},
        title = "{Baade's window and APOGEE. Metallicities, ages, and chemical abundances}",
      journal = {\aap},
         year = 2017,
        month = apr,
       volume = {600},
          eid = {A14},
        pages = {A14},
          doi = {10.1051/0004-6361/201630154},
archivePrefix = {arXiv},
       eprint = {1702.01547},
 primaryClass = {astro-ph.GA},
       adsurl = {https://ui.adsabs.harvard.edu/abs/2017A&A...600A..14S}
}

@ARTICLE{siqueira-mello16,
       author = {{Siqueira-Mello}, C. and {Chiappini}, C. and {Barbuy}, B. and {Freeman}, K. and {Ness}, M. and {Depagne}, E. and {Cantelli}, E. and {Pignatari}, M. and {Hirschi}, R. and {Frischknecht}, U. and {Meynet}, G. and {Maeder}, A.},
        title = "{Looking for imprints of the first stellar generations in metal-poor bulge field stars}",
      journal = {\aap},
         year = 2016,
        month = sep,
       volume = {593},
          eid = {A79},
        pages = {A79},
          doi = {10.1051/0004-6361/201628104},
archivePrefix = {arXiv},
       eprint = {1605.08939},
 primaryClass = {astro-ph.SR},
       adsurl = {https://ui.adsabs.harvard.edu/abs/2016A&A...593A..79S}
}

@ARTICLE{smith21,
       author = {{Smith}, Verne V. and {Bizyaev}, Dmitry and {Cunha}, Katia and {Shetrone}, Matthew D. and {Souto}, Diogo and {Allende Prieto}, Carlos and {Masseron}, Thomas and {M{\'e}sz{\'a}ros}, Szabolcs and {J{\"o}nsson}, Henrik and {Hasselquist}, Sten and {Osorio}, Yeisson and {Garc{\'\i}a-Hern{\'a}ndez}, D.~A. and {Plez}, Bertrand and {Beaton}, Rachael L. and {Holtzman}, Jon and {Majewski}, Steven R. and {Stringfellow}, Guy S. and {Sobeck}, Jennifer},
        title = "{The APOGEE Data Release 16 Spectral Line List}",
      journal = {\aj},
         year = 2021,
        month = jun,
       volume = {161},
       number = {6},
          eid = {254},
        pages = {254},
          doi = {10.3847/1538-3881/abefdc},
archivePrefix = {arXiv},
       eprint = {2103.10112},
 primaryClass = {astro-ph.SR},
       adsurl = {https://ui.adsabs.harvard.edu/abs/2021AJ....161..254S}
}

@ARTICLE{sobeck06,
       author = {{Sobeck}, Jennifer S. and {Ivans}, Inese I. and {Simmerer}, Jennifer A. and {Sneden}, Christopher and {Hoeflich}, Peter and {Fulbright}, Jon P. and {Kraft}, Robert P.},
        title = "{Manganese Abundances in Cluster and Field Stars}",
      journal = {\aj},
         year = 2006,
        month = jun,
       volume = {131},
       number = {6},
        pages = {2949-2958},
          doi = {10.1086/503106},
archivePrefix = {arXiv},
       eprint = {astro-ph/0604592},
 primaryClass = {astro-ph},
       adsurl = {https://ui.adsabs.harvard.edu/abs/2006AJ....131.2949S}
}

@ARTICLE{souza23,
       author = {{Souza}, S.~O. and {Ernandes}, H. and {Valentini}, M. and {Barbuy}, B. and {Chiappini}, C. and {P{\'e}rez-Villegas}, A. and {Ortolani}, S. and {Fria{\c{c}}a}, A.~C.~S. and {Queiroz}, A.~B.~A. and {Bica}, E.},
        title = "{Chrono-chemodynamical analysis of the globular cluster NGC 6355: Looking for the fundamental bricks of the Bulge}",
      journal = {\aap},
         year = 2023,
        month = mar,
       volume = {671},
          eid = {A45},
        pages = {A45},
          doi = {10.1051/0004-6361/202245286},
archivePrefix = {arXiv},
       eprint = {2301.05227},
 primaryClass = {astro-ph.GA},
       adsurl = {https://ui.adsabs.harvard.edu/abs/2023A&A...671A..45S}
}

@ARTICLE{souza24a,
       author = {{Souza}, S.~O. and {Libralato}, M. and {Nardiello}, D. and {Kerber}, L.~O. and {Ortolani}, S. and {P{\'e}rez-Villegas}, A. and {Oliveira}, R.~A.~P. and {Barbuy}, B. and {Bica}, E. and {Griggio}, M. and {Dias}, B.},
        title = "{Combined Gemini-South and HST photometric analysis of the globular cluster NGC 6558: The age of the metal-poor population of the Galactic bulge}",
      journal = {\aap},
         year = 2024,
        month = oct,
       volume = {690},
          eid = {A37},
        pages = {A37},
          doi = {10.1051/0004-6361/202450795},
archivePrefix = {arXiv},
       eprint = {2407.15918},
 primaryClass = {astro-ph.GA},
       adsurl = {https://ui.adsabs.harvard.edu/abs/2024A&A...690A..37S}
}

@ARTICLE{timmes95,
       author = {{Timmes}, F.~X. and {Woosley}, S.~E. and {Weaver}, Thomas A.},
        title = "{Galactic Chemical Evolution: Hydrogen through Zinc}",
      journal = {\apjs},
         year = 1995,
        month = jun,
       volume = {98},
        pages = {617},
          doi = {10.1086/192172},
archivePrefix = {arXiv},
       eprint = {astro-ph/9411003},
 primaryClass = {astro-ph},
       adsurl = {https://ui.adsabs.harvard.edu/abs/1995ApJS...98..617T}
}

@ARTICLE{tumlinson10,
       author = {{Tumlinson}, Jason},
        title = "{Chemical Evolution in Hierarchical Models of Cosmic Structure. II. The Formation of the Milky Way Stellar Halo and the Distribution of the Oldest Stars}",
      journal = {\apj},
         year = 2010,
        month = jan,
       volume = {708},
       number = {2},
        pages = {1398-1418},
          doi = {10.1088/0004-637X/708/2/1398},
archivePrefix = {arXiv},
       eprint = {0911.1786},
 primaryClass = {astro-ph.GA},
       adsurl = {https://ui.adsabs.harvard.edu/abs/2010ApJ...708.1398T}
}

@ARTICLE{vandenhoek97,
       author = {{van den Hoek}, L.~B. and {Groenewegen}, M.~A.~T.},
        title = "{New theoretical yields of intermediate mass stars}",
      journal = {\aaps},
         year = 1997,
        month = jun,
       volume = {123},
        pages = {305-328},
          doi = {10.1051/aas:1997162},
       adsurl = {https://ui.adsabs.harvard.edu/abs/1997A&AS..123..305V}
}

@ARTICLE{vanderswaelmen16,
       author = {{Van der Swaelmen}, M. and {Barbuy}, B. and {Hill}, V. and {Zoccali}, M. and {Minniti}, D. and {Ortolani}, S. and {G{\'o}mez}, A.},
        title = "{Heavy elements Ba, La, Ce, Nd, and Eu in 56 Galactic bulge red giants}",
      journal = {\aap},
         year = 2016,
        month = jan,
       volume = {586},
          eid = {A1},
        pages = {A1},
          doi = {10.1051/0004-6361/201525709},
archivePrefix = {arXiv},
       eprint = {1511.03919},
 primaryClass = {astro-ph.GA},
       adsurl = {https://ui.adsabs.harvard.edu/abs/2016A&A...586A...1V}
}

@ARTICLE{wilson19,
       author = {{Wilson}, J.~C. and {Hearty}, F.~R. and {Skrutskie}, M.~F. and {Majewski}, S.~R. and {Holtzman}, J.~A. and {Eisenstein}, D. and {Gunn}, J. and {Blank}, B. and {Henderson}, C. and {Smee}, S. and {Nelson}, M. and {Nidever}, D. and {Arns}, J. and et al.},
        title = "{The Apache Point Observatory Galactic Evolution Experiment (APOGEE) Spectrographs}",
      journal = {\pasp},
         year = 2019,
        month = may,
       volume = {131},
       number = {999},
        pages = {055001},
          doi = {10.1088/1538-3873/ab0075},
archivePrefix = {arXiv},
       eprint = {1902.00928},
 primaryClass = {astro-ph.IM},
       adsurl = {https://ui.adsabs.harvard.edu/abs/2019PASP..131e5001W}
}

@INPROCEEDINGS{woosley95,
       author = {{Woosley}, S.~E. and {Weaver}, T.~A.},
        title = "{Nucleosynthesis and Supernovae in Massive Stars}",
    booktitle = {Nuclei in the Cosmos III},
         year = 1995,
       editor = {{Busso}, Maurizio and {Raiteri}, Claudia M. and {Gallino}, Roberto},
       series = {American Institute of Physics Conference Series},
       volume = {327},
        month = jan,
    publisher = {AIP},
        pages = {365},
          doi = {10.1063/1.47368},
       adsurl = {https://ui.adsabs.harvard.edu/abs/1995AIPC..327..365W}
}

@ARTICLE{yong14,
       author = {{Yong}, David and {Alves Brito}, Alan and {Da Costa}, Gary S. and {Alonso-Garc{\'\i}a}, Javier and {Karakas}, Amanda I. and {Pignatari}, Marco and {Roederer}, Ian U. and {Aoki}, Wako and {Fishlock}, Cherie K. and {Grundahl}, Frank and {Norris}, John E.},
        title = "{Chemical abundances in bright giants of the globular cluster M62 (NGC 6266)}",
      journal = {\mnras},
         year = 2014,
        month = apr,
       volume = {439},
       number = {3},
        pages = {2638-2650},
          doi = {10.1093/mnras/stu118},
archivePrefix = {arXiv},
       eprint = {1401.3784},
 primaryClass = {astro-ph.GA},
       adsurl = {https://ui.adsabs.harvard.edu/abs/2014MNRAS.439.2638Y}
}

@ARTICLE{yoshida08,
       author = {{Yoshida}, Takashi and {Suzuki}, Toshio and {Chiba}, Satoshi and {Kajino}, Toshitaka and {Yokomakura}, Hidekazu and {Kimura}, Keiichi and {Takamura}, Akira and {Hartmann}, Dieter H.},
        title = "{Neutrino-Nucleus Reaction Cross Sections for Light Element Synthesis in Supernova Explosions}",
      journal = {\apj},
         year = 2008,
        month = oct,
       volume = {686},
       number = {1},
        pages = {448-466},
          doi = {10.1086/591266},
archivePrefix = {arXiv},
       eprint = {0807.2723},
 primaryClass = {astro-ph},
       adsurl = {https://ui.adsabs.harvard.edu/abs/2008ApJ...686..448Y}
}

@ARTICLE{TurboNLTE,
       author = {{Gerber}, Jeffrey M. and {Magg}, Ekaterina and {Plez}, Bertrand and {Bergemann}, Maria and {Heiter}, Ulrike and {Olander}, Terese and {Hoppe}, Richard},
        title = "{Non-LTE radiative transfer with Turbospectrum}",
      journal = {\aap},
         year = 2023,
        month = jan,
       volume = {669},
          eid = {A43},
        pages = {A43},
          doi = {10.1051/0004-6361/202243673},
archivePrefix = {arXiv},
       eprint = {2206.00967},
 primaryClass = {astro-ph.SR},
       adsurl = {https://ui.adsabs.harvard.edu/abs/2023A&A...669A..43G}
}

@ARTICLE{Ezzeddine18,
       author = {{Ezzeddine}, R. and {Merle}, T. and {Plez}, B. and {Gebran}, M. and {Th{\'e}venin}, F. and {Van der Swaelmen}, M.},
        title = "{An empirical recipe for inelastic hydrogen-atom collisions in non-LTE calculations}",
      journal = {\aap},
         year = 2018,
        month = oct,
       volume = {618},
          eid = {A141},
        pages = {A141},
          doi = {10.1051/0004-6361/201630352},
       adsurl = {https://ui.adsabs.harvard.edu/abs/2018A&A...618A.141E}
}

@ARTICLE{Ernandes25,
       author = {{Ernandes}, H. and {Sk{\'u}lad{\'o}ttir}, {\'A}. and {Feltzing}, S. and {Feuillet}, D.},
        title = "{Disentangling Milky Way halo populations at low metallicities using [Al/Fe]}",
      journal = {\aap},
         year = 2025,
        month = nov,
       volume = {703},
          eid = {A256},
        pages = {A256},
          doi = {10.1051/0004-6361/202557149},
archivePrefix = {arXiv},
       eprint = {2509.06773},
 primaryClass = {astro-ph.GA},
       adsurl = {https://ui.adsabs.harvard.edu/abs/2025A&A...703A.256E}
}

@ARTICLE{Ardern-Arentsen2024,
       author = {{Ardern-Arentsen}, Anke and {Monari}, Giacomo and {Queiroz}, Anna B.~A. and {Starkenburg}, Else and {Martin}, Nicolas F. and {Chiappini}, Cristina and {Aguado}, David S. and {Belokurov}, Vasily and {Carlberg}, Ray and {Monty}, Stephanie and {Myeong}, GyuChul and {Schultheis}, Mathias and {Sestito}, Federico and {Venn}, Kim A. and {Vitali}, Sara and {Yuan}, Zhen and {Zhang}, Hanyuan and {Buder}, Sven and {Lewis}, Geraint F. and {Oliver}, William H. and {Wan}, Zhen and {Zucker}, Daniel B.},
        title = "{The Pristine Inner Galaxy Survey - VIII. Characterizing the orbital properties of the ancient, very metal-poor inner Milky Way}",
      journal = {\mnras},
         year = 2024,
        month = may,
       volume = {530},
       number = {3},
        pages = {3391-3411},
          doi = {10.1093/mnras/stae1049},
archivePrefix = {arXiv},
       eprint = {2312.03847},
 primaryClass = {astro-ph.GA},
       adsurl = {https://ui.adsabs.harvard.edu/abs/2024MNRAS.530.3391A}
}

@ARTICLE{Ryde2025,
       author = {{Ryde}, N. and {Nandakumar}, G. and {Albarrac{\'\i}n}, R. and {Schultheis}, M. and {Rojas-Arriagada}, A. and {Zoccali}, M.},
        title = "{Chemical abundances in the Milky Way's nuclear stellar disc}",
      journal = {\aap},
         year = 2025,
        month = jul,
       volume = {699},
          eid = {A176},
        pages = {A176},
          doi = {10.1051/0004-6361/202554791},
archivePrefix = {arXiv},
       eprint = {2505.15924},
 primaryClass = {astro-ph.GA},
       adsurl = {https://ui.adsabs.harvard.edu/abs/2025A&A...699A.176R}
}

@ARTICLE{nandakumar2025,
       author = {{Nandakumar}, Govind and {Ryde}, Nils and {Schultheis}, Mathias and {Rich}, R. Michael and {di Matteo}, Paola and {Thorsbro}, Brian and {Mace}, Gregory},
        title = "{The First Chemical Census of the Milky Way's Nuclear Star Cluster}",
      journal = {\apjl},
         year = 2025,
        month = mar,
       volume = {982},
       number = {1},
          eid = {L14},
        pages = {L14},
          doi = {10.3847/2041-8213/adbb6d},
archivePrefix = {arXiv},
       eprint = {2502.17756},
 primaryClass = {astro-ph.GA},
       adsurl = {https://ui.adsabs.harvard.edu/abs/2025ApJ...982L..14N}
}

@article{Hawkins15,
    author = {Hawkins, K. and Jofré, P. and Masseron, T. and Gilmore, G.},
    title = {Using chemical tagging to redefine the interface of the Galactic disc and halo},
    journal = {Monthly Notices of the Royal Astronomical Society},
    volume = {453},
    number = {1},
    pages = {758-774},
    year = {2015},
    month = {08},
    issn = {0035-8711},
    doi = {10.1093/mnras/stv1586},
    url = {https://doi.org/10.1093/mnras/stv1586},
    eprint = {https://academic.oup.com/mnras/article-pdf/453/1/758/4915800/stv1586.pdf},
}



\begin{appendix}

\section{Coordinates and orbital parameters}

We presents the astrometric, kinematic, and orbital properties of the analysed stars. We list Galactic coordinates, distances, proper motions, radial velocities, and the derived orbital parameters used to characterise the spatial distribution and dynamical behaviour of the metal-rich bulge spheroid sample.

\begin{table*}
\centering
\caption[5]{Coordinates, distances from \texttt{StarHorse} from \cite{queiroz21}, proper motions from \textit{Gaia} EDR3, radial velocities and orbit parameters from \cite{queiroz20,queiroz21}.}
\resizebox{0.9\textwidth}{!}{
\begin{tabular}{lc | cc  cc | c | cc | cccccc}
\noalign{\smallskip}
\hline
\noalign{\smallskip}
\hbox{ID} & \hbox{ID}  & l & b & RA & Dec. & d$_{\odot}$ & $\mu_{\alpha^*}^{8}$ & $\mu_{\delta}^{8}$ & RV & r$_{\rm min}$ & r$_{\rm max}$ & |z|$_{\rm max}$ & ecc \\  
\noalign{\smallskip}
int. & 2MASS & ($_{\circ}$) & ($_{\circ}$) & ($_{\circ}$)  & ($_{\circ}$)  & (kpc) & (mas yr$^{-1}$) & (mas yr$^{-1}$) & (km s$^{-1}$) & (kpc) & (kpc) & (kpc) & \\
\hline
\noalign{\smallskip}
%
a1  & 2M16252511-4103018 & 340.4649 & 5.728  & 246.355 & -41.051 & 8.52 & -0.30  & -6.00  & 75.46   $\pm$ 0.02 & 0.46 $\pm$ 0.19 & 3.99 $\pm$ 0.33 & 2.60 $\pm$ 0.31 & 0.80 $\pm$ 0.08 \\
a2  & 2M17031887-3421087 & 350.4563 & 4.394  & 255.829 & -34.352 & 8.31 & -3.96  & -7.59  & 76.58   $\pm$ 0.01 & 0.17 $\pm$ 0.11 & 1.74 $\pm$ 0.40 & 1.00 $\pm$ 0.20 & 0.86 $\pm$ 0.11 \\
a3  & 2M17043110-3433265 & 350.4434 & 4.072  & 256.130 & -34.557 & 10.06 & -1.95 & -5.85  & 122.26  $\pm$ 0.01 & 0.14 $\pm$ 0.06 & 3.40 $\pm$ 0.77 & 2.01 $\pm$ 0.44 & 0.92 $\pm$ 0.05 \\
a4  & 2M17055616-3439207 & 350.5409 & 3.780  & 256.484 & -34.656 & 9.30 & -1.48  & -6.07  & 28.51   $\pm$ 0.02 & 0.12 $\pm$ 0.06 & 2.23 $\pm$ 0.51 & 1.30 $\pm$ 0.34 & 0.90 $\pm$ 0.04 \\
a5  & 2M17133444-3041065 & 354.7044 & 4.834  & 258.394 & -30.685 & 8.97 & -1.10  & -3.24  & -57.59  $\pm$ 0.01 & 0.14 $\pm$ 0.06 & 1.89 $\pm$ 0.44 & 1.12 $\pm$ 0.33 & 0.87 $\pm$ 0.06 \\
a6  & 2M17140245-3106280 & 354.4180 & 4.507  & 258.510 & -31.108 & 6.91 & -5.88  & -5.74  & 39.52   $\pm$ 0.01 & 0.10 $\pm$ 0.05 & 1.67 $\pm$ 0.49 & 0.70 $\pm$ 0.20 & 0.90 $\pm$ 0.07 \\
a7  & 2M17153186-2426492 & 0.0855   & 8.078  & 258.883 & -24.447 & 6.37 & -8.08  & -4.96  & -99.56  $\pm$ 0.02 & 0.14 $\pm$ 0.05 & 2.94 $\pm$ 0.25 & 1.86 $\pm$ 0.23 & 0.91 $\pm$ 0.04 \\
a8  & 2M17163330-2808396 & 357.1634 & 5.774  & 259.139 & -28.144 & 7.83 & -4.80  & -4.77  & 101.64  $\pm$ 0.02 & 0.09 $\pm$ 0.03 & 1.22 $\pm$ 0.15 & 1.00 $\pm$ 0.06 & 0.86 $\pm$ 0.04 \\
a9  & 2M17163774-2824023 & 356.9618 & 5.613  & 259.157 & -28.401 & 7.29 & -3.78  & -4.23  & 129.54  $\pm$ 0.01 & 0.13 $\pm$ 0.05 & 1.61 $\pm$ 0.41 & 1.11 $\pm$ 0.14 & 0.88 $\pm$ 0.04 \\
a10 & 2M17173212-2407066 & 0.6206   & 7.889  & 259.384 & -24.119 & 6.75 & -3.54  & -8.07  & -59.82  $\pm$ 0.02 & 0.13 $\pm$ 0.07 & 2.10 $\pm$ 0.29 & 1.32 $\pm$ 0.08 & 0.88 $\pm$ 0.07 \\
a11 & 2M17192106-2638374 & 358.7585 & 6.120  & 259.838 & -26.644 & 7.11 & -4.66  & -6.97  & 48.49   $\pm$ 0.02 & 0.08 $\pm$ 0.03 & 1.45 $\pm$ 0.15 & 0.92 $\pm$ 0.07 & 0.89 $\pm$ 0.04 \\
a12 & 2M17192497-2650234 & 358.6043 & 5.997  & 259.854 & -26.840 & 5.94 & -7.37  & -7.96  & -177.41 $\pm$ 0.01 & 0.16 $\pm$ 0.06 & 3.87 $\pm$ 0.40 & 2.47 $\pm$ 0.49 & 0.92 $\pm$ 0.04 \\
a13 & 2M17193684-2748495 & 357.8234 & 5.408  & 259.904 & -27.814 & 4.92 & -5.22  & -10.55 & 13.62   $\pm$ 0.01 & 0.16 $\pm$ 0.06 & 3.93 $\pm$ 0.34 & 2.21 $\pm$ 0.52 & 0.92 $\pm$ 0.03 \\
a14 & 2M17251032-2636183 & 359.5251 & 5.065  & 261.293 & -26.605 & 7.00 & -6.32  & -6.72  & 30.15   $\pm$ 0.02 & 0.09 $\pm$ 0.04 & 1.44 $\pm$ 0.42 & 0.80 $\pm$ 0.15 & 0.88 $\pm$ 0.05 \\
a15 & 2M17263539-2035044 & 4.7620   & 8.113  & 261.647 & -20.585 & 4.91 & -3.82  & -10.12 & -68.80  $\pm$ 0.01 & 0.18 $\pm$ 0.09 & 3.83 $\pm$ 0.30 & 2.45 $\pm$ 0.14 & 0.92 $\pm$ 0.04 \\
a16 & 2M17280115-2829180 & 358.3045 & 3.495  & 262.005 & -28.488 & 7.16 & -2.01  & -8.27  & -52.94  $\pm$ 0.02 & 0.09 $\pm$ 0.05 & 1.53 $\pm$ 0.48 & 0.69 $\pm$ 0.17 & 0.87 $\pm$ 0.05 \\
a17 & 2M17285838-2956022 & 357.2130 & 2.523  & 262.243 & -29.934 & 6.08 & -5.22  & -8.84  & -79.69  $\pm$ 0.01 & 0.08 $\pm$ 0.04 & 2.40 $\pm$ 0.40 & 0.41 $\pm$ 0.28 & 0.93 $\pm$ 0.03 \\
a18 & 2M17322148-2419153 & 2.3351   & 4.963  & 263.090 & -24.321 & 5.94 & -5.12  & -8.00  & -12.99  $\pm$ 0.02 & 0.14 $\pm$ 0.11 & 2.81 $\pm$ 0.77 & 0.94 $\pm$ 0.39 & 0.89 $\pm$ 0.07 \\
a19 & 2M17324991-3019162 & 357.3470 & 1.613  & 263.208 & -30.321 & 7.70 & -1.97  & -7.25  & 183.17  $\pm$ 0.02 & 0.08 $\pm$ 0.04 & 1.56 $\pm$ 0.36 & 0.40 $\pm$ 0.18 & 0.91 $\pm$ 0.05 \\
a20 & 2M17344080-3326314 & 354.9402 & -0.411 & 263.670 & -33.442 & 8.17 & -6.52  & -0.70  & 23.80   $\pm$ 0.02 & 0.09 $\pm$ 0.04 & 1.34 $\pm$ 0.39 & 0.96 $\pm$ 0.27 & 0.88 $\pm$ 0.05 \\
a21 & 2M17344126-2652250 & 0.4670   & 3.1414 & 263.672 & -26.874 & 8.14 & -4.28  & -1.27  & -240.81 $\pm$ 0.02 & 0.21 $\pm$ 0.10 & 2.51 $\pm$ 0.19 & 1.16 $\pm$ 0.32 & 0.85 $\pm$ 0.06 \\
a22 & 2M17345655-2646140 & 0.5846   & 3.1491 & 263.736 & -26.771 & 7.62 & -6.14  & -5.29  & -95.22  $\pm$ 0.02 & 0.07 $\pm$ 0.06 & 1.01 $\pm$ 0.35 & 0.58 $\pm$ 0.05 & 0.86 $\pm$ 0.08 \\
a23 & 2M17351303-1836250 & 7.5473   & 7.4627 & 263.804 & -18.607 & 6.35 & -5.04  & -9.61  & -88.53  $\pm$ 0.01 & 0.17 $\pm$ 0.13 & 2.88 $\pm$ 0.30 & 1.87 $\pm$ 0.26 & 0.89 $\pm$ 0.07 \\
a24 & 2M17351307-2717212 & 0.1803   & 2.8178 & 263.804 & -27.289 & 9.19 & -0.93  & -4.92  & -3.94   $\pm$ 0.02 & 0.12 $\pm$ 0.05 & 1.14 $\pm$ 0.25 & 0.64 $\pm$ 0.11 & 0.83 $\pm$ 0.09 \\
a25 & 2M17363449-3926379 & 350.0845 & -3.965 & 264.144 & -39.444 & 10.69 & -2.34 & -2.93  & -21.01  $\pm$ 0.02 & 0.14 $\pm$ 0.06 & 3.98 $\pm$ 0.70 & 2.31 $\pm$ 0.44 & 0.93 $\pm$ 0.04 \\
a26 & 2M17383707-2056139 & 5.9821   & 5.551 & 264.654 & -20.937 & 8.94 & -6.55   & -1.77  & -60.01  $\pm$ 0.01 & 0.41 $\pm$ 0.43 & 2.65 $\pm$ 0.68 & 2.20 $\pm$ 0.58 & 0.72 $\pm$ 0.18 \\
a27 & 2M17384300-3904130 & 350.6229 & -4.117 & 264.679 & -39.070 & 9.25 & -6.91  & -0.61  & -1.97   $\pm$ 0.02 & 0.27 $\pm$ 0.30 & 3.75 $\pm$ 0.72 & 2.77 $\pm$ 0.68 & 0.87 $\pm$ 0.10 \\
a28 & 2M17394883-2208105 & 5.1066   & 4.683 & 264.953 & -22.136 & 7.97 & -4.45   & -6.83  & -51.50  $\pm$ 0.02 & 0.08 $\pm$ 0.03 & 1.08 $\pm$ 0.04 & 0.75 $\pm$ 0.02 & 0.86 $\pm$ 0.05 \\
a29 & 2M17410295-2228195 & 4.9711   &  4.263 & 265.262 & -22.472 & 8.06 & -3.53  & -1.65  & -155.79 $\pm$ 0.01 & 0.12 $\pm$ 0.19 & 1.86 $\pm$ 0.40 & 1.14 $\pm$ 0.41 & 0.89 $\pm$ 0.10 \\
a30 & 2M17412628-2218383 & 5.1563   &  4.271 & 265.360 & -22.311 & 9.28 & -3.18  & -3.79  & 3.06    $\pm$ 0.02 & 0.08 $\pm$ 0.06 & 1.57 $\pm$ 0.43 & 0.78 $\pm$ 0.26 & 0.90 $\pm$ 0.06 \\
a31 & 2M17414603-3335116 & 355.6079 & -1.737 & 265.442 & -33.587 & 8.62 & -2.94  & -2.56  & 8.39    $\pm$ 0.01 & 0.08 $\pm$ 0.04 & 1.08 $\pm$ 0.22 & 0.32 $\pm$ 0.07 & 0.86 $\pm$ 0.05 \\
a32 & 2M17422805-3410327 & 355.1833 & -2.171 & 265.617 & -34.176 & 9.96 & -1.38  & -4.00  & -6.49   $\pm$ 0.01 & 0.11 $\pm$ 0.05 & 1.96 $\pm$ 0.45 & 0.51 $\pm$ 0.38 & 0.91 $\pm$ 0.06 \\
a33 & 2M17430247-3503236 & 354.4950 & -2.734 & 265.760 & -35.057 & 7.45 & -3.15  & -8.40  & 8.46    $\pm$ 0.01 & 0.11 $\pm$ 0.04 & 1.13 $\pm$ 0.28 & 0.41 $\pm$ 0.03 & 0.82 $\pm$ 0.06 \\
a34 & 2M17440199-2238149 & 5.1913   &  3.589 & 266.008 & -22.637 & 9.66 & -3.18  & -2.26  & -18.90  $\pm$ 0.01 & 0.18 $\pm$ 0.08 & 2.39 $\pm$ 0.43 & 1.33 $\pm$ 0.27 & 0.86 $\pm$ 0.07 \\
a35 & 2M17463610-2336189 & 4.6705   &  2.582 & 266.650 & -23.605 & 8.13 & 0.87   & -4.83  & -160.26 $\pm$ 0.01 & 0.12 $\pm$ 0.09 & 1.52 $\pm$ 0.29 & 0.95 $\pm$ 0.36 & 0.86 $\pm$ 0.09 \\
a36 & 2M17474925-2211497 & 6.0229   &  3.069 & 266.955 & -22.197 & 7.10 & -4.29  & -2.69  & -162.59 $\pm$ 0.01 & 0.24 $\pm$ 0.30 & 2.34 $\pm$ 1.26 & 0.85 $\pm$ 0.59 & 0.85 $\pm$ 0.16 \\
a37 & 2M17503867-3706497 & 353.5216 & -5.110 & 267.661 & -37.114 & 8.85 & -4.00  & -1.60  & 2.18    $\pm$ 0.02 & 0.12 $\pm$ 0.04 & 1.85 $\pm$ 0.64 & 1.46 $\pm$ 0.40 & 0.89 $\pm$ 0.03 \\
a38 & 2M17504075-3704209 & 353.5609 & -5.095 & 267.670 & -37.072 & 8.20 & -3.74  & -3.52  & 53.18   $\pm$ 0.01 & 0.12 $\pm$ 0.07 & 1.37 $\pm$ 0.16 & 0.83 $\pm$ 0.11 & 0.85 $\pm$ 0.08 \\
a39 & 2M17505373-1803210 & 9.9599   &  4.565 & 267.724 & -18.056 & 9.53 & -5.62  & -4.24  & -100.58 $\pm$ 0.02 & 0.21 $\pm$ 0.51 & 2.90 $\pm$ 1.37 & 2.05 $\pm$ 0.81 & 0.85 $\pm$ 0.11 \\
a40 & 2M17520538-1822315 & 9.8276   &  4.159 & 268.022 & -18.375 & 7.51 & -4.96  & -4.32  & -78.31  $\pm$ 0.01 & 0.11 $\pm$ 0.06 & 1.71 $\pm$ 0.08 & 0.79 $\pm$ 0.12 & 0.88 $\pm$ 0.06 \\
a41 & 2M17525971-1912041 & 9.2223   &  3.557 & 268.249 & -19.201 & 7.85 & -2.40  & -5.30  & -111.46 $\pm$ 0.01 & 0.14 $\pm$ 0.13 & 1.51 $\pm$ 0.16 & 0.56 $\pm$ 0.08 & 0.84 $\pm$ 0.13 \\
a42 & 2M17530359-2258461 & 5.9688   &  1.629 & 268.265 & -22.979 & 6.66 & -2.87  & -8.32  & -121.52 $\pm$ 0.01 & 0.09 $\pm$ 0.04 & 1.82 $\pm$ 0.07 & 0.32 $\pm$ 0.01 & 0.91 $\pm$ 0.04 \\
a43 & 2M17531682-2104467 & 7.6335   &  2.549 & 268.320 & -21.080 & 8.38 & -5.16  & -5.19  & -102.63 $\pm$ 0.02 & 0.22 $\pm$ 0.08 & 1.29 $\pm$ 0.17 & 0.57 $\pm$ 0.15 & 0.72 $\pm$ 0.11 \\
a44 & 2M17540360-3638369 & 354.2747 & -5.464 & 268.515 & -36.644 & 9.75 & -2.03  & -4.67  & 56.91   $\pm$ 0.01 & 0.12 $\pm$ 0.05 & 1.90 $\pm$ 0.68 & 1.20 $\pm$ 0.26 & 0.87 $\pm$ 0.04 \\
a45 & 2M17591751-3116052 & 359.4944 & -3.731 & 269.823 & -31.268 & 9.11 & -2.85  & -3.44  & 72.27   $\pm$ 0.02 & 0.11 $\pm$ 0.05 & 1.27 $\pm$ 0.48 & 0.80 $\pm$ 0.29 & 0.86 $\pm$ 0.05 \\
a46 & 2M18000162-3047385 & 359.9853 & -3.633 & 270.007 & -30.794 & 8.84 & -3.72  & -2.59  & 149.87  $\pm$ 0.01 & 0.12 $\pm$ 0.08 & 1.52 $\pm$ 0.32 & 0.95 $\pm$ 0.26 & 0.87 $\pm$ 0.09 \\
a47 & 2M18011040-2844047 & 1.9020   & -2.833 & 270.293 & -28.735 & 7.98 & -2.01  & -7.57  & -172.08 $\pm$ 0.01 & 0.07 $\pm$ 0.03 & 1.19 $\pm$ 0.24 & 0.49 $\pm$ 0.05 & 0.88 $\pm$ 0.04 \\
a48 & 2M18013822-2727488 & 3.0591   & -2.294 & 270.409 & -27.464 & 6.25 & -6.01  & -8.47  & -24.28  $\pm$ 0.01 & 0.09 $\pm$ 0.04 & 1.96 $\pm$ 0.57 & 0.36 $\pm$ 0.23 & 0.90 $\pm$ 0.04 \\
a49 & 2M18014518-2752453 & 2.7096   & -2.522 & 270.438 & -27.879 & 8.76 & -3.45  & -6.35  & -185.77 $\pm$ 0.01 & 0.08 $\pm$ 0.04 & 1.43 $\pm$ 0.51 & 0.44 $\pm$ 0.18 & 0.90 $\pm$ 0.04 \\
a50 & 2M18014962-3002151 & 0.8365   & -3.598 & 270.457 & -30.038 & 8.58 & -2.76  & -3.07  & 156.81  $\pm$ 0.01 & 0.10 $\pm$ 0.04 & 1.46 $\pm$ 0.34 & 0.72 $\pm$ 0.29 & 0.89 $\pm$ 0.04 \\
a51 & 2M18022293-2850057 & 1.9452   & -3.113 & 270.596 & -28.835 & 8.98 & -0.25  & -5.00  & -253.62 $\pm$ 0.02 & 0.17 $\pm$ 0.08 & 2.59 $\pm$ 0.73 & 1.73 $\pm$ 0.64 & 0.88 $\pm$ 0.05 \\
a52 & 2M18031846-3000497 & 1.0146   & -3.866 & 270.827 & -30.014 & 9.09 & -3.26  & -4.53  & -261.25 $\pm$ 0.01 & 0.13 $\pm$ 0.06 & 2.53 $\pm$ 1.08 & 1.43 $\pm$ 0.73 & 0.92 $\pm$ 0.05 \\
a53 & 2M18032107-2959127 & 1.0428   & -3.861 & 270.838 & -29.987 & 7.23 & -4.45  & -6.34  & -187.40 $\pm$ 0.01 & 0.11 $\pm$ 0.05 & 1.78 $\pm$ 0.34 & 0.66 $\pm$ 0.12 & 0.89 $\pm$ 0.05 \\
a54 & 2M18034066-3002198 & 1.0319   & -3.948 & 270.919 & -30.039 & 7.41 & -3.58  & -6.44  & -196.85 $\pm$ 0.01 & 0.09 $\pm$ 0.08 & 1.59 $\pm$ 0.85 & 0.66 $\pm$ 0.40 & 0.90 $\pm$ 0.03 \\
a55 & 2M18034548-3122086 & 359.8756 & -4.612 & 270.940 & -31.369 & 8.55 & -1.64  & -3.41  & -263.85 $\pm$ 0.01 & 0.20 $\pm$ 0.15 & 2.87 $\pm$ 0.64 & 1.72 $\pm$ 0.48 & 0.88 $\pm$ 0.09 \\
a56 & 2M18034880-3002202 & 1.0462   & -3.974 & 270.953 & -30.039 & 6.95 & -5.10  & -7.59  & -32.72  $\pm$ 0.01 & 0.11 $\pm$ 0.07 & 1.48 $\pm$ 0.33 & 0.55 $\pm$ 0.03 & 0.87 $\pm$ 0.06 \\
a57 & 2M18041554-3001431 & 1.1024   & -4.053 & 271.065 & -30.029 & 9.29 & -5.10  & -0.84  & 113.43  $\pm$ 0.02 & 0.20 $\pm$ 0.24 & 3.14 $\pm$ 0.88 & 2.33 $\pm$ 0.61 & 0.87 $\pm$ 0.09 \\
a58 & 2M18041724-3132539 & 359.7734 & -4.797 & 271.072 & -31.548 & 8.49 & 1.14   & -3.57  & 120.32  $\pm$ 0.02 & 0.14 $\pm$ 0.07 & 1.62 $\pm$ 0.52 & 1.37 $\pm$ 0.36 & 0.85 $\pm$ 0.06 \\
a59 & 2M18044803-2752467 & 3.0402   & -3.109 & 271.200 & -27.880 & 6.52 & -4.21  & -7.84  & -152.01 $\pm$ 0.02 & 0.09 $\pm$ 0.04 & 2.08 $\pm$ 0.56 & 0.46 $\pm$ 0.29 & 0.91 $\pm$ 0.04 \\
a60 & 2M18050144-3005149 & 1.1316   & -4.226 & 271.256 & -30.087 & 6.33 & -4.43  & -6.94  & -134.30 $\pm$ 0.01 & 0.11 $\pm$ 0.06 & 2.16 $\pm$ 0.62 & 0.65 $\pm$ 0.26 & 0.90 $\pm$ 0.06 \\
a61 & 2M18050235-3002348 & 1.1721   & -4.208 & 271.260 & -30.043 & 7.44 & -3.25  & -9.78  & 180.06  $\pm$ 0.01 & 0.11 $\pm$ 0.03 & 2.18 $\pm$ 0.06 & 0.85 $\pm$ 0.02 & 0.90 $\pm$ 0.03 \\
a62 & 2M18051997-2923441 & 1.7704   & -3.949 & 271.333 & -29.396 & 8.68 & 4.01   & -1.99  & -29.30  $\pm$ 0.01 & 0.21 $\pm$ 0.27 & 3.66 $\pm$ 0.95 & 2.46 $\pm$ 0.67 & 0.88 $\pm$ 0.09 \\
a63 & 2M18052433-3036305 & 0.7146   & -4.551 & 271.351 & -30.608 & 7.65 & -5.44  & -5.16  & -192.49 $\pm$ 0.01 & 0.13 $\pm$ 0.05 & 1.76 $\pm$ 0.36 & 0.99 $\pm$ 0.14 & 0.88 $\pm$ 0.04 \\
a64 & 2M18061581-2748521 & 3.2549   & -3.360 & 271.566 & -27.814 & 9.90 & -3.16  & -2.47  & -202.05 $\pm$ 0.01 & 0.24 $\pm$ 0.14 & 3.40 $\pm$ 0.96 & 2.32 $\pm$ 0.62 & 0.87 $\pm$ 0.07 \\
a65 & 2M18073289-3156045 & 359.7692 & -5.591 & 271.887 & -31.935 & 7.69 & -3.26  & -7.25  & -262.44 $\pm$ 0.01 & 0.19 $\pm$ 0.08 & 2.83 $\pm$ 0.17 & 1.60 $\pm$ 0.10 & 0.87 $\pm$ 0.05 \\
a66 & 2M18074173-3153200 & 359.8246 & -5.596 & 271.924 & -31.889 & 8.53 & -2.62  & -4.59  & 62.66   $\pm$ 0.01 & 0.08 $\pm$ 0.04 & 1.05 $\pm$ 0.26 & 0.89 $\pm$ 0.10 & 0.87 $\pm$ 0.06 \\
a67 & 2M18074493-3145155 & 359.9488 & -5.542 & 271.937 & -31.754 & 9.95 & -0.53  & -4.50  & 77.42   $\pm$ 0.01 & 0.15 $\pm$ 0.06 & 2.55 $\pm$ 1.06 & 1.86 $\pm$ 0.52 & 0.89 $\pm$ 0.04 \\
a68 & 2M18075097-3202528 & 359.6999 & -5.701 & 271.962 & -32.048 & 9.61 & -2.83  & -4.17  & 6.94    $\pm$ 0.01 & 0.11 $\pm$ 0.06 & 1.63 $\pm$ 0.59 & 1.15 $\pm$ 0.21 & 0.88 $\pm$ 0.05 \\
a69 & 2M18080123-3143583 & 359.9955 & -5.582 & 272.005 & -31.733 & 8.16 & -2.91  & -6.84  & 185.71  $\pm$ 0.02 & 0.12 $\pm$ 0.07 & 1.89 $\pm$ 0.27 & 1.12 $\pm$ 0.11 & 0.88 $\pm$ 0.05 \\
a70 & 2M18084466-2529453 & 5.5551   & -2.722 & 272.186 & -25.496 & 6.01 & -1.67  & -9.35  & -50.35  $\pm$ 0.01 & 0.13 $\pm$ 0.05 & 2.27 $\pm$ 0.59 & 0.53 $\pm$ 0.37 & 0.90 $\pm$ 0.03 \\
a71 & 2M18084995-3124422 & 0.3620   & -5.581 & 272.208 & -31.412 & 9.73 & -0.85  & -3.52  & -47.46  $\pm$ 0.02 & 0.13 $\pm$ 0.10 & 1.90 $\pm$ 0.87 & 1.46 $\pm$ 0.44 & 0.88 $\pm$ 0.05 \\
a72 & 2M18093136-2559533 & 5.1992   & -3.117 & 272.381 & -25.998 & 7.12 & -5.50  & -8.02  & -67.81  $\pm$ 0.01 & 0.08 $\pm$ 0.05 & 1.36 $\pm$ 0.43 & 0.48 $\pm$ 0.12 & 0.88 $\pm$ 0.05 \\
a73 & 2M18112625-2527164 & 5.8842   & -3.235 & 272.859 & -25.455 & 6.97 & -6.98  & -6.00  & -73.10  $\pm$ 0.01 & 0.11 $\pm$ 0.07 & 1.59 $\pm$ 0.45 & 0.71 $\pm$ 0.25 & 0.87 $\pm$ 0.05 \\
a74 & 2M18125891-2610195 & 5.4187   & -3.883 & 273.245 & -26.172 & 9.68 & -4.06  & -2.67  & -134.81 $\pm$ 0.01 & 0.17 $\pm$ 0.12 & 2.82 $\pm$ 0.84 & 2.11 $\pm$ 0.41 & 0.89 $\pm$ 0.10 \\
a75 & 2M18163725-3253337 & 359.8229 & -7.731 & 274.155 & -32.893 & 7.28 & -0.90  & -9.01  & 9.31    $\pm$ 0.00 & 0.12 $\pm$ 0.06 & 1.51 $\pm$ 0.29 & 1.36 $\pm$ 0.05 & 0.86 $\pm$ 0.07 \\
a76 & 2M18191892-2203579 & 9.7281   & -3.214 & 274.829 & -22.066 & 7.56 & -0.79  & -6.45  & -127.82 $\pm$ 0.01 & 0.15 $\pm$ 0.10 & 1.75 $\pm$ 0.12 & 0.62 $\pm$ 0.06 & 0.82 $\pm$ 0.09 \\
a77 & 2M18203837-2235146 & 9.4107   & -3.729 & 275.160 & -22.587 & 7.66 & -4.73  & -6.10  & -72.35  $\pm$ 0.02 & 0.09 $\pm$ 0.04 & 1.53 $\pm$ 0.18 & 0.58 $\pm$ 0.02 & 0.89 $\pm$ 0.04 \\
a78 & 2M18365761-2322171 & 10.4317  & -7.429 & 279.240 & -23.371 & 7.21 & -2.00  & -5.96  & -129.66 $\pm$ 0.02 & 0.15 $\pm$ 0.05 & 2.43 $\pm$ 0.28 & 1.50 $\pm$ 0.14 & 0.88 $\pm$ 0.04 \\
\hline
\noalign{\smallskip}
\label{dynamics}
\end{tabular}}
\end{table*}

\section{Derived abundances}

We provide the individual chemical abundances derived in this work for all stars in the sample. The table includes the final abundance ratios adopted in the analysis of the abundance trends discussed in this work.

\begin{table*}[]
\small
\scalefont{0.45}
\caption{Abundances derived in the present work.}
\label{tab:abonds}
\resizebox{0.95\textwidth}{!}{
\centering
\begin{tabular}{lccccccccccccc}
\hline
ID & [Fe/H] & [C/Fe] & [N/Fe] & [O/Fe] & [Al/Fe]$_{LTE}$ & [Al/Fe]$_{non-LTE}$  &  [P/Fe] & [S/Fe] & [K/Fe] &  [Mn/Fe] &  [Ce/Fe] \\
\hline
a1  & $-$0.36 & +0.10 & +0.00 & +0.25 & 0.29 & -0.14 & +0.25 & +0.15 & +0.15  & -0.35 & +0.25 \\
a2  & $-$0.73 & +0.35 & +0.40 & +0.65 & 0.18 & -0.20 & +0.80 & +0.40 & +0.18  & -0.12 & +0.15 \\
a3  & $-$0.71 & +0.25 & +0.45 & +0.65 & 0.35 & -0.24 & ---   & +0.50 & +0.25  & -0.35 & +0.40 \\
a4  & $-$0.73 & +0.35 & +0.40 & +0.65 & 0.22 & -0.14 & +0.90 & +0.45 & +0.19  & -0.20 & +0.50 \\
a5  & $+$0.40 & +0.38 & +1.00 & +0.75 & 0.31 & -0.30 & +0.25 & ---   & +0.55  & +0.35 & -0.28 \\
a6  & $-$0.28 & +0.15 & +0.40 & +0.50 & 0.32 & -0.18 & +0.30 & +0.35 & +0.23  & +0.00 & +0.00 \\
a7  & $-$0.63 & +0.40 & +0.30 & +0.70 & 0.28 & -0.18 & ---   & +0.40 & ---    & -0.04 & +0.50 \\
a8  & $-$0.73 & +0.15 & +0.90 & +0.55 & 0.25 & -0.20 & +0.60 & +0.60 & +0.20  & -0.20 & +0.80 \\
a9  & $-$0.71 & +0.25 & +0.25 & +0.70 & 0.25 & -0.15 & ---   & +0.30 & +0.32  & -0.30 & +0.50 \\
a10 & $-$0.77 & +0.27 & +0.10 & +0.55 & 0.29 &  ---  & +0.30 & +0.40 & +0.05  & -0.30 & +0.40 \\
a11 & $-$0.56 & +0.17 & +0.30 & +0.55 & 0.39 &  ---  & +0.20 & +0.35 & +0.10  & -0.25 & +0.25 \\
a12 & $-$0.38 & +0.30 & +0.20 & +0.45 & 0.28 & -0.23 & ---   & +0.22 & +0.10  & -0.07 & +0.15 \\
a13 & $-$0.66 & +0.30 & +0.20 & +0.60 & 0.32 &  ---  & ---   & +0.35 & +0.07  & -0.28 & +0.20 \\
a14 & $-$0.26 & +0.30 & +0.20 & +0.50 & 0.23 & -0.30 & +0.60 & +0.30 & +0.14  & -0.12 & +0.25 \\
a15 & $-$0.72 & +0.30 & +0.40 & +0.55 & 0.32 & -0.20 & +0.40 & +0.55 & +0.18  & -0.25 & +0.55 \\
a16 & $-$0.79 & +0.40 & +0.30 & +0.75 & 0.00 & -0.30 & +0.45 & +0.60 & +0.21  & -0.20 & +0.30 \\
a17 & $+$0.03 & +0.10 & +0.10 & +0.30 & 0.15 & -0.26 & +0.35 & +0.25 & +0.05  & +0.00 & +0.30 \\
a18 & $-$0.48 & +0.13 & +0.45 & +0.50 & 0.37 & -0.13 & +0.25 & +0.40 & +0.15  & -0.20 & +0.20 \\
a19 & $-$0.61 & +0.25 & +0.10 & +0.50 & 0.00 & -0.38 &  ---  & +0.20 & +0.08  & -0.15 & +0.00 \\
a20 & $-$0.55 & +0.28 & +0.20 & +0.45 & 0.10 & -0.37 & +0.35 & +0.35 & +0.15  & -0.15 & +0.15 \\
a21 & $-$0.31 & +0.20 & +0.25 & +0.35 & 0.25 & -0.18 & +0.70 & +0.20 & -0.06  & -0.20 & +0.10 \\
a22 & $-$0.64 & +0.30 & +0.20 & +0.50 & 0.18 & -0.29 & ---   & +0.45 & +0.15  & -0.35 & +0.40 \\
a23 & $-$0.72 & +0.42 & +0.30 & +0.64 & 0.28 & -0.30 & +0.60 & +1.00 & +0.35  & +0.00 & +0.50 \\
a24 & $-$0.21 & +0.35 & +0.20 & +0.30 & 0.00 & -0.37 & +0.50 & +0.24 & +0.00  & -0.03 & +0.10 \\
a25 & $-$0.29 & +0.07 & +0.10 & +0.60 & 0.07 & -0.30 & +0.45 & +0.25 & +0.12  & -0.15 & +0.20 \\
a26 & $+$0.01 & +0.11 & +0.00 & +0.32 & 0.20 & -0.31 & +0.10 & +0.35 & +0.00  & +0.00 & +0.20 \\
a27 & $-$0.50 & +0.40 & +0.10 & +0.45 & 0.19 & -0.38 & +0.85 & +0.15 & +0.05  & -0.20 & +0.15 \\
a28 & $-$0.81 & +0.21 & +0.00 & +0.40 & 0.20 &  ---  & +0.45 & +0.20 & +0.15  & -0.30 & +0.15 \\
a29 & $+$0.24 & +0.25 & +0.00 & +0.20 & 0.20 & -0.28 & ---   & +0.35 & +0.00  & +0.00 & -0.10 \\
a30 & $-$0.58 & +0.30 & +0.20 & +0.40 & 0.22 & -0.14 & +0.75 & +0.28 & +0.16  & -0.25 & +0.25 \\
a31 & $+$0.26 & +0.17 & +0.15 & +0.20 & 0.17 &   --- & +0.15 & +0.15 & +0.08  & +0.00 & -0.10 \\
a32 & $+$0.51 & +0.36 & +0.52 & +0.50 & 0.15 & -0.33 &  ---  & ---   & +0.25  & +0.15 & -0.25 \\
a33 & $-$0.21 & +0.20 & +0.00 & +0.20 & 0.21 & -0.31 &$<$0.40& +0.10 & +0.00  & +0.00 & +0.25 \\
a34 & $+$0.22 & +0.40 & +0.00 & +0.50 & 0.24 & -0.38 & ---   & +0.70 & +0.00  & +0.00 & -0.10 \\
a35 & $-$0.25 & +0.25 & +0.10 & +0.40 & 0.36 & -0.23 & ---   & +0.30 & +0.06  & -0.15 & +0.30 \\
a36 & $+$0.42 & +0.35 & +0.40 & +0.50 & 0.23 & -0.34 & ---   & ---   & +0.03  & +0.05 & -0.30 \\
a37 & $-$0.65 & +0.19 & +0.00 & +0.40 & 0.25 & -0.13 & +0.55 & +0.60 & +0.28  & -0.25 & +0.10 \\
a38 & $-$0.17 & +0.22 & +0.30 & +0.30 & 0.15 &  ---  & +0.40 & +0.00 & +0.04  & -0.05 & +0.10 \\
a39 & $-$0.71 & +0.15 & +0.40 & +0.60 & 0.29 & -0.16 & ---   & +0.50 & -0.05  & -0.30 & +0.40 \\
a40 & $+$0.02 & +0.15 & +0.05 & +0.20 & 0.23 & -0.28 & +0.25 & +0.15 & +0.05  & +0.00 & +0.00 \\
a41 & $+$0.32 & +0.23 & +0.23 & +0.18 & 0.24 & -0.34 &  ---  & ---   & +0.08  & +0.23 & -0.15 \\
a42 & $+$0.41 & +0.30 & +0.20 & +0.60 & 0.15 & -0.39 & ---   & ---   & +0.05  & +0.18 & -0.35 \\
a43 & $-$0.74 & +0.35 & +0.40 & +0.60 & 0.14 & -0.35 & +0.40 & +0.65 & +0.00  & -0.20 & +0.50 \\
a44 & $-$0.27 & +0.20 & +0.00 & +0.35 & 0.18 & -0.29 &  ---  & +0.10 & +0.17  & -0.24 & +0.15 \\
a45 & $-$0.52 & +0.22 & +0.00 & +0.45 & 0.36 & -0.07 & +0.80 & +0.18 & +0.15  & -0.30 & +0.25 \\
a46 & $+$0.14 & +0.25 & +0.20 & +0.50 & 0.28 & -0.27 &  ---  & +0.45 & +0.25  & +0.10 & +0.20 \\
a47 & $-$0.66 & +0.25 & +0.00 & +0.40 & 0.30 & -0.18 & ---   & +0.80 & +0.25  & -0.25 & +0.25 \\
a48 & $-$0.78 & +0.25 & +0.00 & +0.50 & 0.15 & -0.26 & +0.00 & +0.34 & +0.12  & -0.30 & +0.35 \\
a49 & $-$0.73 & +0.35 & +0.70 & +0.70 &-0.00 & ---   & ---   & +0.50 & +0.08  & -0.26 & +0.50 \\
a50 & $+$0.22 & +0.25 & +0.00 & +0.60 & 0.35 & -0.20 & +0.10 & +0.50 & +0.45  & +0.21 & -0.15 \\
a51 & $-$0.36 & +0.20 & +0.00 & +0.30 & 0.33 & -0.15 & +1.00 & +0.30 & +0.15  & -0.24 & +0.10 \\
a52 & $-$0.07 & +0.25 & +0.00 & +0.35 & 0.35 & -0.20 & ---   & +0.40 & +0.00  & -0.06 & +0.00 \\
a53 & $+$0.07 & +0.15 & +0.00 & +0.20 & 0.25 & -0.23 & ---   & +0.15 & +0.15  & +0.00 & +0.10 \\
a54 & $+$0.32 & +0.43 & +0.68 & +0.58 & 0.23 & -0.28 & ---   & ---   & +0.00  & +0.36 & -0.20 \\
a55 & $+$0.29 & +0.46 & +0.31 & +0.62 & 0.26 & -0.25 & ---   & +0.20 & +0.21  & +0.11 & +0.11 \\
a56 & $+$0.15 & +0.15 & +0.20 & +0.20 & 0.10 & -0.30 & +0.15 & +0.17 & -0.05  & +0.00 & +0.15 \\
a57 & $-$0.61 & +0.20 & +0.00 & +0.40 & 0.34 & -0.18 & +0.45 & +0.60 & +0.12  & -0.25 & +0.30 \\
a58 & $-$0.50 & +0.32 & +0.20 & +0.40 & 0.38 & -0.08 & +0.25 & +0.25 & +0.10  & -0.18 & +0.20 \\
a59 & $-$0.53 & +0.10 & +0.20 & +0.35 & 0.22 & -0.18 & ---   & +0.30 & +0.15  & -0.27 & +0.15 \\
a60 & $-$0.45 & +0.00 & +0.20 & +0.30 & 0.10 & -0.30 & ---   & +0.28 & +0.08  & -0.19 & +0.15 \\
a61 & $-$0.13 & +0.15 & +0.00 & +0.20 & 0.24 & -0.36 & +0.35 & +0.22 & +0.08  & -0.20 & +0.10 \\
a62 & $+$0.18 & +0.20 & +0.10 & +0.20 & 0.00 & -0.22 & +0.45 & +0.15 & +0.05  & +0.00 & -0.40 \\
a63 & $+$0.28 & +0.35 & +0.20 & +0.35 & 0.31 & -0.15 & ---   &  ---  & +0.18  & +0.32 & -0.35 \\
a64 & $-$0.57 & +0.15 & +0.20 & +0.45 & 0.27 & -0.19 & ---   & +0.45 & +0.00  & -0.34 & +0.00 \\
a65 & $-$0.17 & +0.25 & +0.20 & +0.30 & 0.51 & -0.13 & ---   & +0.30 & +0.15  & -0.10 & +0.00 \\
a66 & $-$0.70 & +0.25 & +0.45 & +0.65 & 0.20 & -0.25 & +0.55 & +0.35 & +0.025 & -0.07 & +0.55 \\
a67 & $-$0.21 & +0.25 & +0.20 & +0.30 & 0.33 & -0.23 & +0.05 & -0.35 & +0.20  & +0.00 & +0.20 \\
a68 & $+$0.15 & +0.25 & +0.00 & +0.20 & 0.08 & -0.38 & +0.25 & +0.40 & +0.00  & +0.00 & +0.00 \\
a69 & $-$0.57 & +0.20 & +0.00 & +0.40 & 0.38 & -0.10 & +0.05 & +0.30 & +0.05  & -0.35 & +0.10 \\
a70 & $-$0.45 & +0.30 & +0.00 & +0.40 & 0.36 & -0.14 & +0.10 & +0.35 & +0.25  & -0.20 & +0.20 \\
a71 & $-$0.69 & +0.25 & +0.30 & +0.65 & 0.50 & -0.05 & +0.60 & +0.60 & +0.30  & -0.20 & +0.50 \\
a72 & $-$0.73 & +0.43 & +0.20 & +0.70 & 0.31 & -0.20 & +0.25 & +0.70 & +0.25  & -0.11 & +0.65 \\
a73 & $-$0.51 & +0.25 & +0.00 & +0.40 & 0.30 & -0.08 & +0.35 & +0.30 &$-$0.7  & -0.18 & +0.15 \\
a74 & $+$0.17 & +0.20 & +0.00 & +0.30 & 0.29 & -0.20 & +0.15 & ---   & +0.12  & +0.00 & +0.10 \\
a75 & $-$0.79 & +0.20 & +0.10 & +0.45 & 0.20 & -0.18 & +0.20 & +0.29 & +0.10  & -0.35 & +0.40 \\
a76 & $+$0.00 & +0.20 & +0.00 & +0.20 & 0.13 & -0.25 & +0.15 & +0.00 & +0.00  & +0.00 & +0.00 \\
a77 & $-$0.55 & +0.15 & +0.00 & +0.42 & 0.19 & -0.23 & +0.45 & +0.50 & +0.15  & -0.30 & +0.10 \\
a78 & $-$0.72 & +0.35 & +0.40 & +0.70 & 0.32 & -0.35 & +0.60 & +0.70 & +0.23  & -0.20 & +0.55 \\    
\hline         
    \end{tabular}}
\end{table*}

\end{appendix}

\end{document}